\documentclass[final,twocolumn,natbib]{svjour3}

\usepackage{graphicx}
\usepackage{mathptmx}      

\newcommand{\lb}{\label}
\newcommand{\be}{\begin{equation}}
\newcommand{\ee}{\end{equation}}
\newcommand{\ber}{\begin{eqnarray}}
\newcommand{\eer}{\end{eqnarray}}
\newcommand{\bers}{\begin{eqnarray*}}
\newcommand{\eers}{\end{eqnarray*}}
\newcommand{\bea}{\begin{eqnarray}}
\newcommand{\eea}{\end{eqnarray}}
\def\bratio#1#2{\left(\frac{#1}{#2}\right)}
\newcommand{\beq}{\begin{equation}}
\newcommand{\eeq}{\end{equation}}
\newcommand{\D}{\Delta}
\newcommand{\p}{\partial}

\journalname{Space Sci Rev}

\begin{document}

\title{Turbulence, Magnetic Reconnection in Turbulent Fluids and Energetic
       Particle Acceleration}

\titlerunning{Fast Reconnection in Turbulence and Acceleration}

\author{A. Lazarian \and L. Vlahos \and G. Kowal \and H. Yan \and A. Beresnyak \and E.~M. de Gouveia Dal Pino}

\authorrunning{Lazarian et al.}

\institute{A. Lazarian \at
           Department of Astronomy, University of Wisconsin-Madison,
           475 N. Charter St., Madison, WI, 53706, USA \\
              Tel.: +1-608-262-1715\\
              Fax: +1-608-263-6386\\
           \email{lazarian@astro.wisc.edu}
           \and
           L. Vlahos \at
           Department of Physics, University of Thessaloniki,
           54124 Thessaloniki, Greece\\
           \email{vlahos@astro.auth.gr}
           \and
           G. Kowal \at
           Instituto de Astronomia, Geof\'\i sica e Ci\^encias Atmosf\'ericas,
           Universidade de S\~ao Paulo,
           Rua do Mat\~ao, 1226 -- Cidade Universit\'{a}ria, CEP 05508-090,
           S\~ao Paulo/SP, Brazil\\
           \email{kowal@astro.iag.usp.br}
           \and
           H. Yan \at
           Kavli Institute, Peking University, 5 Yi He Yaun Rd, Bejing, 100871 China\\
           \email{hryan@pku.edu.cn}
           \and
           A. Beresnyak \at
           Los Alamos Laboratory, Los Alamos, NM, 87545, USA\\
           and\\
           Ruhr-Universitat University, Bochum,  44789, Bochum, Germany\\
           \email{beresnyak@wisc.edu}
           \and
           E.~M. de Gouveia Dal Pino \at
           Instituto de Astronomia, Geof\'\i sica e Ci\^encias Atmosf\'ericas,
           Universidade de S\~ao Paulo,
           Rua do Mat\~ao, 1226 -- Cidade Universit\'{a}ria, CEP 05508-090,
           S\~ao Paulo/SP, Brazil\\
           \email{dalpino@astro.iag.usp.br}
}

\date{Received: date / Accepted: date}

\maketitle

\begin{abstract}
Turbulence is ubiquitous in astrophysics.  It radically changes many
astrophysical phenomena, in particular, the propagation and acceleration of
cosmic rays.  We present the modern understanding of compressible
magnetohydrodynamic (MHD) turbulence, in particular its decomposition into
Alfv\'en, slow and fast modes, discuss the density structure of turbulent
subsonic and supersonic media, as well as other relevant regimes of
astrophysical turbulence.  All this information is essential for understanding
the energetic particle acceleration that we discuss further in the review.  For
instance, we show how fast and slow modes accelerate energetic particles through
the second order Fermi acceleration, while density fluctuations generate
magnetic fields in pre-shock regions enabling the first order Fermi acceleration
of high energy cosmic rays.  Very importantly, however, the first order Fermi
cosmic ray acceleration is also possible in sites of magnetic reconnection.  In
the presence of turbulence this reconnection gets fast and we present numerical
evidence supporting the predictions of the Lazarian \& Vishniac (1999) model of
fast reconnection.  The efficiency of this process suggests that magnetic
reconnection can release substantial amounts of energy in short periods of time.
As the particle tracing numerical simulations show that the particles can be
efficiently accelerated during the reconnection, we argue that the process of
magnetic reconnection may be much more important for particle acceleration than
it is currently accepted.  In particular, we discuss the acceleration arising
from reconnection as a possible origin of the anomalous cosmic rays measured by
Voyagers as well as the origin cosmic ray excess in the direction of Heliotail.
\keywords{Turbulence \and Magnetic reconnection \and Acceleration \and Cosmic rays}
\end{abstract}

\section{Introduction}
\label{sec:intro}

\begin{figure}
\centering
\includegraphics[width=\columnwidth]{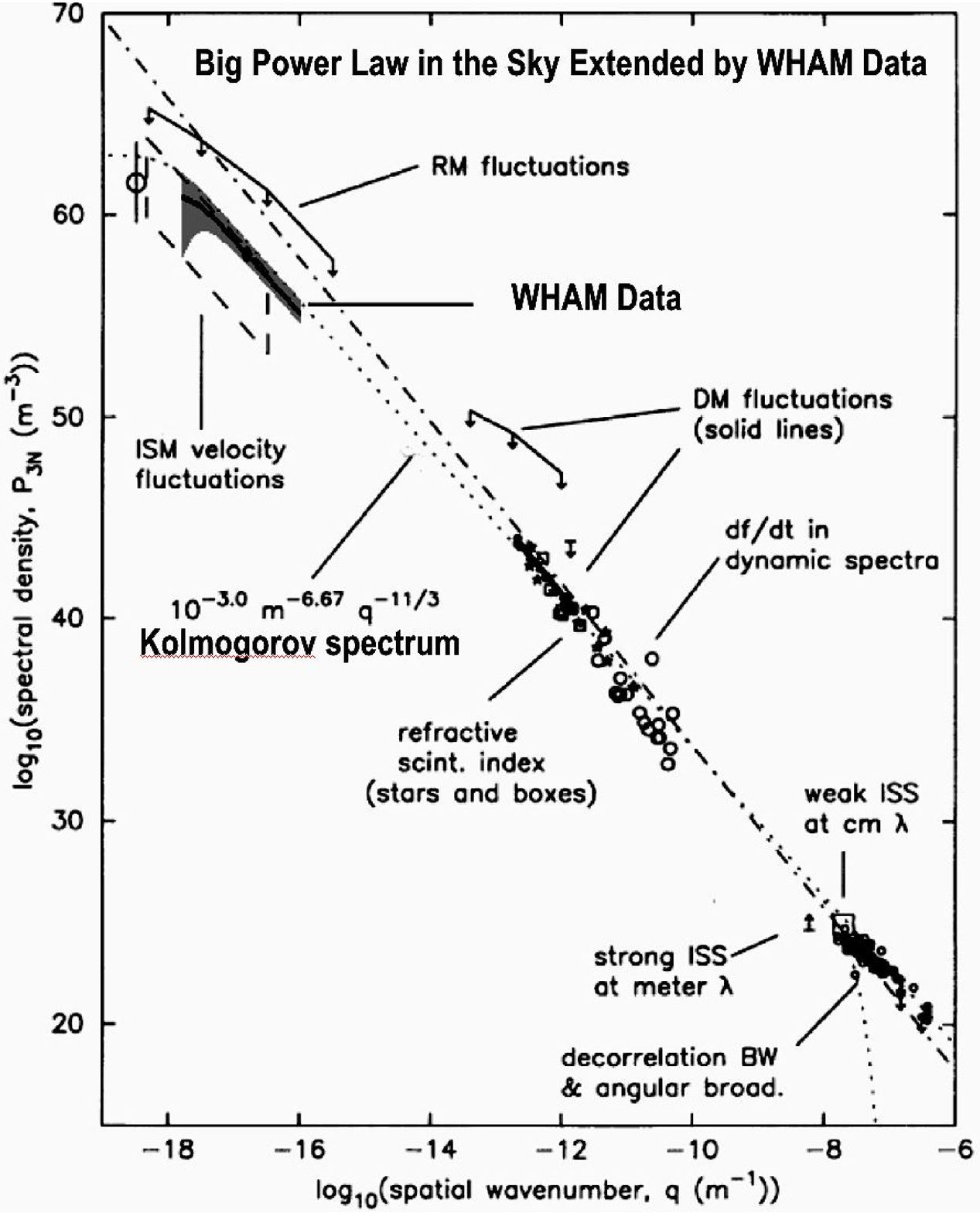}
\caption{Turbulence in the interstellar gas as revealed by electron density
fluctuations.  "Big Power Law in the Sky" in \cite{armstrong95} extended using
WHAM data.  The slope corresponds to that of Kolmogorov turbulence.  Modified
from \cite{chepurnov10a}.\label{fig:big_power_law}}
\end{figure}

It is well known that astrophysical fluids are magnetized and turbulent
\cite[see][and references
therein]{armstrong81,armstrong95,verdini07,lazarian09}.  For interstellar medium
Fig.~\ref{fig:big_power_law} illustrates the turbulent power density plotted
against the inverse of the scale length, with data at large scales, i.e. at
small wave numbers $q$ expanded using the Wisconsin H$_{\alpha}$ Mapper (WHAM)
data on electron density fluctuations\footnote{A more direct evidence comes from
the observations of spectral lines.  Apart from showing non-thermal Doppler
broadening \cite[see][]{larson81}, they also reveal spectra of supersonic
turbulent velocity fluctuations when analyzed with techniques like Velocity
Channel Analysis (VCA) of Velocity Coordinate Spectrum (VCS) developed
\cite[see][]{lazarian00,lazarian04,lazarian06,lazarian08} and applied to the
observational data \cite[see][]{padoan04,padoan09,chepurnov10b}.}
\citep{chepurnov10a}.  A similar Kolmogorov-type power law is measured with in
situ measurements in Solar wind \cite{coleman68, matthaeus82a, matthaeus82b,
leamon98}\footnote{More discussion of the solar wind turbulence can be found in
\cite{bruno05, petrosyan10}}.  On much larger scales, clusters of galaxies also
reveal turbulent spectrum \citep{ensslin06}.

The fact that turbulence is ubiquitous in astrophysical environments is the most
natural and expected.  Magnetized astrophysical plasmas generally have very
large Reynolds numbers due to the large length scales involved, as well as the
fact that the motions of charged particles in the direction perpendicular to
magnetic fields are constrained.  Laminar plasma flows at these high Reynolds
numbers $R=VL_f/\nu$, where $V$ and $L_f$ are the velocity and the scale of the
flow, $\nu$ is fluid viscosity,  are prey to numerous linear and
finite-amplitude instabilities, from which turbulent motions readily develop.

For interstellar medium the drivers of turbulence include supernovae explosions
that shape the interstellar medium \citep{mckee77,nakamura06}, accretion flows
\citep{klessen10}, magneto-rotational instability in the galactic disk
\citep{sellwood99}, thermal instability \cite[see][]{kritsuk02,koyama02},
collimated outflows \cite[see][]{nakamura07}, etc..  Cluster mergers, jets of
radio galaxies and turbulent galaxy wakes are among the drivers of turbulence in
clusters of galaxies \citep{brunetti09}.  Similarly, the fast plasma flow and
plasma instabilities provide the natural environment for turbulence to develop
in the solar wind.  In more special circumstances, instabilities in the fluid of
cosmic rays as well as energy injection via magnetic reconnection can serve as
the source of turbulence.  The latter provide special cases that we also
consider in the paper, but the generic set up is that both the evolution of
cosmic rays and reconnection is considered in the presence of {\it pre-existing}
turbulence which is created by external sources\footnote{The relevant analogy:
turbulence that is experienced by the passengers of an  airplane does not arise
from the airplane motion, but originates due to complex processes taking place
in the atmosphere.}.

The model that has been shown to successfully reproduce the properties of
incompressible MHD turbulence was proposed by \citet[][henceforth
GS95]{goldreich95}.  Numerical studies of MHD turbulence in \cite{cho02,cho03}
allowed to extend the latter model to compressible fluids.  This allowed to
change our understanding of the effects of turbulence on the scattering and
acceleration of cosmic rays \citep{yan02,yan04}.

Turbulence is known to affect most of the properties of fluids, e.g. thermal
conductivity, propagation of waves and energetic particles, magnetic field
generation etc. \cite[see reviews and books by][and references
therein]{moffatt78,dmitruk01,schlickeiser03,vishniac03,cranmer05,longair10}.
For instance, the cosmic ray streaming instability that is the basic textbook
process is being suppressed in the presence of turbulence in the interstellar
medium \citep{yan02,farmer04,beresnyak08}.  Similarly, \citet[][henceforth
LV99]{lazarian99} shows that the constrains on the classical Sweet-Parker
reconnection are being lifted in the presence of turbulence and the reconnection
rate becomes fast, i.e. independent on resistivity.

Fast magnetic reconnection predicted in the presence of turbulence in
\cite{lazarian99} entails many astrophysical consequences \cite[see e.g.][for
the consequences for star formation]{lazarian11}.  It entails, for instance, the
efficient acceleration of cosmic rays in reconnection regions as discussed in
\cite{degouveia05} \cite[see also][]{lazarian05}.

In what follows, we discuss the nature of MHD turbulence in \S 2, its
consequences for magnetic reconnection in \S 3 and for cosmic ray acceleration
in \S 4. We describe the first order Fermi acceleration that takes place in the
reconnection sites in \S 5. The discussion and summary are presented in \S 6 and
\S 7, respectively.

\section{MHD turbulence in astrophysical plasmas}

MHD turbulence is widely spread process that is essential both for astrophysical
reconnection and cosmic ray physics.  The balanced MHD turbulence is relatively
well explored subject, but other regimes of MHD turbulence, e.g. imbalanced MHD
turbulence, viscosity dominated MHD turbulence are still insufficiently studied
and the consequences of these regimes are poorly explored.

\subsection{Applicability of MHD approximation}

The paper is focused on the properties of MHD turbulence as we claim that the
small scale plasma effects are of secondary importance for both the dynamics of
turbulence at sufficiently large scales as well as for reconnection and
acceleration of cosmic rays in the presence of turbulence.

The applicability of MHD description to plasmas has been revisited recently in
\citet[][henceforth ELV11]{eyink11}.  There three characteristic length-scales
were considered: the ion gyroradius $\rho_i,$ the ion mean-free-path length
$\ell_{mfp,i}$, and the scale $L$ of large-scale variation of magnetic and
velocity fields.  Astrophysical plasmas are in many cases ``strongly
collisional'' in the sense that $\ell_{mfp,i}\ll \rho_i,$ which is the case for
the interiors of stars and accretion disks.  In such cases, a fluid description
of the plasma is valid.  In the ``weakly collisional''  $\ell_{mfp,i}\gg
\rho_i,$.  The ratio
\begin{equation}
\frac{\ell_{mfp,i}}{\rho_i}\propto \frac{\Lambda}{\ln\Lambda}\frac{v_A}{c},
\end{equation}
follows from the standard formula for the Coulomb collision frequency
\cite[see][Eq.~1.25]{fitzpatrick08}.  Here $\Lambda=4\pi n\lambda_D^3$ is the
plasma parameter, or the number of particles within the Debye screening sphere.
When astrophysical plasmas are very weakly coupled (hot and rarefied), then
$\Lambda$ is large, e.g. of the order of $10^9$ or more for the warm component
of the interstellar medium or solar wind (see Table~\ref{tab:parameters}).  For
such ratio the expansion over small ion Larmor radius $\rho_i$ provides
``kinetic MHD equations'' which differ from the standard MHD by having
anisotropic pressure tensor.

\begin{table*}[t]
\caption{Parameters for Some Astrophysical Plasmas \cite[adopted from][]{eyink11}. \label{tab:parameters}}
\centering
\begin{tabular}{llll}
\hline
\hline
Parameter & warm ionized & post-CME & solar wind at  \\
& ISM${\,\!}^a$ & current sheets${\,\!}^b$ & magnetosphere${\,\!}^c$ \\
\hline
density $n,\,cm^{-3}$ & .5 & $7\times 10^7$ & 10 \\
temperature $T,\, eV$ & .7 & $10^3$ &10 \\
plasma parameter $\Lambda$ & $4\times 10^9$ & $2\times 10^{10}$ & $5\times 10^{10}$ \\
ion thermal velocity $v_{th,i},\,cm/s$ & $10^6$ & $3\times 10^7$ & $5\times 10^6$ \\
ion mean-free-path $\ell_{mfp,i},\,cm$ & $6\times 10^{11}$ & $10^{10}$ & $7\times 10^{12}$ \\
magnetic diffusivity $\lambda,\, cm^2/s$ & $10^7$ & $8\times 10^2$ & $6\times 10^5$ \\
\hline
magnetic field $B,\, G$ & $10^{-6}$ & 1 & $10^{-4}$ \\
plasma beta $\beta$ & 14 & 3 & 1 \\
Alfv\'en speed $v_A,\,cm/s$ & $3\times 10^5$ & $3\times 10^7$ & $7\times 10^6$ \\
ion gyroradius $\rho_i,\,cm$ & $10^8$ & $3\times 10^3$ & $6\times 10^6$ \\
\hline
large-scale velocity $U,\,cm/s$ & $10^6$ & $4\times 10^6$ & $5\times 10^6$ \\
large length scale $L,\,cm$ & $10^{20}$ & $5\times 10^{10}$ & $10^8$ \\
Lundquist number $S_L=\frac{v_A L}{\lambda}$ & $3\times 10^{18}$ & $2\times 10^{15}$ & $10^9$ \\
resistive length${\,\!}^*$ $\ell_\eta^\perp,\, cm$ & $5\times 10^5$ & 1 & 20\\
\hline
& & & \\
\multicolumn{4}{l}{\footnotesize{${\,\!}^a$\cite{norman96,ferriere01} \,\,\,\, ${\,\!}^b$\cite{bemporad08}
\,\,\,\, ${\,\!}^c$\cite{zimbardo10}}}\\
\multicolumn{4}{l}{\footnotesize{*This nominal resistive scale is calculated from $\ell_\eta^\perp \simeq L (v_A/U)
S_{L}^{-3/4}$, assuming GS95 turbulence holds}}\\
\multicolumn{4}{l}{\footnotesize{down to that scale, and should not be taken literally
when $\ell_\eta^\perp<\rho_i.$}}\\
\end{tabular}
\end{table*}

Plasmas that are not strongly collisional further subdivided into two classes:
``collisionless'' plasmas for which $\ell_{mfp,i}\gg L,$ the largest scales of
interest, and ``weakly collisional'' plasmas for which $L\gg \ell_{mfp,i}.$  In
the latter case the``kinetic MHD'' description can be further reduced in
complexity at scales greater than $\ell_{mfp,i}$ (see ELV11).  This reproduces a
fully hydrodynamic MHD description at those scales, with anisotropic transport
behavior associated to the well-magnetized limit.  Among our examples in Table 1
above, the warm ionized ISM is ``weakly collisional'', while post-CME current
sheets and the solar wind impinging on the magnetosphere are close to being
``collisionless.''

Additional important simplifications occur if the following assumptions are
satisfied:  turbulent fluctuations are small compared to the mean magnetic
field, have length-scales parallel to the mean field much larger than
perpendicular length-scales, and have frequencies low compared to the ion
cyclotron frequency.  These are standard assumptions of the \cite{goldreich95}
theory of MHD turbulence.  They are the basis of the ``gyrokinetic
approximation'' \citep{schekochihin07,schekochihin09}.  At length-scales larger
than the Larmor radius $\rho_i$, another reduction takes place.  The
incompressible shear-Alfv\'en wave modes exhibit dynamics independent of
compressive motions and can be described by the ``Reduced MHD'' (RMHD) equations
\cite[see][]{strauss76,zank92a,goldreich95,cho03}.  This fact is essential for
the LV99 justifying the use of the treatment based on an incompressible MHD
model.

\subsection{MHD turbulence: cascades of Alf\'enic, slow and fast modes}
\label{sec:mhd_waves}

Magnetized turbulence is a tough and complex problem with many excellent
monographs and reviews devoted to different aspect of it \cite[see][and
references therein]{biskamp03}.  A broad outlook on the astrophysical
implications of the turbulence can be found in a review by \cite{elmegreen04},
while the effects of turbulence on molecular clouds and star formation are well
reviewed in \cite{mckee07}.  However, the issues of turbulence spectrum and its
anisotropies, we feel, are frequently given less attention than  they deserve.

We deal with magnetohydrodynamic (MHD) turbulence which provides a correct
fluid-type description of plasma turbulence at large scales.  Astrophysical
turbulence is a direct consequence of large scale fluid motions experiencing low
friction.  The Reynolds numbers are typically very large in astrophysical flows
as the scales are large.  As magnetic fields decrease the viscosity for the
plasma motion perpendicular to their direction, $Re$ numbers get really
astronomically large.  For instance, $Re$ numbers of $10^{10}$ and larger are
very common for astrophysical flows.  For so large $Re$ the inner degrees of
fluid motion get excited and a complex pattern of motion develops.

While turbulence is an extremely complex chaotic non-linear phenomenon, it
allows for a remarkably simple statistical description \cite[see][]{biskamp03}.
If the injections and sinks of the energy are correctly identified, we can
describe turbulence for {\it arbitrary} $Re$ and $Rm$.  The simplest description
of the complex spatial variations of any physical variable, $X({\bf r})$, is
related to the amount of change of $X$ between points separated by a chosen
displacement ${\bf l}$, averaged over the entire volume of interest.  Usually
the result is given in terms of the Fourier transform of this average, with the
displacement ${\bf l}$ being replaced by the wavenumber ${\bf k}$ parallel to
${\bf l}$ and $|{\bf k}|=1/|{\bf l}|$.  For example, for isotropic turbulence
the kinetic energy spectrum, $E(k)dk$, characterizes how much energy resides at
the interval $k, k+dk$.  At some large scale $L$ (i.e., small $k$), one expects
to observe features reflecting energy injection.  At small scales, energy
dissipation should be seen.  Between these two scales we expect to see a
self-similar power-law scaling reflecting the process of non-linear energy
transfer.

Thus, in spite of its complexity, the turbulent cascade is self-similar over its
inertial range.  The physical variables are proportional to simple powers of the
eddy sizes over a large range of sizes, leading to scaling laws expressing the
dependence of certain non-dimensional combinations of physical variables on the
eddy size.  Robust scaling relations can predict turbulent properties on the
whole range of scales, including those that no large-scale numerical simulation
can hope to resolve.  These scaling relations are extremely important for
obtaining an insight of the processes at the small scales.

The presence of a magnetic field makes MHD turbulence anisotropic \citep[][see
\citeauthor{oughton03}, \citeyear{oughton03} for a
review]{montgomery81,matthaeus83,shebalin83,higdon84,goldreich95}.  The relative
importance of hydrodynamic and magnetic forces changes with scale, so the
anisotropy of MHD turbulence does too.  Many astrophysical results, e.g. the
dynamics of dust, scattering and acceleration of energetic particles, thermal
conduction, can be obtained if the turbulence spectrum and its anisotropy are
known.  As we discuss below, additional important insight can be obtained if we
know turbulence intermittency.

Estimates of turbulence anisotropy obtained in relation to the observations of
magnetic fluctuation of the outer heliosphere and solar wind \cite[see][and
references therein]{zank92b} provided, for an extended period of time, the only
guidance for theoretical advances.  This resulted in a picture of MHD turbulence
consisting of 2D "reduced MHD" perturbations carrying approximately 80\% of
energy and "slab" Alfv\'enic waves carrying the remaining 20\% of energy
\cite[see][and references therein]{matthaeus02}.  In other words, in the
suggested picture the MHD turbulence was presented by two anisotropic
components, one having wave vectors mostly perpendicular to magnetic field (the
2D one), the other having them mostly parallel to magnetic field (the slab one).
This model became a default one for many calculations of the propagation of
cosmic rays \cite[see][]{bieber88,bieber94}.  On the contrary, guided mostly by
compressible MHD numerical simulations, the interstellar community adopted a
model of the MHD turbulence where the basic MHD modes, i.e. slow, fast and
Alfv\'enic are well coupled together and efficiently dissipate energy in shocks
\citep{stone98,maclow99}.  Little cross-talk between the two communities did not
stimulate the interdisciplinary debates on the nature of MHD turbulence, which
was regretful, as the heliospheric community has the advantage of the in-situ
spacecraft measurements.

In spite of the intrinsic limitations of the "brute force" approach, we feel
that reliable results can be obtained numerically if the studies are  focused on
a particular property of turbulence in order to get a clear picture of the
underlying physics occurring on small scales (``microphysics'') that cannot be
resolved in ``global'' interstellar simulations\footnote{By contrast, numerical
simulations that deal with many physical conditions simultaneously cannot
distinguish between the effects of different processes.  Moreover, they
inevitably have a more restricted interval of scales on which energy is injected
by numerics, initial conditions, or boundary conditions.  Their results are,
therefore, difficult to interpret in physical terms.}.

For instance, numerical studies in \cite{cho02,cho03} showed that the Alfv\'enic
turbulence develops an independent cascade which is marginally affected by the
fluid compressibility.  This observation corresponds to theoretical expectations
of the GS95 theory that we briefly describe below \cite[see also][]{lithwick01}.
In this respect we note that the MHD approximation is widely used to describe
the actual magnetized plasma turbulence over scales that are much larger than
both the mean free path of the particles and their Larmor radius \cite[see][and
references therein]{kulsrud83,kulsrud05}.  More generally, the most important
incompressible Alfv\'enic part of the plasma motions can described by MHD even
below the mean free path but on the scales larger than the Larmor radius.

We claim that, while having a long history of ideas, the theory of MHD
turbulence has become testable recently due to the advent of numerical
simulations \cite[see][]{biskamp03} which confirmed \cite[see][and references
therein]{cho05} the prediction of magnetized Alfv\'enic eddies being elongated
in the direction of magnetic field \cite[see][]{shebalin83,higdon84} and
provided results consistent with the quantitative relations for the degree of
eddy elongation obtained in GS95.  Indeed, GS95 made predictions regarding
relative motions parallel and perpendicular to {\bf B} for Alfv\'enic
turbulence.  The model did not predict the generation of any "slab" modes and,
instead of pure 2D Alfv\'enic modes, predicted that most of the Alfv\'enic
energy is concentrated in the modes with a so-called "critical balance" between
the parallel and perpendicular motions.  The latter can be understood within
intuitive picture where eddies mixing magnetic field perpendicular to its
local\footnote{The notion of the direction being local is critical.  Small
eddies are affected by magnetic field in their vicinity, rather than a global
field.  No universal scalings are possible to obtain in the frame of  the mean
magnetic field.} direction induce Alfv\'enic waves with the period equal to the
period of the eddy rotation.  This results in the scale-dependent anisotropy of
velocity and magnetic perturbations, with the anisotropy being larger for
smaller eddies.

\begin{table*}[t]
\caption{Regimes and ranges of MHD turbulence. \label{tab:regimes}}
\centering
\begin{tabular}{lllll}
\hline
\hline
Type          & Injection &  Range   & Motion & Ways\\
of MHD turbulence & velocity & of scales & type & of study\\
\hline
Weak & $V_L<V_A$ & $[L, l_{trans}]$ & wave-like & analytical\\
\hline
Strong &  & & & \\
subAlfv\'enic& $V_L<V_A$ & $[l_{trans}, l_{min}]$ & eddy-like & numerical \\
\hline
Strong &  &  &  & \\
superAlfv\'enic & $V_L> V_A$ & $[l_A, l_{min}]$ & eddy-like & numerical \\
\hline
& & & \\
\multicolumn{5}{l}{\footnotesize{$L$ and $l_{min}$ are injection and dissipation scales}}\\
\multicolumn{5}{l}{\footnotesize{$l_{trans}$  and $l_{A}$ are given by Eq. (\ref{trans}) and
Eq. (\ref{alf}), respectively.}}\\
\end{tabular}
\end{table*}

The hydrodynamic counterpart of the MHD turbulence theory is the famous
\cite{kolmogorov41} theory of turbulence.  In the latter theory energy is
injected at large scales, creating large eddies which correspond to large $Re$
numbers and therefore do not dissipate energy through
viscosity\footnote{Reynolds number $Re\equiv L_fV/\nu=(V/L_f)/(\nu/L^2_f)$ which
is the ratio of an eddy turnover rate $\tau^{-1}_{eddy}=V/L_f$ and the viscous
dissipation rate $\tau_{dis}^{-1}=\eta/L^2_f$.  Therefore large $Re$ correspond
to negligible viscous dissipation of large eddies over the cascading time
$\tau_{casc}$ which is equal to $\tau_{eddy}$ in Kolmogorov turbulence.} but
transfer energy to smaller eddies.  The process continues until the cascade
reaches the scales which are small enough that the energy is dissipated during
one turnover time of the eddies at the corresponding scales.  In the absence of
compressibility the hydrodynamic cascade of energy is $\sim v^2_l/\tau_{casc, l}
=const$, where $v_l$ is the velocity at the scale $l$ and the cascading time for
the eddies of size $l$ is $\tau_{cask, l}\approx l/v_l$.  From this the well
known relation $v_l\sim l^{1/3}$ follows.

A frequent mental picture that astrophysicists have of the Alfv\'enic turbulence
is based of Alfv\'en waves with wave vectors along the magnetic field.  This is
not true for the strong Alfv\'enic turbulence which, similar to its hydrodynamic
counterpart, can be described in terms of eddies\footnote{The description in
terms of interacting wave packets or modes is also possible with the
corresponding wave vectors tending to get more and more perpendicular to the
magnetic field as the cascade develops.}.  However, contrary  to Kolmogorov
turbulence, in the presence of dynamically important magnetic field eddies
become anisotropic.  At the same time, one can imagine eddies mixing magnetic
field lines perpendicular to the direction of magnetic field.

The nature of Alfv\'enic cascade is expressed through the critical balance
condition in GS95 model of strong turbulence, namely,
\begin{equation}
l_{\|}^{-1}V_A\sim l_{\bot}^{-1}v_l,
\label{crit}
\end{equation}
where $v_l$ is the eddy velocity, while the $l_{\|}$ and $l_{\bot}$ are,
respectively, eddy scales parallel and perpendicular to the local direction of
magnetic field.  The critical balance condition states that the parallel size of
an eddy is determined by the distance Alfv\'enic perturbation can propagate
during the eddy turnover.  The notion of {\it local} is important\footnote{To
stress the difference between local and global systems here we do not use the
language of $k$-vectors.  Wave vectors parallel and perpendicular to magnetic
fields can be used, if only the wave vectors are understood in terms of a
wavelet transform defined with the local reference system rather than ordinary
Fourier transform defined with the mean field system.}, as no universal
relations exist if eddies are treated with respect to the global mean magnetic
field \cite[LV99;][]{cho00,maron01,lithwick01,cho02}.

The critical balance is the feature of the strong turbulence, which is the case
when the turbulent energy is injected at $V_A$.  If the energy is injected at
velocities lower than $V_A$ the cascade is weak with $l_{\bot}$ of the eddies
increasing while $l_{\|}$ staying the same \citep{ng96,lazarian99,galtier02}.
In other words, as a result of the weak cascade the eddies get thinner, but
preserve the same length along the local magnetic field.  This decreases
$l_{\bot}$ and eventually makes Eq.~(\ref{crit}) satisfied.  If the injection
velocity is $V_L$ and turbulent injection scale is $L$, the transition to the
strong MHD turbulence happens at the scale $l(v_l/V_A)^2$ and the velocity at
this scale is $V_{strong}=V_A (v_l/V_A)^2$ \cite[LV99;][]{lazarian06a}.  Thus
the weak turbulence has a limited, i.e. $[l, l(v_l/V_A)^2]$ inertial interval
and get strong at smaller scales.

While GS95 assumed that the turbulent energy is injected at $V_A$ at the
injection scale $l$, LV99 provided general relations for the turbulent scaling
at small scales for the case that the injection velocity $V_L$ is less than or
equal to $V_A$, which can be written in terms of $l_{\|}$ and $l_{\bot}$:
\begin{equation}
l_{\|}\approx l \left(\frac{l_{\bot}}{l}\right)^{2/3}
\left(\frac{V_A}{V_L}\right)^{4/3}
\label{Lambda}
\end{equation}
\begin{equation}
v_{l}\approx V_L \left(\frac{l}{L}\right)^{1/3} \left(\frac{V_L}{V_A}\right)^{1/3}
\label{vl}
\end{equation}

It is important to stress that the scales $l_{\bot}$ and $l_{\|}$ are measured
with respect to the system of reference related to the direction of the local
magnetic field "seen" by the eddy.  This notion was not present in the original
formulation of the GS95 theory and was added to it in LV99.  The local system of
reference was later used in numerical studies in \cite{cho00}, \cite{maron01},
and \cite{cho02a} testing GS95 theory.  In terms of mixing motions, it is rather
obvious that the free Kolmogorov-type mixing is possible only with respect to
the local magnetic field of the eddy rather than the mean magnetic field of the
flow.

The original GS95 picture deals with transAlfv\'enic turbulence, i.e. with
$V_A\sim V_L$.  For low Alfv\'enic Mach numbers, i.e. for $V_A\gg V_L$ at large
scales $\sim L$ the turbulence is weak \cite[see][]{ng97,lazarian99,galtier00}
and magnetic fields are slightly perturbed by propagating Alfv\'en waves.  The
wave packets in weak turbulence evolve changing their perpendicular scale
$l_{\bot}$, while their scale $l_{\|}$ along the magnetic field does not change.
As scaling of weak turbulence predicts $V_l\sim V_L(l_\bot/L)^{1/2}$ (LV99), at
the scale
\begin{equation}
l_{trans}\sim L(V_L/V_A)^2\equiv LM_A^2
\label{trans}
\end{equation}
the critical balance condition $l_{\|}/V_A\approx l_{\bot}/V_l$ is getting
satisfied making turbulence strong.  It is easy to see that the velocity
corresponding to $l_{trans}$ is $V_{trans}\sim V_L (V_L/V_A)$.

For superAlfv\'enic turbulence the situation is somewhat different.  Magnetic
field gets dynamically important as soon as its energy density exceeds the
energy of eddies at  the Ohmic dissipation scale, which translates into
Alfv\'enic velocity getting larger than the velocity of eddies at the Ohmic
dissipation scale or the ion Larmor radius, whichever is larger.  As the
velocity in Kolmogorov turbulence scale as $v_l\sim l^{1/3}$, it is clear that
even weak magnetic field can make a significant impact on the dynamics of the
smallest eddies.  In view of that it is advantageous to introduce a scale at
which the magnetic field gets dynamically important and the nature of the
turbulence changes from hydrodynamic to MHD \cite[see][]{lazarian06a}, namely,
\begin{equation}
l_A=L(V_A/V_L)^3=LM_A^{-3}
\label{alf}
\end{equation}

The relations predicted in GS95 were confirmed numerically for incompressible
\citep{cho00,maron01,cho02a,beresnyak09,beresnyak10,beresnyak11a} and
compressible MHD turbulence\footnote{Some studies of MHD compressible
turbulence, e.g. \cite{vestuto03} did not perform a decomposition of MHD
perturbations into Alfv\'en, slow and fast modes as it is done in
\cite{cho02,cho03}.  They did not use local system of reference for which the
GS95 scaling is formulated.  Therefore a direct comparison of their results with
the GS95 predictions is difficult.} \citep[][see also \citeauthor{cho03a}
\citeyear{cho03a}, for a review]{cho02,cho03,kowal10}.  They are in good
agreement with observed and inferred astrophysical spectra.  A remarkable fact
revealed in \cite{cho02a} is that fluid motions perpendicular to {\bf B} are
identical to hydrodynamic motions.  This provides an essential physical insight
and explains why in some respects MHD turbulence and hydrodynamic turbulence are
similar, while in other respects they are different.

GS95 provided theoretical arguments in favor of weak coupling between fast and
Alfv\'en modes, and low impact of slow modes to Alfv\'en modes \cite[see
also][]{lithwick01}.  This challenged the paradigm accepted by the interstellar
community.  While the decomposition of MHD perturbations into fundamental MHD
waves was widely used in the literature \cite[see][]{dobrowolny80} it was
usually assumed that the Alfv\'enic waves exist and interact with other waves
for many periods \cite[see a discussion in][]{zweibel03}.  This is not the case
of the GS95 model of turbulence, where the Alfv\'en modes non-linearly decay
within one wave period.  This reduces the time of interaction and therefore the
coupling.  Interestingly enough, in GS95 model, the Alfv\'enic modes can affect
slow modes, but the opposite is not true.  These results were successfully
tested in \cite{cho02,cho03} and \cite{kowal10}.

\begin{figure*}[t]
\includegraphics[width=6.5in]{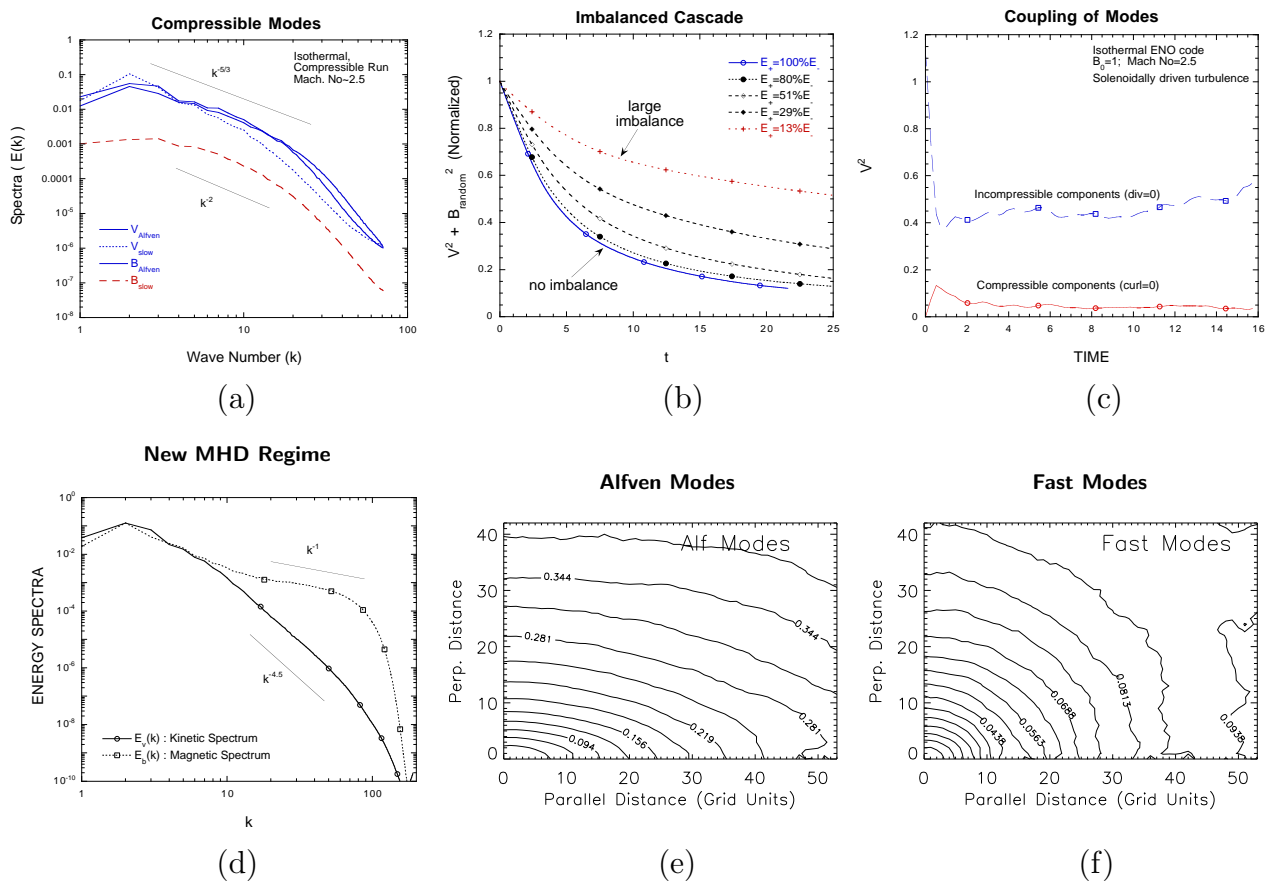}
\caption{{\it Statistics of MHD Turbulence:}  (a). Scaling of compressible
motions for plasma with magnetic pressure ten times the gas pressure. This
regime is important for molecular clouds (from CL02). The velocity (solid line)
and {\bf B} (dashed line) spectra are plotted against $k$ ($\equiv$ 1/(eddy
size)).  They show well defined statistical properties that allow further
fruitful applications.  (b). The simulations of decaying turbulence show that
the rate of the decay of the total energy is a strong function of the imbalance
of the energy contained in waves initially moving in opposite directions. In the
lowest curve, the waves have the same amplitude. The energy decays more slowly
with imbalance \citep{cho02a}. (c). The evolution of kinetic energy in 3-D
compressible turbulence, initially started with $\nabla\cdot v=0$ as if the gas
were incompressible. The dashed curve shows the evolution of the $\nabla\cdot
v=0$ motions. We see that Alfv\'enic turbulence creates only a marginal amount
of compressible motions, suggesting that Alfv\'enic modes should evolve
independently of the compressible cascade (CL02).  (d). Magnetic fluctuations
persist beyond the turbulent damping scale at large $k$, while hydrodynamic
fluctuations damp out in partially ionized gas \citep{lazarian04a}. This
viscosity-dominated regime of turbulence may dominate small scale structure of
partially ionized gas. (e-f). Isocontours of equal correlation for Alfv\'en and
fast modes (CL02). (e) The Alfv\'enic motions are much more correlated along
{\bf B} than perpendicular to it. (f) In contrast, fast magnetosonic
fluctuations show essentially circular (isotropic) isocontours of correlation.
\label{spectrum}}
\end{figure*}

Some of the relevant results are illustrated in Figure~\ref{spectrum}.  Contrary
to the common expectation, the modes exhibited nice scaling laws that allow
further analytical and numerical applications.  For instance, numerical studies
in \cite{cho02,cho03} revealed that the GS95 scalings are valid for the
Alfv\'enic part of the turbulence cascade even in the highly compressible regime
\cite[see also][]{beresnyak06}.

\begin{figure*}[t]
\centering
\includegraphics[width=0.3\textwidth]{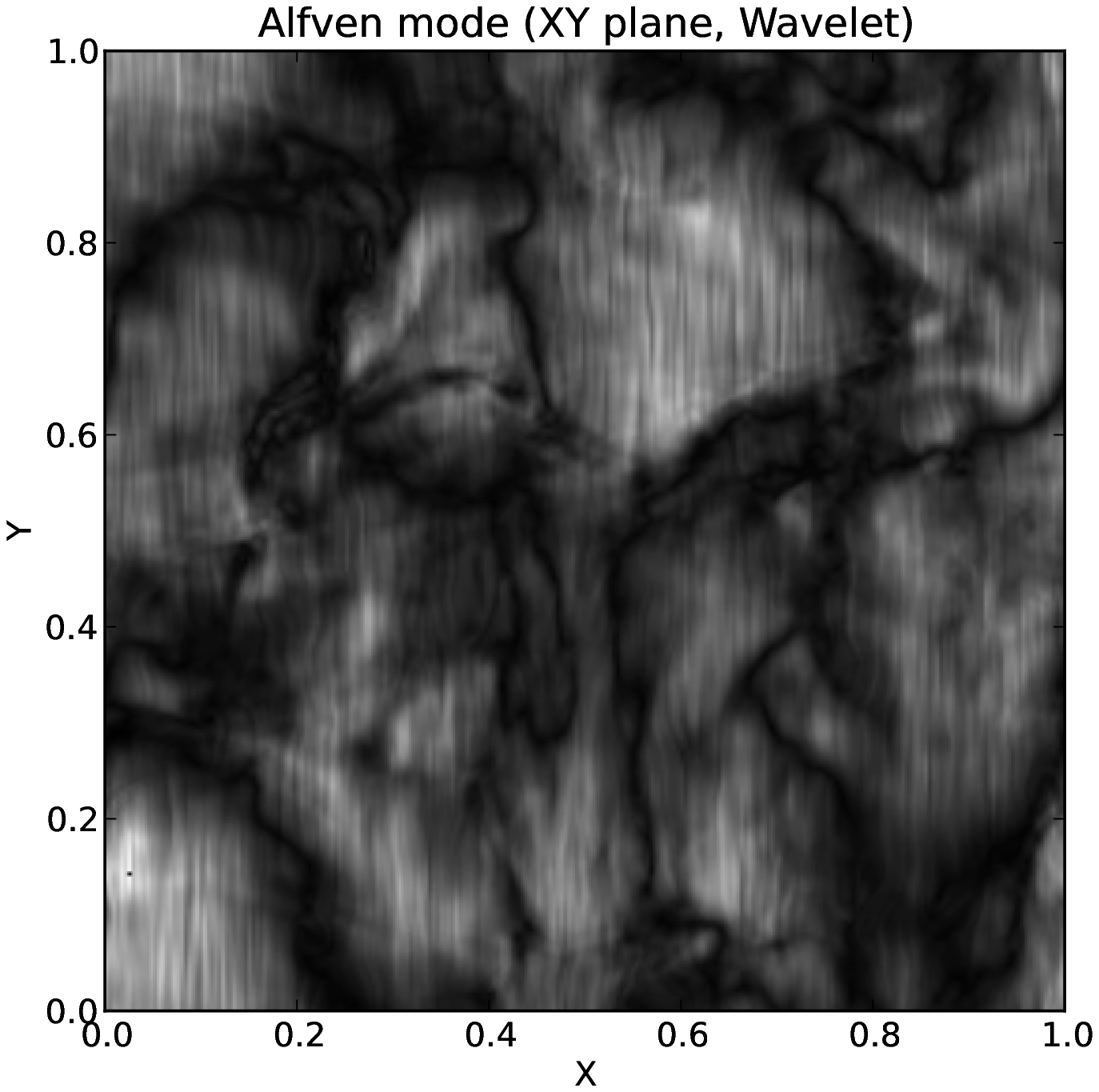}
\includegraphics[width=0.3\textwidth]{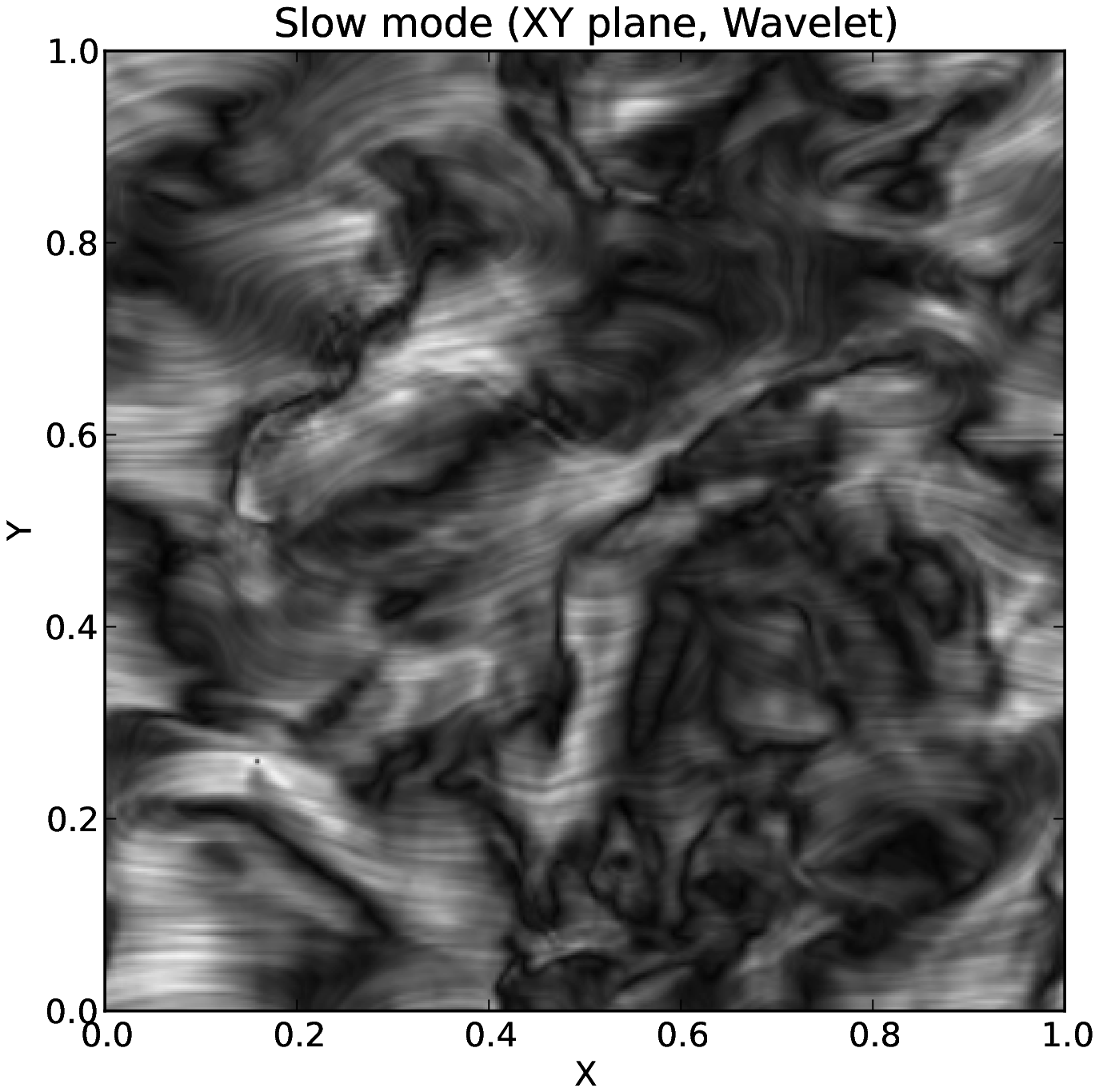}
\includegraphics[width=0.3\textwidth]{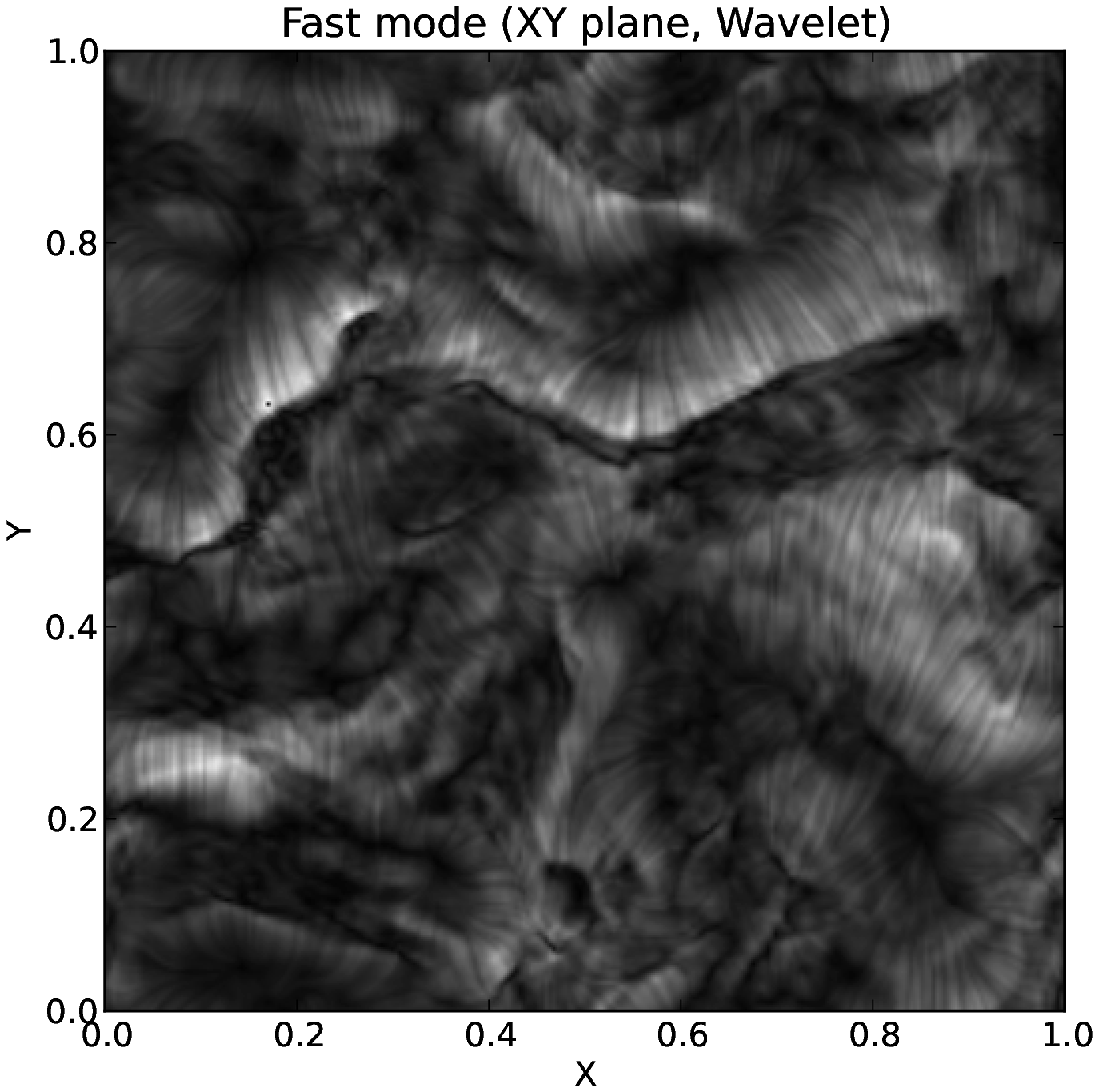} \\
\includegraphics[width=0.3\textwidth]{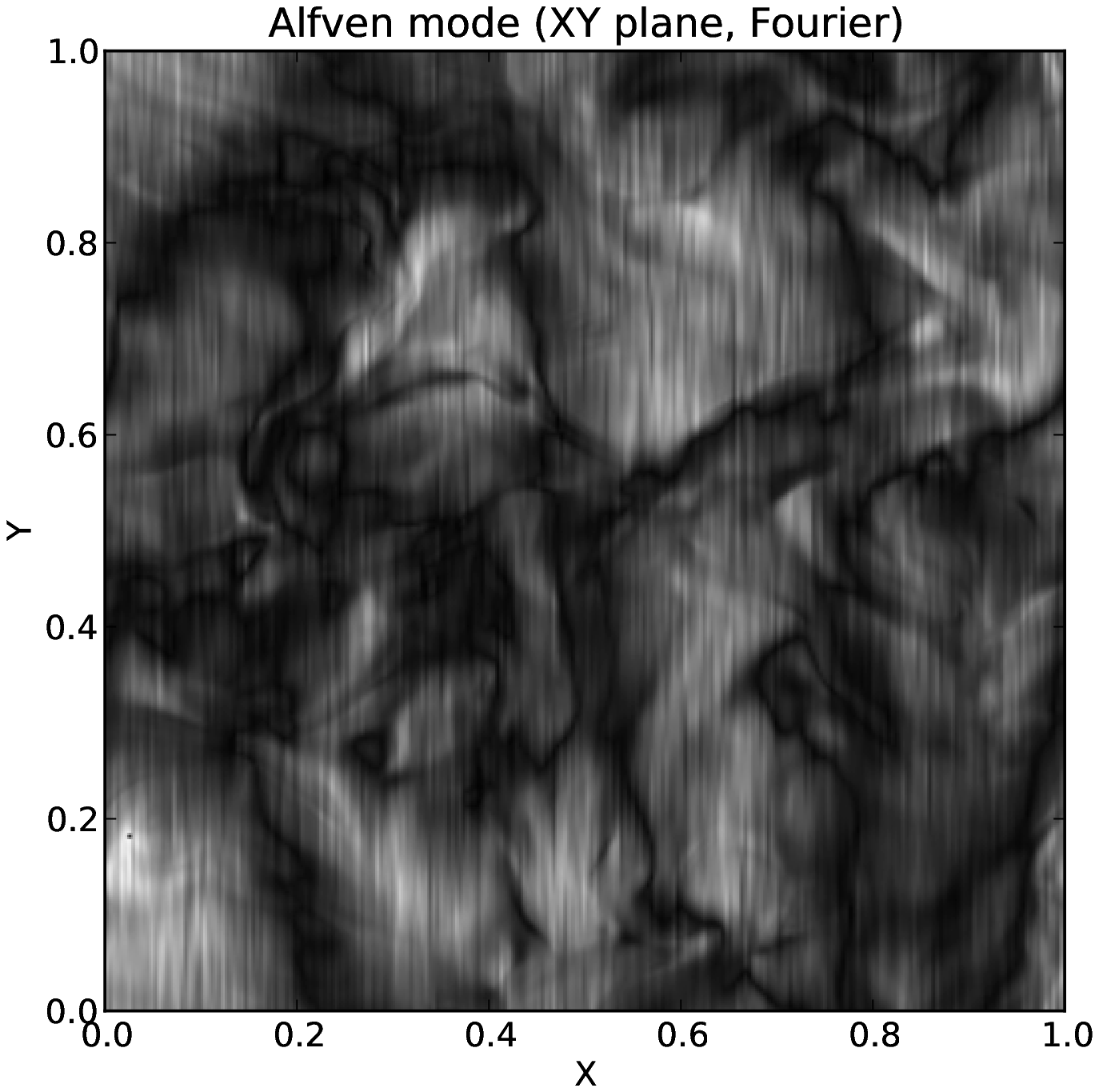}
\includegraphics[width=0.3\textwidth]{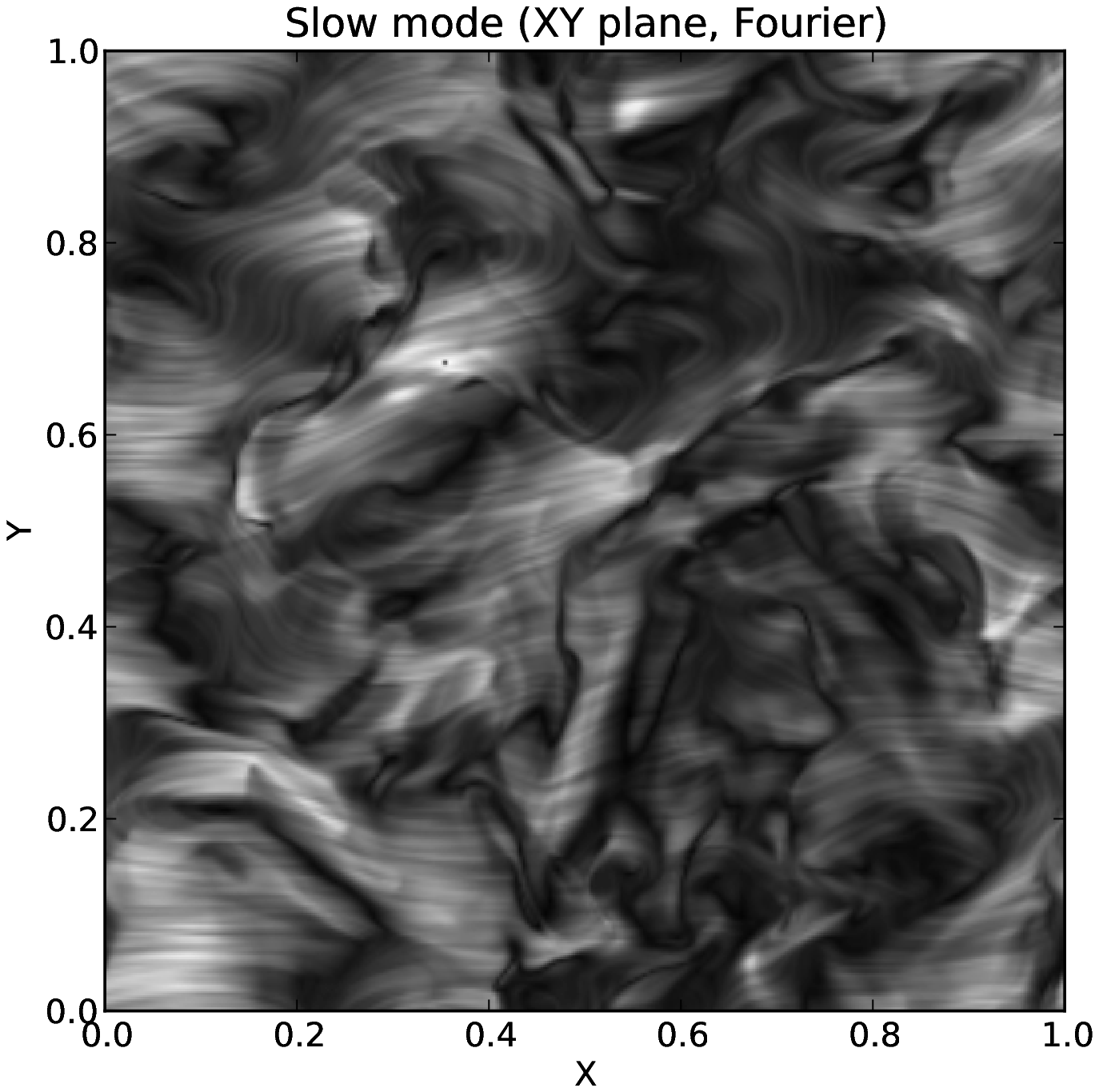}
\includegraphics[width=0.3\textwidth]{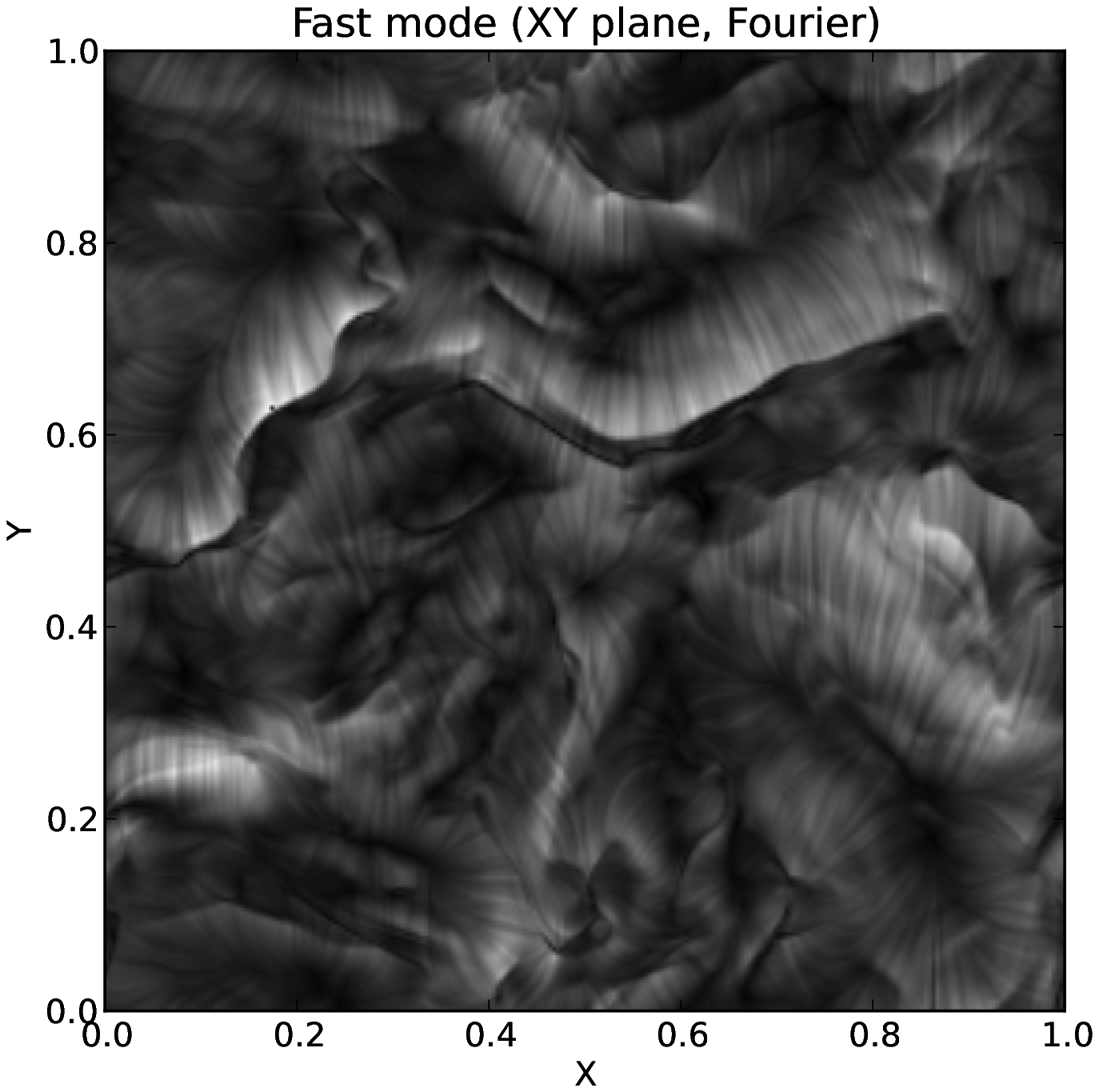}
\caption{The topology of the Alfv\'en, slow and fast mode (left, middle, and
right column, respectively) in the plane parallel to the mean field.  The upper
and lower rows show the MHD modes separated using the wavelet \citep{kowal10}
and Fourier \citep{cho03} methods, respectively.  Texture shows the direction of
velocity field multiplied by the local amplitude. \label{fig:decomposition}}
\end{figure*}

The statistical decomposition of MHD turbulence into Alfv\'en, slow and fast
modes suggested in \cite{cho02} was successfully tested for slow modes of
magnetically dominated plasmas in \cite{cho03}.  The decomposition of MHD
turbulent motions into the discussed three cascades of fundamental modes have
been confirmed by \cite{kowal10}, where in order to improve the decomposition,
the wavelet transformation were used.  The large scale components (large with
respect to the size of the local wavelet), which determine the vector base of
mode projection in Fourier space, are obtained by averaging over the size of the
wavelet.  Therefore, the wavelet transformations allow to approach closer to the
decomposition in the local system of reference than the earlier statistical
approach based on Fourier analysis could do.  In Figure~\ref{fig:decomposition}
we show that the new wavelet decomposition results (upper row) are very similar
to the ones obtained with the Fourier method (lower row).  At the same time, the
study of spectra and anisotropy of velocity, magnetic field reveals the
advantages of the wavelets in comparison with the Fourier technique for studying
turbulence with weak mean field.  For turbulence with $B_{mean}\sim \delta B$
the obtained results are consistent with the \cite{cho02,cho03} studies.

Nevertheless, it would be wrong to say that we have a complete understanding of
the scaling of MHD modes and their interactions.  First of all, one should
distinguish weak and strong Alfv\'enic turbulence.  The weak turbulence is
essentially 2D\footnote{Practical studies of non-linearity of turbulence in
Solar wind are discussed in \cite{sahraoui10, ghosh11}.} , with the turbulent
cascade creating more structure perpendicular to magnetic field as the
turbulence cascades \cite[see][]{galtier00}.  Such a cascade emerges when the
driving of turbulence at the outer scale is weak, i.e. the injection velocity is
much less than the Alfv\'en velocity.  Although the weak turbulence picture
corresponds to the early representation of MHD turbulence (see discussion in \S
2.1), one should keep in mind that the strength of Alfv\'enic interactions
increases with the decrease of the scale along the cascade.  Therefore the
Alfv\'enic turbulence gets eventually strong, while both the inertial range and
the astrophysical utility of the weak Alfv\'enic cascade are limited. The
interaction of weak Alfv\'enic turbulence with fast modes has dependences on the
angle between ${\bf B}$ and the wave vector \citep{chandran05}.

As the Mach number of turbulence increases, shocks should play more important
role in the dynamics of compressible motions.  The transition from the regime
when compressible motions can be well described by the fast and slow mode
cascades versus the situation when shocks carry an appreciable part of the
energy deserves more studies.  A study in \citet[][henceforth
BLC05]{beresnyak05} shows that in the space between shocks the description of
the perturbations with slow modes is valid for all the Mach numbers studied (up
to 10).

\begin{table*}[th]
\caption{Percentage Amount of the Kinetic Energy Contained Within Each Velocity
Component in \cite{kowal10}.  Errors correspond to a measure of the time
variation.\label{tab:energies}}
\begin{center}
\begin{tabular}{cc|cc|ccc}
  ${\cal M}_{s}$ & ${\cal M}_{A}$ & $V_\mathrm{incomp.}$ & $V_\mathrm{comp.}$ & $V_A$ & $V_s$ & $V_f$ \\
\hline
  $\sim 0.8$ & $\sim 0.7$ & 96.5$^{\pm0.8}$ &  3.3$^{\pm0.8}$ & 58$^{\pm4}$ & 37$^{\pm3}$ & 4.8$^{\pm0.7}$ \\
  $\sim 2.5$ & $\sim 0.6$ & 93$^{\pm2}$     &  7$^{\pm2}$     & 58$^{\pm5}$ & 33$^{\pm4}$ & 9$^{\pm2}$ \\
  $\sim 7.5$ & $\sim 0.5$ & 92$^{\pm2}$     &  7$^{\pm2}$     & 56$^{\pm4}$ & 36$^{\pm4}$ & 8.0$^{\pm0.7}$ \\
\\
  $\sim 0.7$ & $\sim 2.1$ & 95$^{\pm2}$ & 5$^{\pm2}$  & 52$^{\pm4}$ & 42$^{\pm4}$ & 6.2$^{\pm0.8}$ \\
  $\sim 2.5$ & $\sim 2.1$ & 86$^{\pm1}$ & 14$^{\pm2}$ & 47$^{\pm3}$ & 37$^{\pm4}$ & 16$^{\pm2}$    \\
  $\sim 7.5$ & $\sim 1.9$ & 84$^{\pm2}$ & 16$^{\pm2}$ & 47$^{\pm4}$ & 33$^{\pm4}$ & 20$^{\pm2}$
\end{tabular}
\end{center}
\end{table*}

The total velocity field contains two components: solenoidal, which is
equivalent to the incompressible part, and potential, which contains the
compressible part of the field and the remaining part which is curl and
divergence free.  In Table~\ref{tab:energies}, we show the percentage
contribution of each component to the total velocity field.  We see that the
compressible part constitutes only a fraction of the total field.  However, the
magnitude of this fraction is different for sub and supersonic models.  In the
case of sub-Alfv\'{e}nic turbulence, it is about 3\% in subsonic models and
about 7\% in supersonic models, which confirms a higher efficiency of the
compression in the presence of supersonic flows.  Furthermore, the fraction also
changes when we compare models with strong and weak magnetic fields.  The
velocity field, in the presence of a weak magnetic field, contains about 5\% of
the compressible part in the model with ${\cal M}_s \sim 0.7$ and even up to
16\% in models with ${\cal M}_s>1$.  The consequence of the presence of a strong
magnetic field results in a reduction of the compressible part of the velocity
field by a factor of 2.  This indicates a substantial role of the magnetic field
in the damping of the generation of the compressible flows.

The wavelet decomposition, important for MHD turbulence, separates the velocity
field into three different MHD waves: an incompressible Alfv\'{e}n wave and slow
and fast magneto acoustic waves, of both which are compressible.  In
Table~\ref{tab:energies}, we included the percentage amount of these components
in the total velocity field.  As we see, most of the energy is contained in the
Alfv\'{e}n wave.  It is almost 60\% in the case of sub-Alfv\'{e}n turbulence,
and about 50\% for super-Alfv\'{e}nic turbulence.  The slow wave contains
approximately 1/3 of the total energy.  However, for the super-Alfv\'{e}nic
case, this amount is slightly higher.  Table~\ref{tab:energies} suggests that
the slow wave is weaker when the turbulence becomes supersonic.  We do not see a
similar behavior for the Alfv\'{e}n wave in the case of models with a strong
magnetic field.  This effect could also take place in the super-Alfv\'{e}nic
models, but it is weakened by relatively large errors.  An interesting
dependence is observed in the case of the fast wave.  Although the fast wave is
the weakest among all MHD waves, it strongly depends on the regime of
turbulence.  Similarly to the compressible part of the velocity field, it is
stronger for models with a weak magnetic field.  In addition, it is much
stronger when turbulence is supersonic, but this strength seems to be weakly
dependent on the sonic Mach number.

\subsection{Density structure of MHD turbulent flows}

Density structure of turbulence is usually discussed in relation to star
formation studies. However, we shall show in \S\ref{sec:acceleration} that
fluctuations of density play important role for energetic particle acceleration.
Thus it is important to survey the basic properties of the turbulent density
field.

The power spectrum of density fluctuations is an important property of a
compressible flow.  In some cases, the spectrum of density can be derived
analytically.  For nearly incompressible turbulent motions in the presence of a
strong magnetic field, the spectrum of density scales similarly to the pressure,
i.e. $E_{\rho}(k)\sim k^{-7/3}$ if we consider the polytropic equation of state
$p=a\rho^{\gamma}$ \citep{biskamp03}.  In weakly magnetized nearly
incompressible MHD turbulence, however, velocities convect density fluctuations
passively inducing the spectrum $E_{\rho}(k)\sim k^{-5/3}$ \citep{montgomery87}.
In supersonic flows, these relations are not valid anymore because of shocks
accumulating matter into the local and highly dense structures.  Due to the high
contrast of density, the linear relation $\delta p = c_s^2 \delta \rho$ is no
longer valid, and the spectrum of density cannot be related to pressure
straightforwardly.

\begin{figure*}
\center
\includegraphics[width=0.45\textwidth]{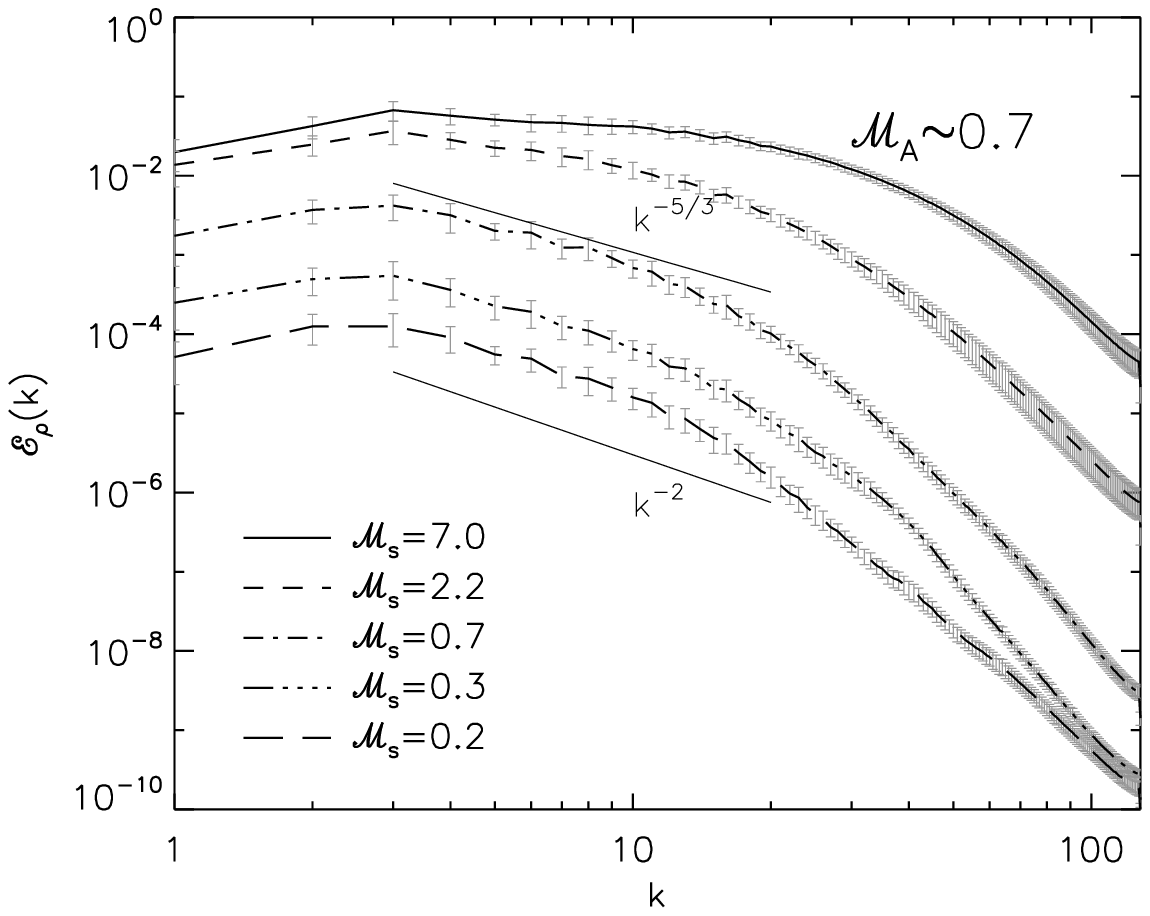}
\includegraphics[width=0.45\textwidth]{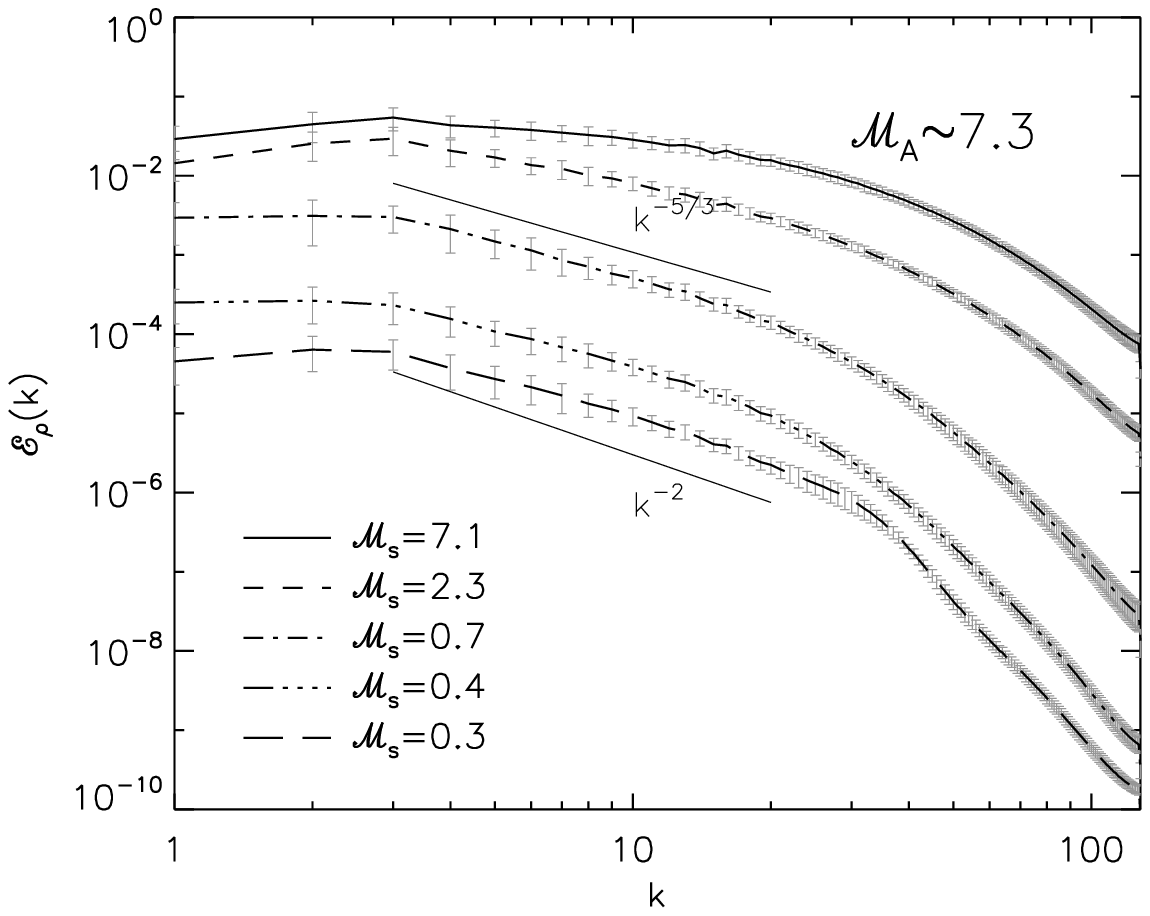}
\caption{Spectra of density for experiments with different values of ${\cal
M}_s$ and with ${\cal M}_A\sim0.7$ ({\em left panel}) and ${\cal M}_A\sim7.3$
({\em right panel}) for models with medium resolution ($256^3$).  Grey error
bars signify the variance of spectra with time.  The solid lines with slopes
$-5/3$ and $-2$ cover the inertial range used to estimate the spectral indices
of $\rho$ spectra. From \cite{kowal07}. \label{fig:spectra}}
\end{figure*}

In Figure~\ref{fig:spectra} we present the power spectra of fluctuations of
density for models with different ${\cal M}_s$.  As expected, we note a strong
growth of the amplitude of density fluctuations with the sonic Mach number at
all scales.  This behavior is observed both in sub-Alfv\'{e}nic as well as in
super-Alfv\'{e}nic turbulence (see Fig. \ref{fig:spectra}).  In Table
\ref{tab:slopes} the spectral index of density $\alpha_\rho$ and the logarithm
of density $\alpha_{\log \rho}$ is presented.  The width of the inertial range
is shown by the range of solid lines with slopes $-5/3$ and $-2$ in all spectra
plots.  Table \ref{tab:slopes} also shows the errors of estimation which combine
the error of the fitting of the spectral index at each time snapshot and the
standard deviation of variance of $\alpha_{\rho, \log \rho}$ in time.  The
slopes of the density spectra do not change significantly with ${\cal M}_s$ for
subsonic experiments and correspond to analytical estimations (about $-2.2$,
which is slightly less than $-7/3$, for turbulence with ${\cal M}_A\sim0.7$ and
about $-1.7$, which is slightly more than $-5/3$, for weakly magnetized
turbulence with ${\cal M}_A\sim7.3$).  Such an agreement confirms the validity
of the theoretical approximations.  Those, nevertheless, do not cover the entire
parameter space.  While the fluid motions become supersonic, they strongly
influence the density structure, making the small-scale structures more
pronounced, which implies flattening of the spectra of density fluctuations
\cite[see values for ${\cal M}_s>1.0$ in Table \ref{tab:slopes}, see
also][]{beresnyak05}.

\begin{table*}[t]
\caption{Slopes of the Power Spectrum of Density and the Logarithm of Density
Fluctuations in \cite{kowal07}. \label{tab:slopes}}
\begin{center}
{\small
\begin{tabular}{ccc|ccc}
\noalign{\smallskip}
 {} & {${\cal M}_{A}\sim0.7$}  & {} & {} & {${\cal M}_{A}\sim7$} & {} \\
 {${\cal M}_{s}$} & {$\alpha_\rho$}  & {$\alpha_{\log\rho}$} & {${\cal M}_{s}$} & {$\alpha_\rho$}  & {$\alpha_{\log\rho}$}\\
\noalign{\smallskip}
\hline
\noalign{\smallskip}
0.23$^{\pm0.01}$ & -2.3$^{\pm0.3}$ & -2.3$^{\pm0.3}$ & 0.26$^{\pm0.03}$ & -1.7$^{\pm0.3}$ & -1.7$^{\pm0.3}$\\
0.33$^{\pm0.01}$ & -2.2$^{\pm0.3}$ & -2.2$^{\pm0.3}$ & 0.36$^{\pm0.04}$ & -1.7$^{\pm0.3}$ & -1.7$^{\pm0.3}$\\
0.68$^{\pm0.03}$ & -2.0$^{\pm0.3}$ & -2.1$^{\pm0.3}$ & 0.74$^{\pm0.06}$ & -1.6$^{\pm0.2}$ & -1.6$^{\pm0.3}$\\
2.20$^{\pm0.03}$ & -1.3$^{\pm0.2}$ & -2.0$^{\pm0.2}$ & 2.34$^{\pm0.08}$ & -1.2$^{\pm0.2}$ & -1.6$^{\pm0.2}$\\
7.0$^{\pm0.3}$   & -0.5$^{\pm0.1}$ & -1.7$^{\pm0.2}$ & 7.1$^{\pm0.3}$   & -0.6$^{\pm0.2}$ & -1.5$^{\pm0.2}$\\
\noalign{\smallskip}
\end{tabular}
}
\\
{\footnotesize
The values have been estimated within the inertial range for models with ${\cal
M}_A\sim0.7$ ({\em left}) and ${\cal M}_A\sim7$ ({\em right}).  Errors of
spectral indices combine the errors of estimation at each time snapshot and the
standard deviation of variance in time.  Errors for sonic Mach numbers are the
standard deviation of their variance in time calculated over the period starting
from $t \ge 5$ to the last available snapshot.
}
\end{center}
\end{table*}

Another question is the anisotropy of density and the logarithm of density
structures.  For subsonic turbulence it is natural to assume that the density
anisotropies will mimic velocity anisotropies in the GS95 picture.  This was
confirmed in \cite{cho03a}, who observed that for supersonic turbulence the
contours of density isocorrelation becomes round, corresponding to isotropy.
\cite{beresnyak05}, however, showed that anisotropies restore the GS95 form if
instead of density one studies the {\it logarithm of density}.  This is due to
the suppression of the influence of the high density peaks, which arise from
shocks.  It is these peaks that mask the anisotropy of weaker, but more widely
spread density fluctuations.

\begin{figure*}
\centering
\includegraphics[width=0.8\columnwidth]{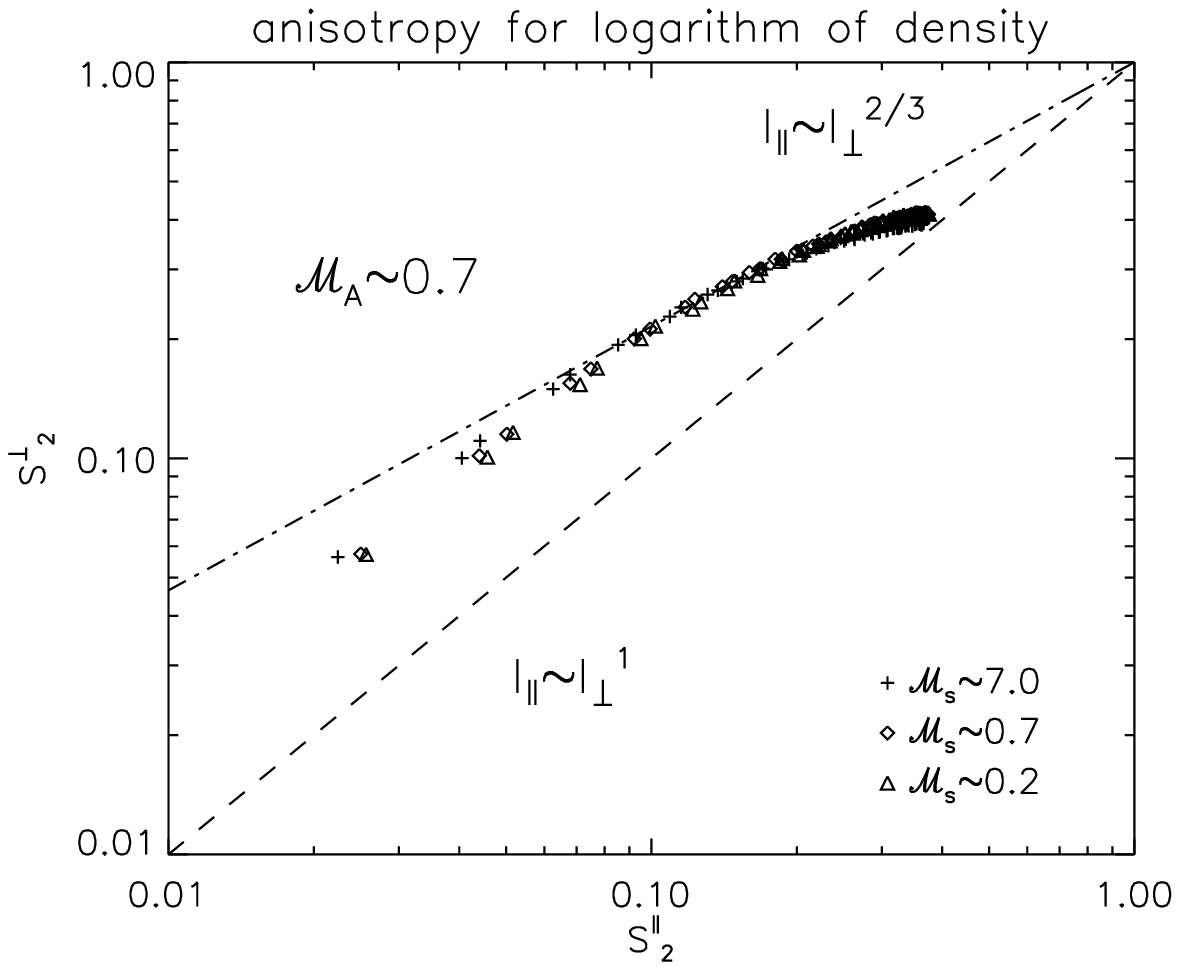}
\includegraphics[width=0.8\columnwidth]{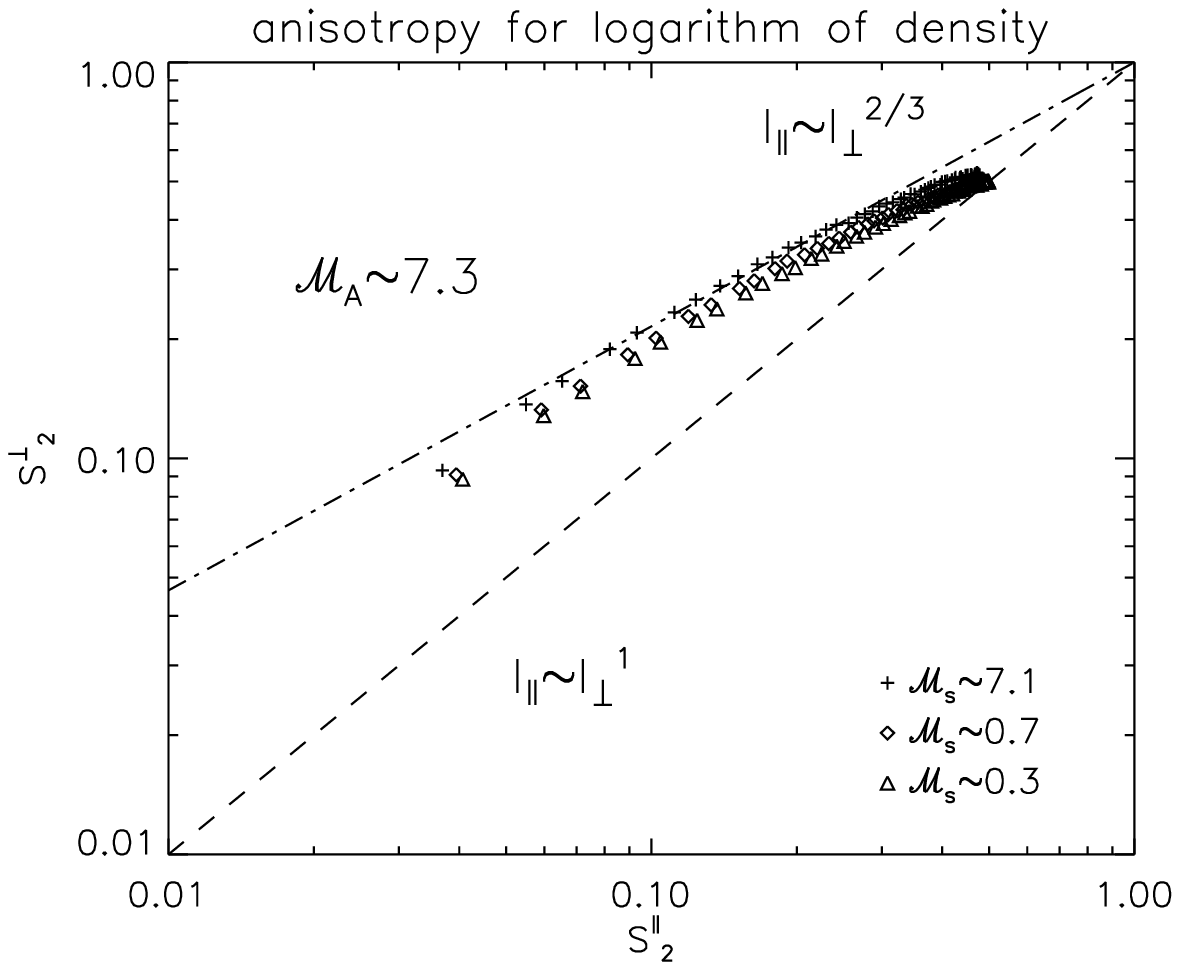}
\caption{{\it Left}: Anisotropy for the 2$^{nd}$-order SF for the logarithm of
density for ${\cal M}_A\sim7.3$;
{\it Right}: the same is for ${\cal M}_A\sim7.3$. The logarithm of density
traces the anisotropies of turbulent magnetic field. From \cite{kowal07}.}
\label{fig:anisotropy}
\end{figure*}

In Figure~\ref{fig:anisotropy} we show lines that mark the corresponding
separation lengths for the second-order structure functions parallel and
perpendicular to the local mean magnetic field.\footnote{The local mean magnetic
field was computed using the procedure of smoothing by a 3D Gaussian profile
with the width equal to the separation length.  Because the volume of smoothing
grows with the separation length $l$, the direction of the local mean magnetic
field might change with $l$ at an arbitrary point.  This is an extension of the
procedures employed in \cite{cho02a}.}  In the case of subAlfv\'{e}nic
turbulence, the degree of anisotropy for density is very difficult to estimate
due to the high dispersion of points.  However, rough estimates suggest more
isotropic density structures, because the points extend along the line
$l_\parallel \sim l_\perp^1$.  For models with ${\cal M}_A \sim 7.3$ the points
in Figure~\ref{fig:anisotropy} have lower dispersion, and the anisotropy is more
like the type from GS95, i.e. $l_\parallel \sim l_\perp^{2/3}$.  In both
Alfv\'{e}nic regimes, the anisotropy of density does not change significantly
with ${\cal M}_s$.  Plots for the logarithm of density show more smooth
relations between parallel and perpendicular structure functions.  The
dispersion of points is very small.  Moreover, we note the change of anisotropy
with the scale.  Lower values of structure functions correspond to lower values
of the separation length (small-scale structures), so we might note that the
logarithm of density structures are more isotropic than the GS95 model at small
scales, but the anisotropy grows a bit larger than the GS95 prediction at larger
scales.  This difference is somewhat larger in the case of models with stronger
external magnetic field (compare plots in the left and right columns of
Figure~\ref{fig:anisotropy}), which may signify their dependence on the strength
of $B_\mathrm{ext}$.  The anisotropy of $\log \rho$ structures is marginally
dependent on the sonic Mach number, similar to the density structures.  All
these observations allow us to confirm the previous studies
\cite[see][]{beresnyak05} that suggest that the anisotropy depends not only on
the scale but on ${\cal M}_A$.

Probability distribution functions (PDFs) of density and column density give us
information about the fraction of the total volume occupied by a given value of
a measured quantity.  For the case of compressible turbulence in which evolution
is described by the Navier-Stokes equation and the isothermal equation of state,
the PDF of density obeys a lognormal form \citep{passot98}.  If we select a
point in space and assume that the density at this point results from subsequent
events perturbing the previous density, then the final density is a product
$\rho_0\Pi_i(1+\delta_i)$, where $\rho_0$ is the initial density at the selected
point and $\delta_i$ is a small compression/rarefaction factor.  By the power of
the Central Limit Theorem, the logarithm of the resulting density, $\log
(\rho/\rho_0)=\sum_i \log (1+\delta_i)$, should obey a Gaussian distribution.

It was discussed in \cite{vazquez01} that a parameter,  which we for our
convenience denote $\aleph$, can determine the form of the column density PDF in
molecular clouds. This parameter is the ratio of the cloud size to the
decorrelation length of the density field. They defined the decorrelation length
as the lag at which the density autocorrelation function (ACF) has decayed to
its 10\% level. If the density perturbation events are uncorrelated for $\aleph
> 1$, large values of the ratio $\aleph$ imply that the Central Limit Theorem
can be applied to those events.

The decorrelation length estimated from the ACF of the density for our models
ranges from about 20 cell sizes of the computational mesh for supersonic models
to about 50 cell sizes for subsonic models, which corresponds to $5 < \aleph <
13$ if we take the size of the computational box as the column length (in the
case of medium resolution, it is equal to 256 cell sizes).  Found values of
$\aleph$ signify that in our models at least partial convergence to a Gaussian
PDF should occur.

\begin{figure*}
 \centering
 \includegraphics[width=0.8\columnwidth]{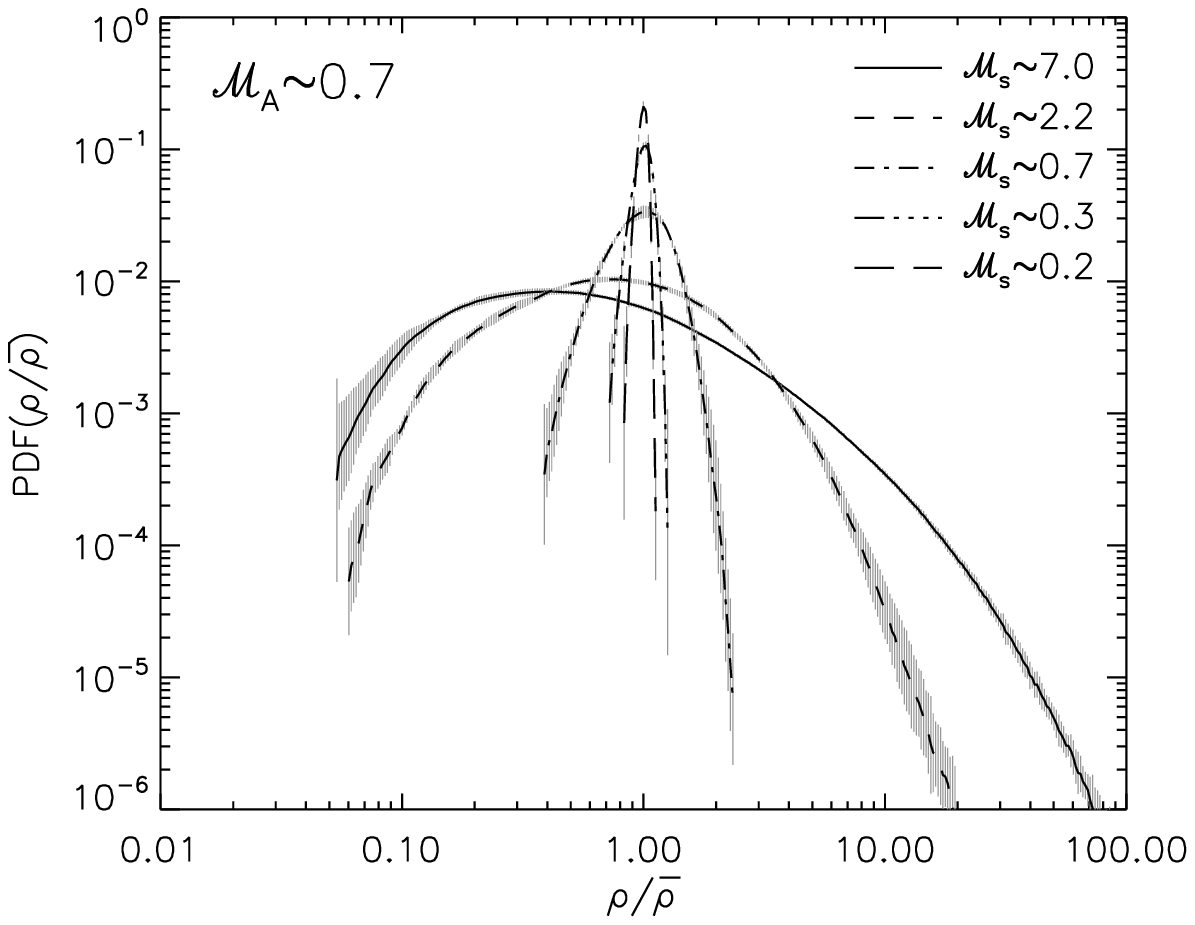}
 \includegraphics[width=0.8\columnwidth]{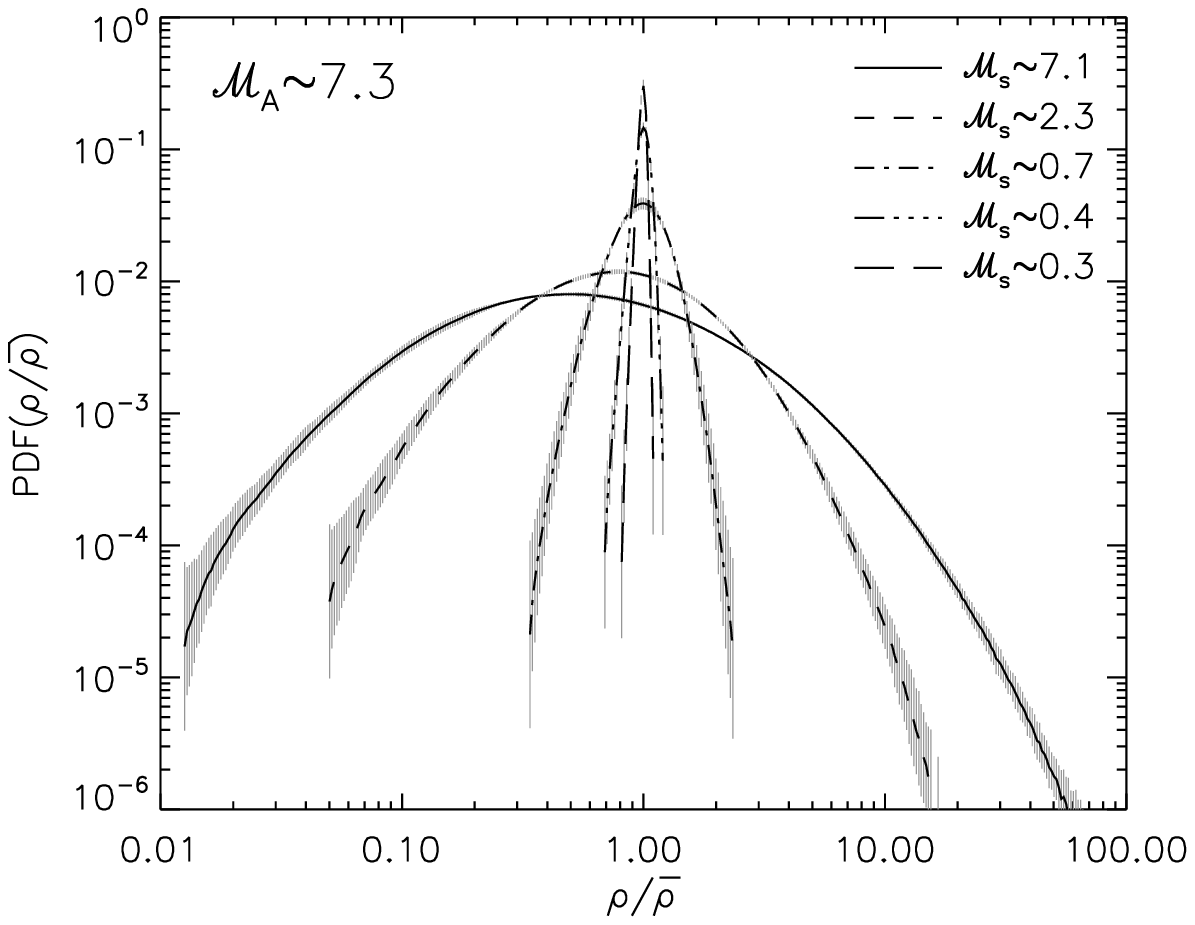}
 \caption{Normalized probability distribution functions (PDFs) of density
$\rho/\bar{\rho}$ ({\em top row}) for models with different values of ${\cal
M}_s$ and for ${\cal M}_A\sim0.7$ ({\em left column}) and ${\cal M}_A\sim7.3$
({\em right column}). Grey error bars signify the standard deviation of the
PDFs, showing the strength of the departure at a single time snapshot from its
mean profile averaged over time from $t\ge5$ to the last available snapshot.
From \cite{kowal07}. \label{fig:dens_pdfs}}
\end{figure*}

In the top row of Figure~\ref{fig:dens_pdfs} we show PDFs of density normalized
by its mean value for all models with medium resolution (256$^3$). The plot on
the left shows results obtained from sub-Alfv\'{e}nic experiments (${\cal
M}_A\sim0.7$), and the plot on the right results from super-Alfv\'{e}nic models
(${\cal M}_A\sim7.3$).

Our plots confirm the strong dependence of PDFs of density on the sonic Mach
number, an already known and well-studied property of density fluctuations in
compressible turbulence \citep[see][and references
therein]{vazquez01,ostriker03}. For most of the models PDFs are lognormal
functions. For super-Alfv\'{e}nic turbulence PDFs are very symmetric about a
vertical line crossing their maxima. However, for models with a strong external
magnetic field (sub-Alfv\'{e}nic turbulence, ${\cal M}_A\sim0.7$) and very low
pressure (supersonic turbulence, ${\cal M}_s > 1.0$), the shape of the density
PDFs is significantly deformed, and its lower value arm ends in higher densities
than in the case with a weak magnetic field (compare models for ${\cal
M}_s\sim7$ drawn with solid lines in the left and right plots of
Figure~\ref{fig:dens_pdfs}). This supports a hypothesis that the gauge symmetry
for $\log \rho$ that exists in Navier-Stokes equations is broken in MHD
equations because of the magnetic tension term \citep{beresnyak05}, which
physically manifests itself by preventing the formation of highly underdense
regions. In the higher density part of the distribution we do not see a similar
effect, because the highly dense structures are created mainly due to shocks,
and in this case, the magnetic tension is too weak to prevent the condensation.

In Figure~\ref{fig:dens_pdfs} we included the degree of variation of PDFs in
time as grey error bars. We note that the departure of PDFs from their mean
profiles is very small in the middle part around the mean value. The strongest
time changes are observed in the low- and high-density tails, but the PDFs for
different models are still separable.

\begin{figure*}
 \centering
 \includegraphics[width=0.66\columnwidth]{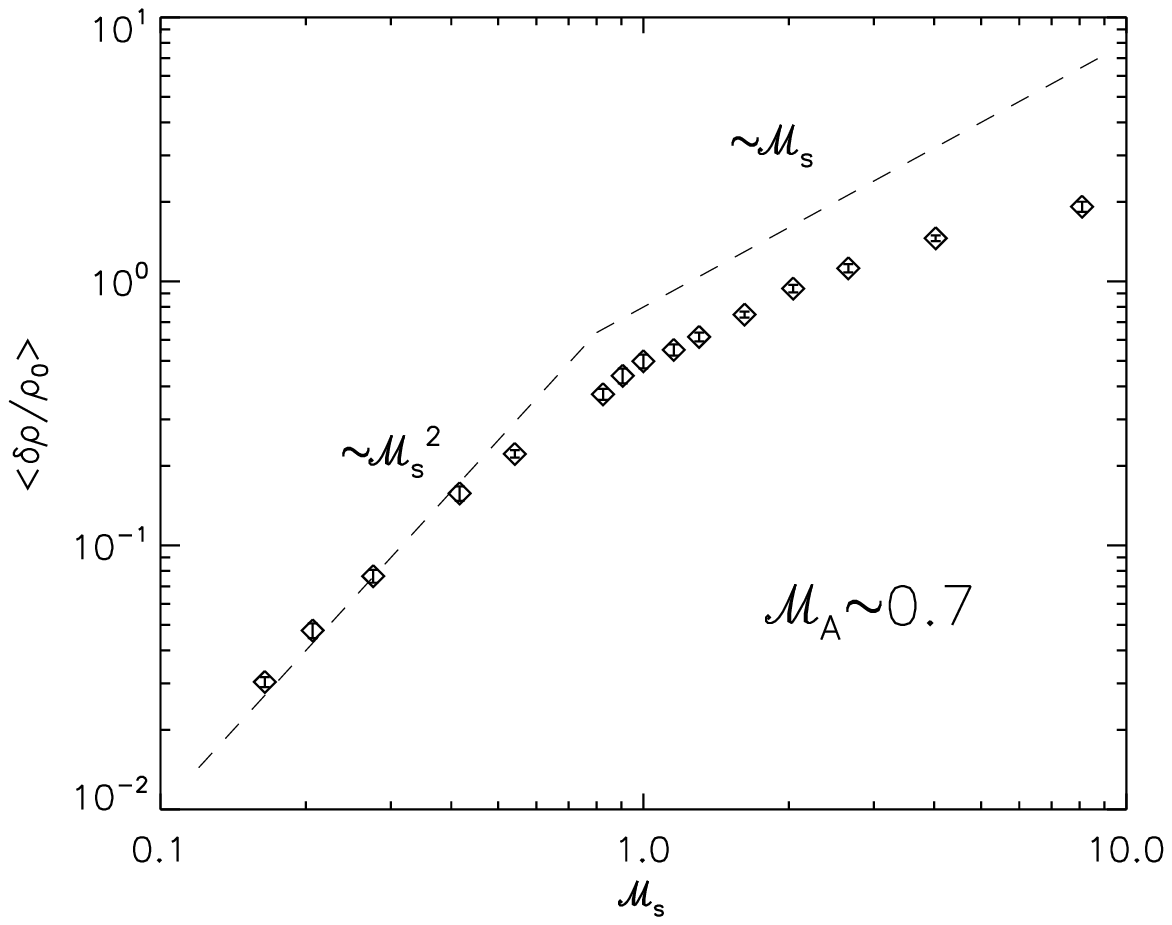}
 \includegraphics[width=0.66\columnwidth]{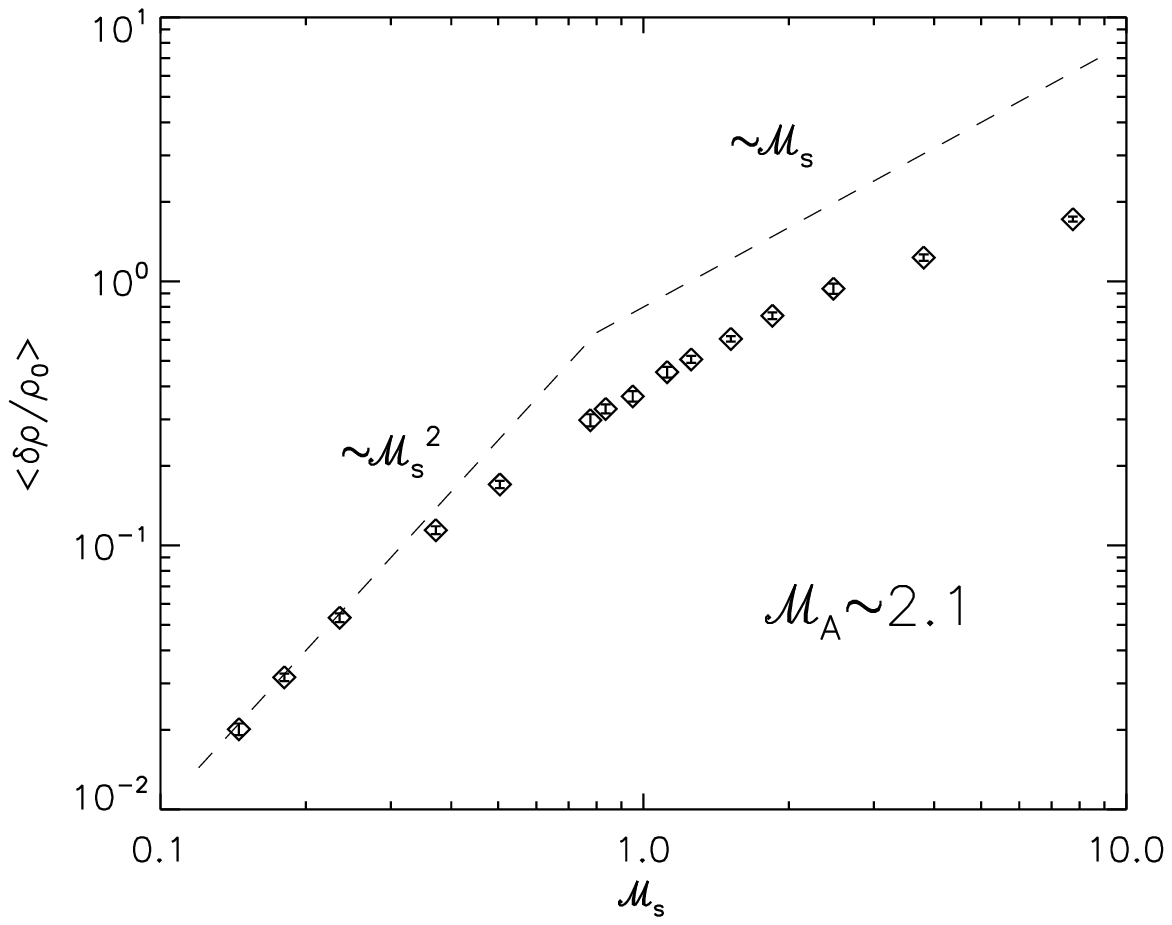}
 \includegraphics[width=0.66\columnwidth]{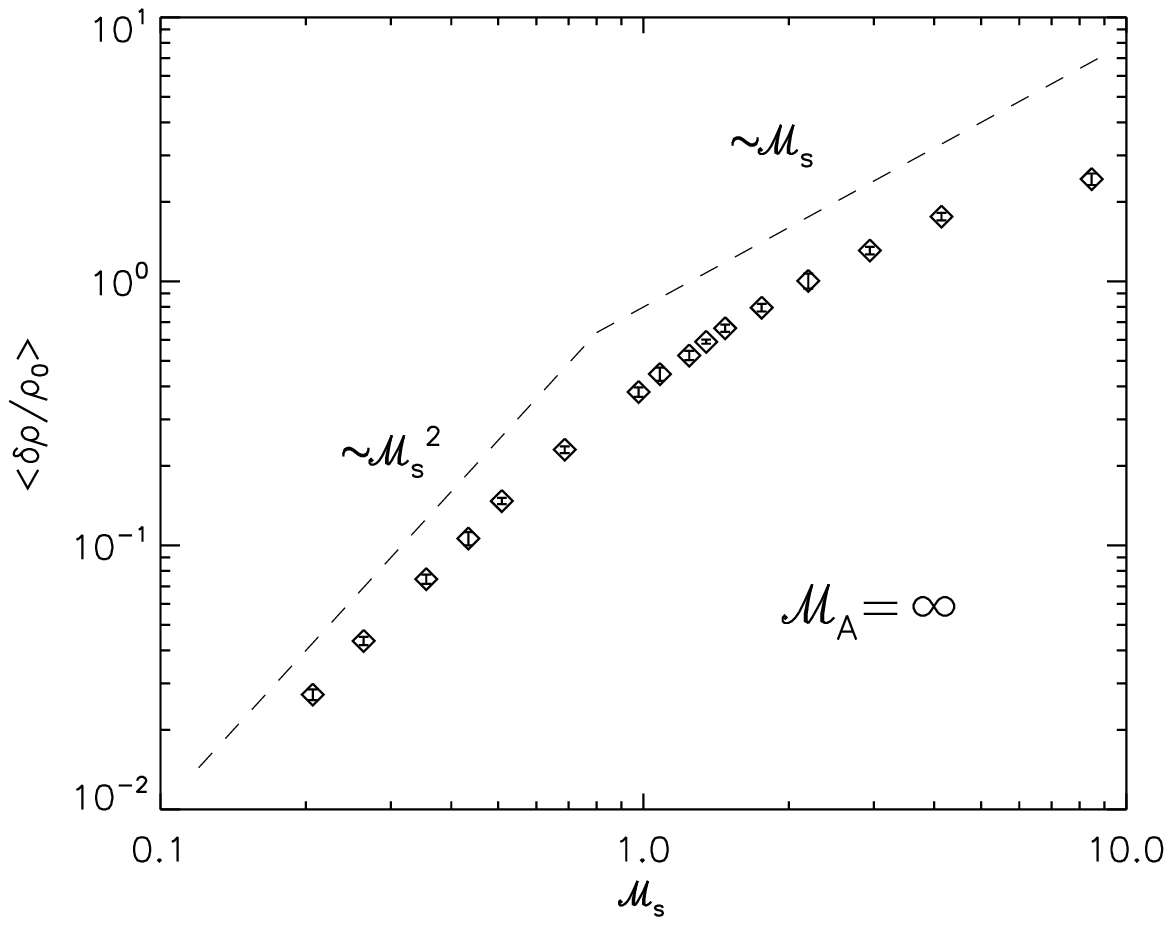}
 \caption{Relation between the mean value of the standard deviation of density
fluctuation $\langle \delta \rho/\rho_0 \rangle$ and the sonic Mach number
${\cal M}_s$ for models with low resolution. Two lines show the analytical
relations derived for magnetic-pressure-dominated turbulence ($\sim {\cal M}_s$)
and the gas-pressure-dominated turbulence ($\sim {\cal M}_s^2$).
\label{fig:var_mach}}
\end{figure*}

According to \cite{cho03}, the relations between the variance of density
fluctuations and the sonic Mach number are $\delta \rho/\rho_0 \sim {\cal M}_s$
when the magnetic pressure dominates and $\delta \rho/\rho_0 \sim {\cal M}_s^2$
when the gas pressure dominates. The former case is observed when ${\cal M}_s
\gg {\cal M}_A$. Indeed, in Figure~\ref{fig:var_mach} we see that the mean
standard deviation of the density fluctuations $\langle \delta \rho/\rho_0
\rangle$ scales with ${\cal M}_s$ almost linearly when ${\cal M}_s > {\cal
M}_A$. When ${\cal M}_s < {\cal M}_A$, the relation is much steeper. A similar
behavior is observed in the case of super-Alfv\'{e}nic turbulence
(Fig.~\ref{fig:var_mach}, {\em right}) although the gas pressure dominates in
all models. For models with a very small value of the sonic Mach number, the
relation is different. The value of $\langle \delta \rho/\rho_0 \rangle$ depends
much less on ${\cal M}_s$.

\section{Outstanding issues of MHD turbulence theory}

\subsection{The actual spectral slope of MHD turbulence}

Recently there has been a debate concerning an asymptotic scaling of MHD
turbulence. While the GS95 theory predicts Kolmogorov's -5/3 scaling, the
modification of theory in \cite{boldyrev05,boldyrev06} claims\footnote{The model
adds to GS95 the effect of the alignment of velocity and magnetic field, e.g.
\cite{ting86} and which has been first numerically quantified in 3D numerical
studies \cite{beresnyak06}. It requires that the dynamical alignment should
increase with the decrease of the scale, which, however, contradicts to
numerical calculations in \cite{beresnyak12}.},  the power law index of -3/2.
\cite{beresnyak09,beresnyak10} showed that MHD turbulence is less local then its
hydrodynamic counterpart and therefore the shallow spectrum in simulations is
expected as a result of an extended bottleneck intrinsic to MHD turbulence.
Below we discuss more recent numerical tests related to this controversy.

One of the way to discriminate different models is by means of direct numerical
simulations. A quantitative method called a resolution study is a traditional
way to establish a correspondance between numerics and theory. This method has
been used in the studies of hydrodynamic turbulence, e.g.
\cite{yeung97,gotoh02}, etc.  We perform several simulations with different
Reynolds numbers. If we believe that turbulence is universal and the scale
separation between forcing scale and dissipation scale is large enough, the
properties of small scales should not depend on how turbulence was driven. This
is because MHD equations and their numerical approximations contain no scale
explicitly, save for the grid scale, so simulation with a smaller dissipation
scale could be considered, as a simulation with the same dissipation scale,
which is a multiple of grid scale, but larger driving scale. E.g., the small
scale statistics in a $1024^3$ simulation should look the same as small-scale
statistics in $512^3$, if the physical size of the elementary cell is the same
and the dissipation scale is the same.

The above scaling argument requires that the geometry of the elementary cells is
the same and the actual numerical scheme used to solve the equations is the
same. The scaling argument does not require high precision on the dissipation
scale or a particular form of dissipation, either explicit or numerical. This is
because we only need the small-scale statistics to be similar in the particular
series of simulations. This is achieved by applying the same numerical
procedure. On the other hand, if turbulence is local, this ensures that small
scales are only marginally affected by large scales.  In practice, the scaling
argument, which is also known as a resolution study, is done in the following
way: the averaged spectra in two simulations are expressed in dimensionless
units corresponding to the expected scaling, for example a
$E(k)k^{5/3}\epsilon^{-2/3}$ is used for hydrodynamics, and plotted versus
dimensionless wavenumber $k\eta$, where dissipation scale $\eta$ correspond to
the same model, e.g. $\eta=(\nu^3/\epsilon)^{1/4}$ is used for scalar second
order viscosity $\nu$ and Kolmogorov phenomenology. As long as the spectra are
plotted this way and the scaling is correct, the curves obtained in simulations
with different resolutions have to collapse into a single plot on small scales.
Not only the spectrum, but any other statistical property of turbulence can be
treated this way.  When the resolution is increasing, the convergence is
supposed to become better. This is because of the increased separation of scale
between driving and dissipation. The optimum strategy for our MHD study is to
perform the largest resolution simulations possible and do the scaling study.

\begin{figure*}
\includegraphics[width=0.9\textwidth]{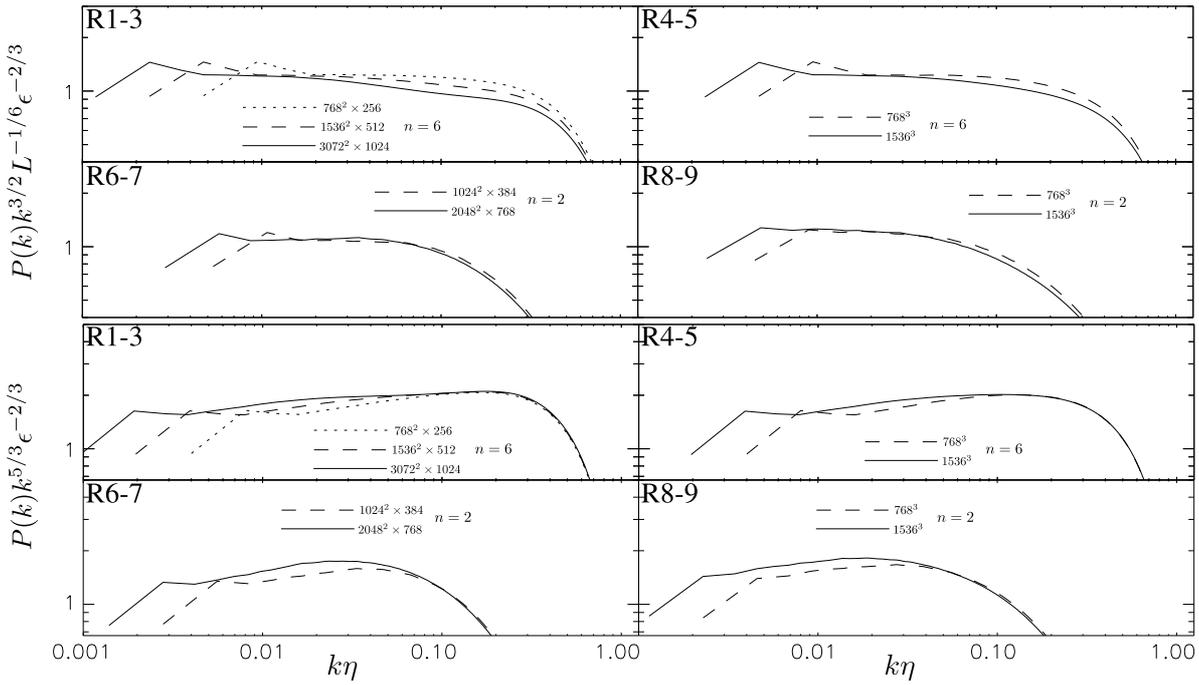}
\caption{Numerical convergence of spectra in 3D reduced MHD simulations. Two
upper rows are used to study convergence assuming Boldyrev model and two bottom
rows -- assuming Goldreich-Sridhar model. $P(k)$ is so-called one-dimensional
spectrum. Note that definition of dissipation scale $\eta$ depends on the model,
and although this difference is tiny in hyperviscous simulations R1-5, but
significant in viscous simulations R6-9. Numerical convergence require that
spectra will be similar on small scales, including the dissipation scale, see,
e.g. \cite{gotoh02}. As we see from the plots, numerical convergence is absent
for Boldyrev model. For Goldreich-Sridhar model the convergence is reached at
the dissipation scale. Higher-resolution simulations are required to demonstrate
convergence in the inertial range. Plotted with data in \cite{beresnyak10}.
\label{converg}}
\end{figure*}

Figure~\ref{converg} shows that the convergence study is suggestive of $-5/3$
spectral slope. The claims related to a more shallow slope are likely to be
related to the lesser degree of locality of MHD turbulence, which results in the
extended bottleneck region \citep{beresnyak10}. The question definitely requires
further studies, but we believe that at the moment there is not enough evidence
to claim that GS95 model contradicts to simulations.

\subsection{Extended current sheets in viscosity dominated turbulence}

\begin{figure*}
\hbox{%
\includegraphics[width=1.9 in]{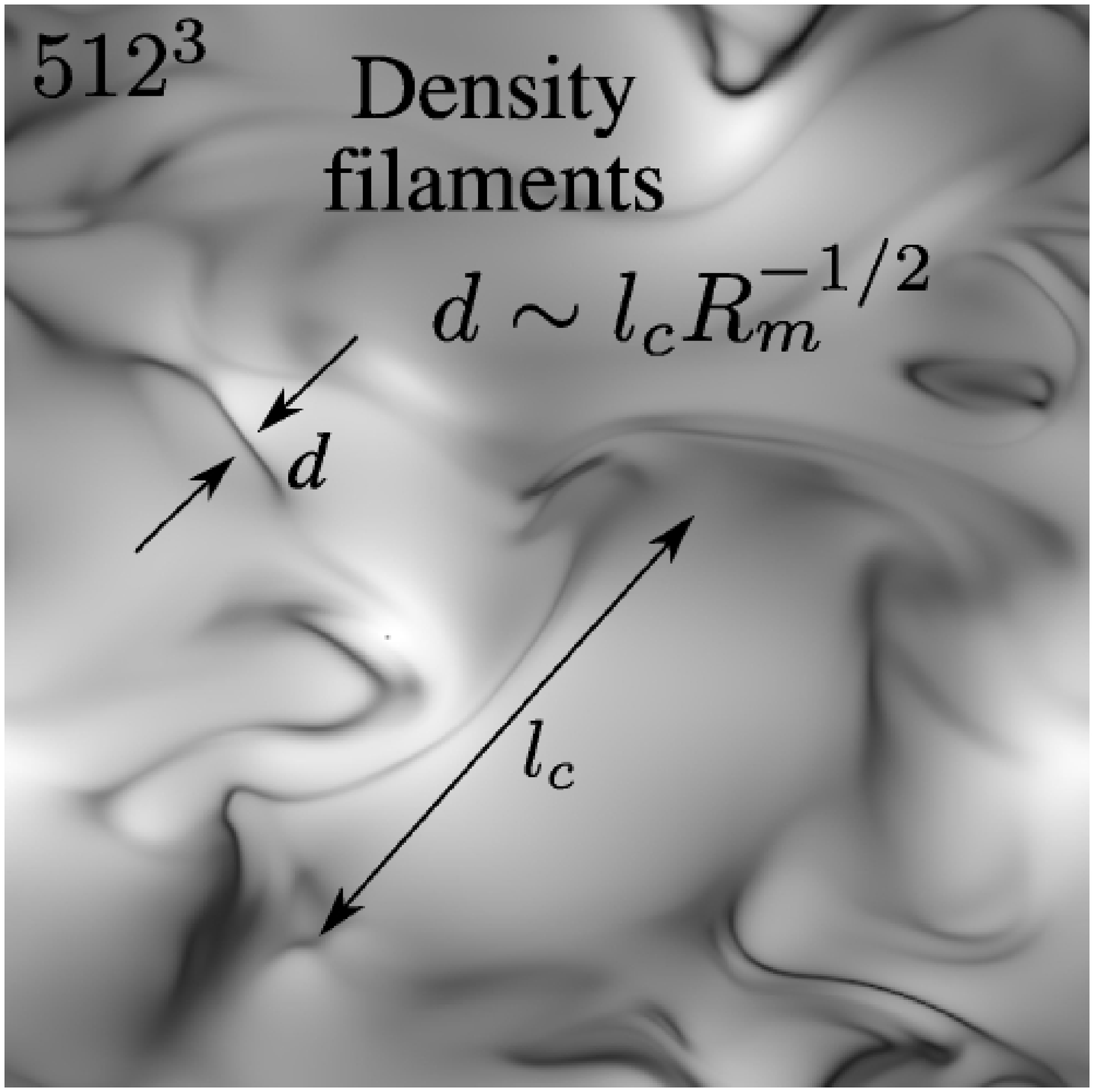}
\hfill
\includegraphics[width=2.1 in]{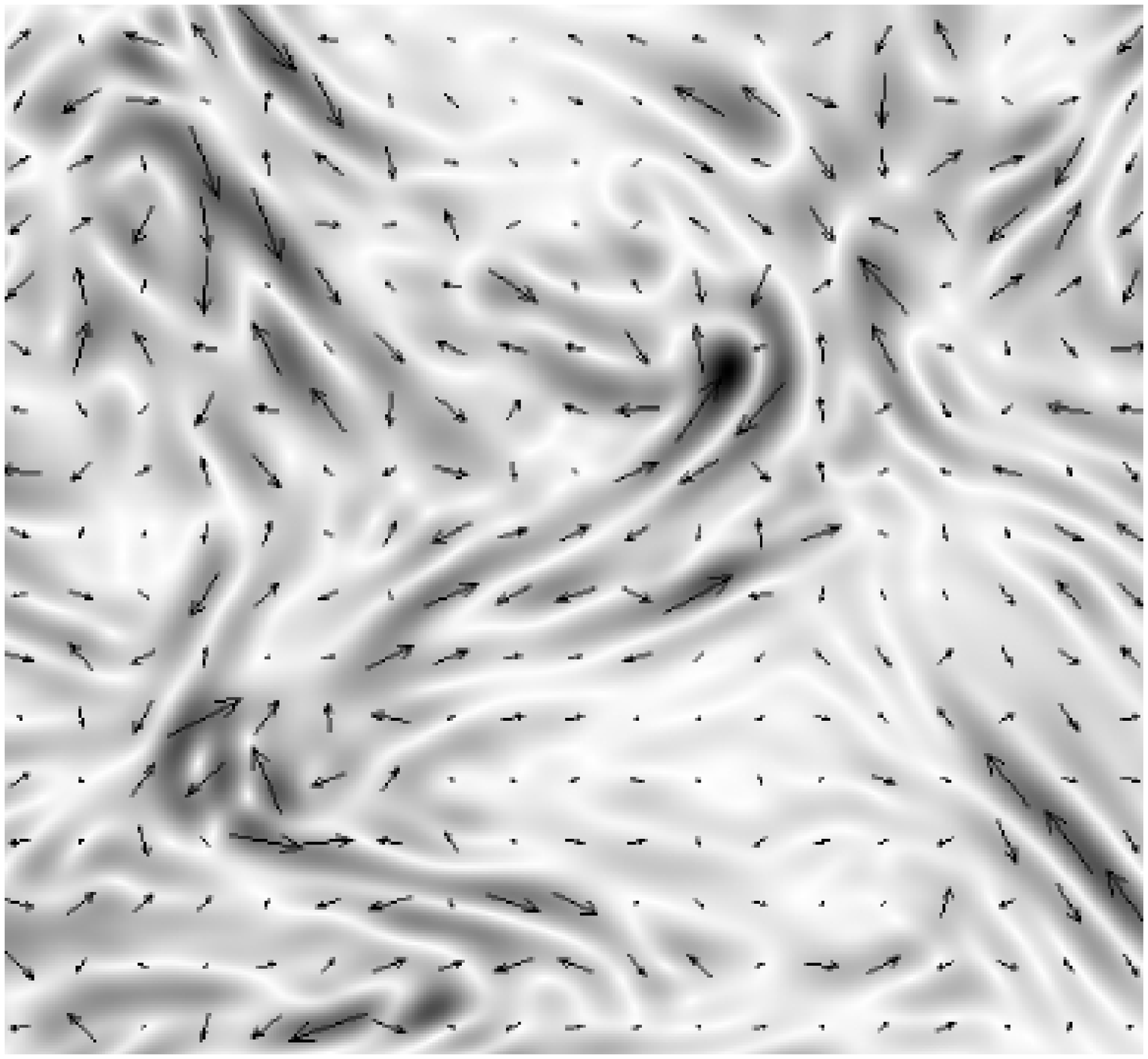}
\hfill
\includegraphics[width=2.5 in]{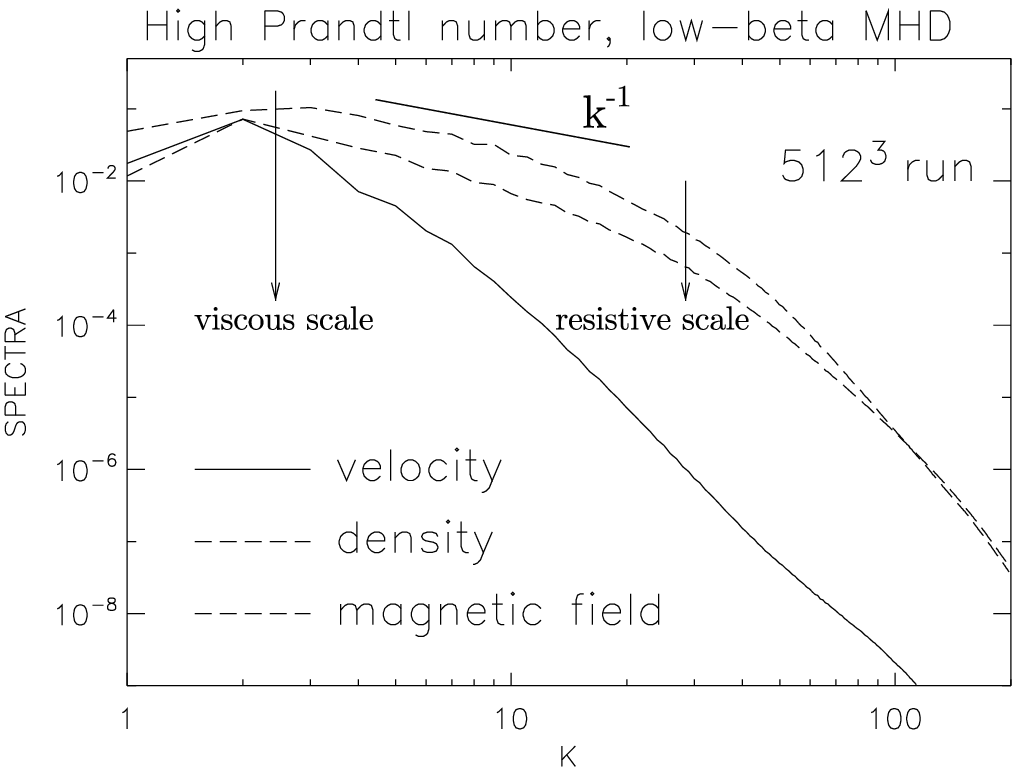}
}
\caption{\small {\it Left}: Filaments of density created by magnetic compression
of the gas in the viscosity-damped regime of MHD turbulence. Darker regions
correspond to higher density. The viscous damping scale $l_c$ is much larger
than the current sheet thickness $d$.  This creates large {\it observed} density
contrasts. {\it Center}: Magnetic reversals (in the plane $\bot$ to mean
$\langle{\bf B}\rangle$) that create compressions of density. Darker regions
correspond to higher magnetic field. {\it Right}: Spectra of density and
magnetic field are similar, while velocity is damped. The resistive scale in
this regime is not $L/Rm$ but $L {Rm}^{-1/2}$.  From Beresnyak \& Lazarian
(preprint).} \label{visc}
\end{figure*}

For high $Pr_m$ fluid \cite{cho02a} reported a new regime of MHD turbulence.
\cite{lazarian04a} showed that while the spectrum of volume-averaged magnetic
fluctuations scales as $E_b(k)\sim k^{-1}$ (see Figure~\ref{visc}), the pressure
within intermittent magnetic structures increases with the decrease of the scale
$\propto k$ and the filling factor decreases $\propto k^{-1}$. The magnetic
pressure compresses the gas as demonstrated in Figure~\ref{visc}. More
importantly, extended current sheets that naturally emerge as magnetic field
fluctuates in the plane perpendicular to the mean magnetic field (see
Figure~\ref{visc}). It was speculated in \cite{lazarian07} that these current
sheets can account for the origin of the small ionized and neutral structures
(SINS) on AU spatial scales \cite{dieter76,heiles97,stanimirovic04}.  These
current sheets may present sites of cosmic ray acceleration.

\cite{goldreich06} appealed not to high $Pr_m$ MHD turbulence per se, but to the
generation of the magnetic field in the turbulent plasma
\cite[see][]{schekochihin04} to account for the high amplitude, but small scale
fluctuations of plasma density observed in the direction of the Galactic center.
We believe that the regime of dynamo in \cite{schekochihin04} and the turbulence
in \cite{lazarian04a} have similarities in terms of the density enhancement that
are created.  Although in the case of magnetic turbulence with sufficiently
strong mean magnetic field, global reversals, that \cite{goldreich06} appeal to
in compressing plasma, do not happen, the reversals of the magnetic field
direction occur in the direction perpendicular to the mean magnetic field. As
the mean magnetic field goes to zero, the two regimes get indistinguishable. The
regime of viscosity damped turbulence requires further systematic studies.

\subsection{Imbalanced MHD turbulence}

\def\L{{\Lambda}}
\def\l{{\lambda}}

MHD turbulence in the presence of sources and sinks gets {\it imbalanced}, in
the sense that the flow of energy in one direction is larger than the flow of
energy in the opposite direction. Solar wind presents a vivid example of
imbalanced turbulence with most waves near the Sun moving in the direction away
from the Sun. While theories of balanced MHD turbulence enjoyed much attention,
the theory of imbalanced turbulence\footnote{Another name for imbalanced
turbulence is a turbulence with non-zero cross-helicity.} attracted less work,
unfortunately. Important papers on imbalanced turbulence include \cite[see][and
references therein]{ting86, matthaeus83, biskamp03}. In terms of Solar wind
observations one may mentioned studies by \cite{roberts87a, roberts87b}, which
showed that the imbalance of turbulence is not increasing contrary to the
idealized theoretical expectations. The analytical results were obtained for
{\it weak imbalanced turbulence} \citep{galtier02,lithwick03} and they are
applicable in a rather narrow range of imbalance ratios.  Some earlier
simulations of strong imbalanced turbulence were limited to rather idealized set
ups \citep{maron01,cho02a}, i.e. for the initial state the results of the
simulations of strong balanced turbulence were used, but the amplitudes of waves
moving in one direction were reduced. This did not allow making definitive
conclusions about properties of imbalanced turbulence.

Attempts to construct the model of stationary {\it strong} imbalanced turbulence
were done in \cite{lithwick07,beresnyak08,chandran08,perez09}.  Below we discuss
only the model in \cite{beresnyak08} as this is the only model that agrees with
numerical simulations performed so far \citep{beresnyak10}.  This model can be
viewed as an extension of GS95 model into the imbalanced regime.
\begin{figure*}
\includegraphics{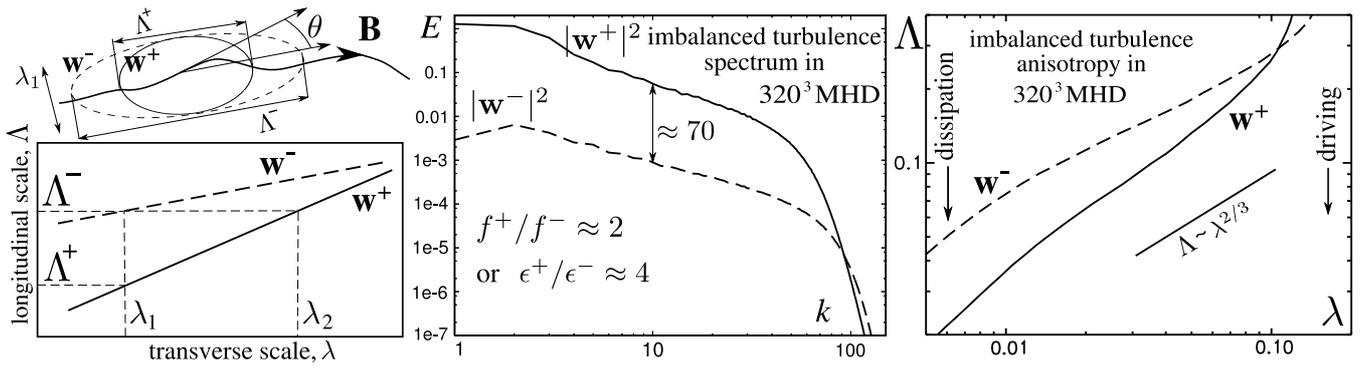}
\caption{Left upper: ${\bf w}^+$ wavepacket, produced by cascading by ${\bf
w}^-$ wavepacket is aligned with respect to ${\bf w}^-$ wavepacket, but
disaligned with respect to the local mean field on scale $\lambda_1$, by the
angle $\theta$. Left lower: the longitudinal scale $\L$ of the wavepackets, as a
function of their transverse scale, $\l$; Middle: the power spectrum of energies
for both waves in an imbalanced forced incompressive $320^3$ numerical
simulation. Right: the $\L(\l)$ dependence in the same simulation, the length
scales are in the units of the cube size. From \cite{beresnyak08}.}
\label{imbalance}
\end{figure*}

While the classic formulation of the GS95 critical balance, based on causality,
is unable to describe consistently the imbalanced case, \cite{beresnyak08}
proposed a new way to relate parallel and perpendicular scales of Alfv\'enic
modes. The new condition was obtained appealing to the process termed
"propagation balance". In the case of the balanced turbulence the "old" and
"new" critical balance condition results in the same GS95 scaling. However, in
the case of imbalanced turbulence the new formulation predicts {\it smaller}
anisotropy for the stronger wave (which directly contradicts old causal critical
balance) is consistent with simulations. The \cite{beresnyak08} model predicts
that the anisotropies of the two waves are different and this should be taken
into account for calculating the scattering and acceleration arising from the
particle-wave interactions.

\section{Magnetic reconnection in astrophysical fluids}

\subsection{Problem of magnetic reconnection}

A magnetic field embedded in a perfectly conducting fluid is generally believed
to preserves its topology for all time \citep{parker79}.  Although ionized
astrophysical objects, like stars and galactic disks, are almost perfectly
conducting, they show indications of changes in topology, ``magnetic
reconnection'', on dynamical time scales \citep{parker70,lovelace76,priest02}.
Reconnection can be observed directly in the solar corona
\citep{innes97,yokoyama95,masuda94}, but can also be inferred from the existence
of large scale dynamo activity inside stellar interiors
\citep{parker93,ossendrijver03}.  Solar flares \citep{sturrock66} and
$\gamma$-ray busts \citep{fox05,galama98} are usually associated with magnetic
reconnection.  A lot of work in the field has concentrated on showing how
reconnection can be rapid in plasmas with very small collisional rates
\citep{shay98,drake01,drake06,daughton06}, which substantially constrains
astrophysical applications of the corresponding reconnection models. The
disregard of pre-existing turbulence is another shortcoming of these models.

A picture of two flux tubes of different directions which get into contact in 3D
space is a generic framework to describe magnetic reconnection. The upper panel
of Figure~\ref{fig_rec} illustrates why reconnection is so slow in the textbook
Sweet-Parker model. Indeed, the model considers magnetic fields that are laminar
and therefore the frozen-in condition for magnetic field is violated only over a
thin layer dominated by plasma resistivity. The scales over which the resistive
diffusion is important are microscopic and therefore the layer is very thin,
i.e. $\Delta\ll L_x$, where $L_x$ is the scale at which magnetic flux tubes come
into contact. The latter is of the order of the diameter of the flux tubes and
typically very large for astrophysical conditions. During the process of
magnetic reconnection all the plasma and the shared magnetic
flux\footnote{Figure~\ref{fig_rec} presents a cross section of the 3D
reconnection layer. A guide field is present in the generic 3D configurations of
reconnecting magnetic flux tubes.} arriving over an astrophysical scale $L_x$
should be ejected through a microscopic slot of thickness $\Delta$. As the
ejection velocity of magnetized plasmas is limited by Alfv\'en velocity $V_A$,
this automatically means that the velocity in the vertical direction, which is
reconnection velocity, is much smaller than $V_A$.

We note that if magnetic reconnection is slow in some astrophysical
environments, this automatically means that the results of present day numerical
simulations in which the reconnection is inevitably fast due to numerical
diffusivity do not correctly represent magnetic field dynamics in these
environments. If, for instance, the reconnection were slow in collisional media
this would entail the conclusion that the entire crop of interstellar,
protostellar and stellar MHD calculations would be astrophysically irrelevant.
To make the situation more complicated, solar flares demonstrate both periods of
slow reconnection when the accumulation of magnetic flux takes place and periods
of fast reconnection when the flare actually occurs. This bimodal character is
another property that a successful model of reconnection should address.

In this section we focus our attention on the reconnection that takes place in
turbulent astrophysical fluids.  The reconnection in such environments was
described in \cite{lazarian99}.  The latter model identified magnetic field
wandering as the key process that induces fast, i.e. independent of fluid
resistivity, magnetic reconnection.  The predicted dependences of magnetic
reconnection on the properties of turbulence, i.e. on the intensity and
injection scale of turbulence, have been successfully tested in
\cite{kowal09,kowal12b}.  The deep consequence of the model is the violation of
the magnetic flux freezing in turbulent fluids \cite[e.g. the explicit claim of
this in][]{vishniac99}.  The connection of the LV99 model with more recent ideas
on the properties of magnetic field in turbulent plasmas are analyzed in
\cite{eyink11}.

While the discussion of possible ways how turbulence can enhance reconnection
rates was not unprecedented, the LV99 model was radically different from what
was proposed earlier.  For instance, the closest to the spirit of LV99, are two
papers by \cite{matthaeus85,matthaeus86}, where the authors performed 2D
numerical simulations of turbulence and provided arguments in favor of magnetic
reconnection getting fast. However, the physics of the processes that they
considered was very different from that in LV99.  For instance, the key process
of field wandering of the LV99 model has not been considered by
\cite{matthaeus85,matthaeus86}.  On the contrary, the components of their
approach, e.g. X-point and possible effects of heating and compressibility are
not ingredients of the LV99 model.  Other papers, e.g. \cite{speiser70} and
\cite{jacobson84} are even more distant in terms of the effects that they
explored, namely, they studied the changes of the microscopic properties of the
plasma induced by turbulence and considered how these changes can accelerate
magnetic reconnection. At the same time, LV99 shows that the microscopic plasma
properties are irrelevant for their model of reconnection.

Below (\S\ref{sec:lv99_test}) we present numerical evidence, based on three
dimensional simulations, that reconnection in a turbulent fluid occurs at a
speed comparable to the rms velocity of the turbulence, regardless of either the
value of the resistivity or degree of collisionality.  In particular, this is
true for turbulent pressures much weaker than the magnetic field pressure so
that the magnetic field lines are only slightly bent by the turbulence.  These
results are consistent with the proposal by LV99 that reconnection is controlled
by the stochastic diffusion of magnetic field lines, which produces a broad
outflow of plasma from the reconnection zone.  This implies that reconnection in
a turbulent fluid typically takes place in approximately a single eddy turnover
time, with broad implications for dynamo activity
\citep{parker70,parker93,stix00} and particle acceleration throughout the
universe \citep{degouveia03,degouveia05,lazarian05,drake06}.

\subsection{Rate of reconnection}

Astrophysical plasmas are often highly ionized and highly magnetized
\citep{parker70}.  The evolution of the magnetic field in a highly conducting
fluid can be described by a simple version of the induction equation
\begin{equation}
\frac{\partial \vec{B}}{\partial t} = \nabla \times \left( \vec{v} \times \vec{B} - \eta \nabla \times \vec{B} \right) ,
\end{equation}
where $\vec{B}$ is the magnetic field, $\vec{v}$ is the velocity field, and
$\eta$ is the resistivity coefficient.  Under most circumstances this is
adequate for discussing the evolution of magnetic field in an astrophysical
plasma.  When the dissipative term on the right hand side is small, as is
implied by simple dimensional estimates, the magnetic flux through any fluid
element is constant in time and the field topology is an invariant of motion. On
the other hand, reconnection is observed in the solar corona and chromosphere
\citep{innes97,yokoyama95,masuda94,ciaravella08}, its presence is required to
explain dynamo action in stars and galactic disks \citep{parker70,parker93}, and
the violent relaxation of magnetic fields following a change in topology is a
prime candidate for the acceleration of high energy particles
\citep{degouveia03,degouveia05,lazarian05,drake06,lazarian09,drake10} in the
universe.   Quantitative general estimates for the speed of reconnection start
with two adjacent volumes with different large scale magnetic field directions
\citep{sweet58,parker57}.

\begin{figure}[!t]
 \begin{center}
\includegraphics[width=1.0 \columnwidth]{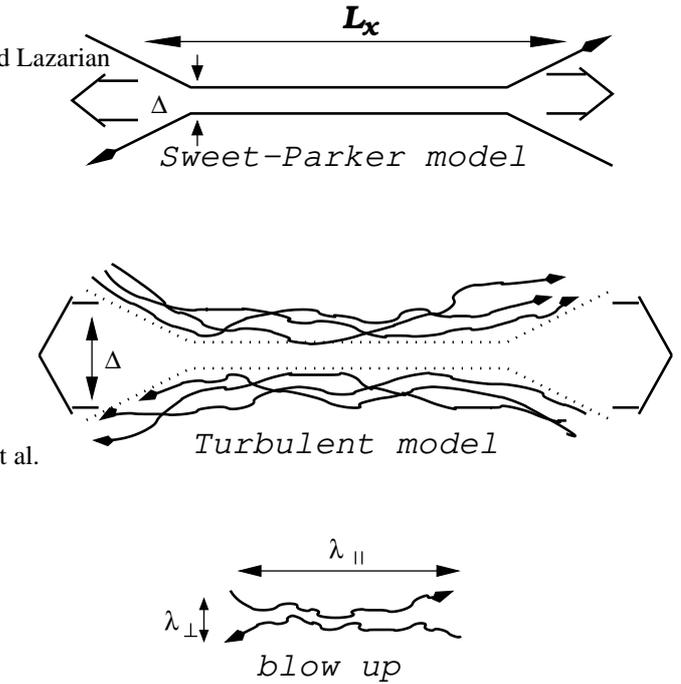}
\caption{
{\it Upper plot}: Sweet-Parker model of reconnection. The outflow is limited by
a thin slot $\Delta$, which is determined by Ohmic diffusivity. The other scale
is an astrophysical scale $L\gg \Delta$.
{\it Middle plot}: Reconnection of weakly stochastic magnetic field according to
LV99. The model that accounts for the stochasticity of magnetic field lines. The
outflow is limited by the diffusion of magnetic field lines, which depends on
field line stochasticity.
{\it Low plot}: An individual small scale reconnection region. The reconnection
over small patches of magnetic field determines the local reconnection rate. The
global reconnection rate is substantially larger as many independent patches
come together. From \cite{lazarian04a}.
} \label{fig_rec}
 \end{center}
\end{figure}

The speed of reconnection, i.e. the speed at which inflowing magnetic field is
annihilated by Ohmic dissipation, is roughly $\eta/\Delta$, where $\Delta$ is
the width of the transition zone (see Figure~\ref{fig_rec}).  Since the
entrained plasma follows the local field lines, and exits through the edges of
the current sheet at roughly the Alfv\'en speed, $V_A$, the resulting
reconnection speed is a tiny fraction of the Alfv\'en speed, $V_A\equiv B/(4\pi
\rho)^{1/2}$ where $L$ is the length of the current sheet.  When the current
sheet is long and the reconnection speed is slow this is referred to as
Sweet-Parker reconnection. Observations require a speed close to $V_A$, so this
expression implies that $L\sim \Delta$, i.e. that the magnetic field lines
reconnect in an ``X point''.

The first model with a stable X point was proposed by \cite{petschek64}.  In
this case the reconnection speed may have little or no dependence on the
resistivity.  The X point configuration is known to be unstable to collapse into
a sheet in the MHD regime \cite[see][]{biskamp96}, but in a collisionless plasma
it can be maintained through coupling to a dispersive plasma mode
\citep{sturrock66}.  This leads to fast reconnection, but with important
limitations.  This process has a limited astrophysical applicability as it
cannot be important for most phases of the interstellar medium \citep[see][for a
list of the idealized phases]{draine98}, not to speak about dense plasmas, such
as stellar interiors and the denser parts of accretion disks.  In addition, it
can only work if the magnetic fields are not wound around each other, producing
a saddle shaped current sheet. In that case the energy required to open up an X
point is prohibitive. The saddle current sheet is generic for not parallel flux
tubes trying to pass through each other. If such a passage is seriously
constrained, the magnetized highly conducting astrophysical fluids should behave
more like Jello rather than normal fluids\footnote{The idea of felt or
Jello-type structure of magnetic fields in interstellar medium was advocated by
Don Cox (private communication).}.

Finally, the traditional reconnection setup does not include ubiquitous
astrophysical turbulence\footnote{The set ups where instabilities play important
role include \cite{shimizu09a,shimizu09b}. For sufficiently large resolution of
simulations those set-ups are expected to demonstrate turbulence. Turbulence
initiation is also expected in the presence of plasmoid ejection
\citep{shibata01}. Numerical viscosity constrains our ability to sustain
turbulence via reconnection, however.}
\cite[see][]{armstrong95,elmegreen04,mckee07,lazarian09,chepurnov10a}.
Fortunately, this approach provides another way of accelerating reconnection.
Indeed, an alternative approach is to consider ways to decouple the width of the
plasma outflow region from $\Delta$.  The plasma is constrained to move along
magnetic field lines, but not necessarily in the direction of the mean magnetic
field.  In a turbulent medium the two are decoupled, and fluid elements that
have some small initial separation will be separated by a large eddy scale or
more after moving the length of the current sheet.  As long as this separation
is larger than the width of the current sheet, the result will not depend on
$\eta$.

LV99 introduced a model that included the effects of magnetic field line
wandering (see Figure~\ref{fig_rec}). The model relies on the nature of
three-dimensional magnetic field wandering in turbulence.  The effects of
compressibility and heating, which were thought to be important in the earlier
studies \citep[see, for example, the discussion in][]{matthaeus85,matthaeus86},
do not play the role for the LV99 model either.  The model is applicable to any
weakly perturbed magnetized fluid, irrespectively, of the degree of plasma being
collisional or collisionless \citep[cf.][]{shay98}.

Two effects are the most important for understanding of the nature of
reconnection in LV99.  First of all, in three dimensions bundles of magnetic
field lines can enter the reconnection region and reconnection there
independently (see Figure~\ref{fig_rec}), which is in contrast to two
dimensional picture where in Sweet-Parker reconnection the process is
artificially constrained. Then, the nature of magnetic field stochasticity and
therefore magnetic field wandering (which determines the outflow thickness, as
illustrated in Figure~\ref{fig_rec}) is very different in 2D and the real 3D
world (LV99). In other words, removing artificial constraints on the
dimensionality of the reconnection region and the magnetic field being
absolutely straight, LV99 explores the real-world astrophysical reconnection.

The scaling relations for Alfv\'enic turbulence allow us to calculate the rate
of magnetic field spreading. This is the effect known to be important for the
cosmic ray propagation \cite{jokipii73} and LV99 quantified it for the case of
GS95 turbulence\footnote{This treatment was used later for dealing with
different problems, from heat propagation in plasmas \cite{narayan01,
lazarian06}  to cosmic ray propagation \cite{yan08a}. }. A bundle of field lines
confined within a region of width $y$ at some particular point will spread out
perpendicular to the mean magnetic field direction as one moves in either
direction following the local magnetic field lines.  The rate of field line
diffusion is given approximately by
\be
{d\langle y^2\rangle\over dx}\sim {\langle y^2\rangle\over \lambda_{\|}},
\ee
where $\lambda_{\|}^{-1}\approx \ell_{\|}^{-1}$, $\ell_{\|}$ is the parallel
scale chosen so that the corresponding vertical scale, $\ell_{\perp}$, is $\sim
\langle y^2\rangle^{1/2}$, and $x$ is the distance along an axis parallel to the
mean magnetic field. Therefore, using equation (\ref{Lambda}) one gets
\be
{d\langle y^2\rangle\over dx}\sim L_i\left({\langle y^2\rangle\over L_i^2}\right)^{2/3}
\left({u_L\over v_A}\right)^{4/3}
\label{eq:diffuse}
\ee
where we have substituted $\langle y^2\rangle ^{1/2}$ for $\ell_{\perp}$.  This
expression for the diffusion coefficient will only apply when $y$ is small
enough for us to use the strong turbulence scaling relations, or in other words
when $\langle y^2\rangle < L_i^2(u_L/v_A)^4$.  Larger bundles will diffuse at a
maximum rate of $L_i(u_L/v_A)^4$.  For $\langle y^2\rangle$ small equation
(\ref{eq:diffuse}) implies that a given field line will wander perpendicular to
the mean field line direction by an average amount
\be
\langle y^2\rangle^{1/2}\approx {x^{3/2}\over L_i^{1/2}} \left({u_L\over v_A}\right)^{2}
\label{eq:diffuse2}
\ee
in a distance $x$.  The fact that the rms perpendicular displacement grows
faster than $x$ is significant.  It implies that if we consider a reconnection
zone, a given magnetic flux element that wanders out of the zone has only a
small probability of wandering back into it.

The advantage of the classical Sweet-Parker scheme of reconnection is that it
naturally follows from the idea of Ohmic diffusion. Indeed, mass conservation
requires that the inflow of matter through the scale of the contact region $L_x$
be equal to the outflow of matter through the diffusion layer $\Delta$, i.e.
\be
v_{rec}= v_A {{\Delta}\over{L_x}}.
\lb{vrec}
\ee

The mean-square vertical distance that a magnetic field-line can diffuse by
resistivity in time $t$ is
\be
\langle y^2(t)\rangle \sim \lambda t.
\lb{diff-dist}
\ee
The field lines are advected out of the sides of the reconnection layer of
length $L_x$ at a velocity of order $v_A.$ Thus, the time that the lines can
spend in the resistive layer is the Alfv\'en crossing time $t_A=L_x/v_A.$ Thus,
field lines can only be merged that are separated by a distance
\be
\Delta = \sqrt{\langle y^2(t_A)\rangle} \sim \sqrt{\lambda t_A} = L_x/\sqrt{S},
\lb{Delta}
\ee
where $S$ is Lundquist number. Combining Eqs. (\ref{vrec}) and (\ref{Delta}) one
gets the famous Sweet-Parker reconnection rate, $v_{rec}=v_A/\sqrt{S}$.

In LV99 magnetic field wandering determines the scale of the outflow $\Delta$
(see Figure~\ref{fig_rec}). Using expressions from the earlier section one can
obtain (LV99):
\be
V_{rec}<v_A\min\left[\left({L_x\over L_i}\right)^{1/2},
\left({L_i\over L_x}\right)^{1/2}\right]
(u_L/V_A)^2.
\label{eq:lim2a}
\ee
This limit on the reconnection speed is fast, both in the sense that it does not
depend on the resistivity, and in the sense that it represents a large fraction
of the Alfv\'en speed. To prove that Eq.~(\ref{eq:lim2a}) indeed constitutes the
reconnection rate LV99 goes through a thorough job of considering all other
possible bottlenecks for the reconnection and shows that they provide higher
reconnection speed.

Below we provide a new derivation of the LV99 reconnection rates which makes
apparent that the LV99 model is a natural generalization of the laminar
Sweet-Parker model to flows with background turbulence. The new argument in
\cite{eyink11} is based on the concept of Richardson diffusion. It is known in
hydrodynamic turbulence that the combination of small scale diffusion and large
scale shear gives rise to Richardson diffusion, where the mean square separation
between two particles grows as $t^3$ once the rms separation exceeds the viscous
damping scale. A similar phenomenon occurs in MHD turbulence.  In both cases the
separation at late times is independent of the microscopic transport
coefficients.  Although the plasma is constrained to move along magnetic field
lines, the combination of turbulence and ohmic dissipation produces a
macroscopic region of points that are "downstream" from the same initial volume,
even in the limit of vanishing resistivity.

Richardson diffusion \citep[see][]{kupiainen03} implies the mean squared
separation of particles $\langle |x_1(t)-x_2(t)|^2 \rangle\approx \epsilon t^3$,
where $t$ is time, $\epsilon$ is the energy cascading rate and
$\langle\cdot\rangle$ denotes an ensemble averaging. For subAlfvenic turbulence
$\epsilon\approx u_L^4/(v_A L_i)$ (see LV99) and therefore analogously to Eq.
(\ref{Delta}) one can write
\be
\Delta\approx \sqrt{\epsilon t_A^3}\approx L(L/L_i)^{1/2}(u_L/V_A)^2
\label{D2}
\ee
where it is assumed that $L<L_i$. Combining Eqs. (\ref{vrec}) and (\ref{D2}) one
gets
\be
V_{rec, LV99}\approx v_A (L/L_i)^{1/2}(u_L/V_A)^2.
\label{LV99}
\ee
in the limit of $L<L_i$. Analogous considerations allow to recover the LV99
expression for $L>L_i$, which differs from Eq.~(\ref{LV99}) by the change of the
power $1/2$ to $-1/2$. These results coincide with those given by
Eq.~(\ref{eq:lim2a}).

It is important to stress that Richardson diffusion ultimately leads to
diffusion over the entire width of large scale eddies once the plasma has moved
the length of one such eddy. The precise scaling exponents for the turbulent
cascade does not affect this result, and all of the alternative scalings
considered in LV99 yield the same behavior.

Other forms of reconnection rate are also useful. For instance, one can take
into account that the velocity of the transition to the strong turbulent regime
is $u_{turb, strong}\sim v_A M_A^2$. Then
\begin{equation}
V_{rec}\approx u_{turb, strong}\left(l/L_i\right)^{1/2},
\label{recon1}
\end{equation}
shows that the reconnection rate scales with the velocity of the strong
turbulence cascade. Similarly, it is useful to rewrite this expression in terms
of the power injection rate $P_{inj}$. As the perturbations on the injection
scale of turbulence are assumed to have velocities $u_l<V_A$, the turbulence is
weak at large scales. Therefore, the relation between the power and the
injection velocities are different from the usual Kolmogorov estimate, namely,
in the case of the weak turbulence $P\sim u_l^4/(lV_A)$ (LV99). Thus we get,
\begin{equation}
V_{rec}\approx \left(\frac{P_{inj}}{LV_A}\right)^{1/2} l,
\label{recon2}
\end{equation}
where $l$ is the length of the turbulent eddies parallel to the large scale
magnetic field lines as well as the injection scale.

If turbulence is superAlfv\'enic, LV99 predicts that the diffusion of magnetic
field happens with the turbulent velocity. This is possible as the subAlfv\'enic
reconnection that happens below the scale $l_A$ given by Eq. (\ref{alf}) does
not present the bottleneck for the reconnection at the larger scales.

The LV99 reconnection velocity (e.g. Eq. (\ref{recon2})) does not depend on
resistivity or plasma effects. Therefore for sufficiently high level of
turbulence we expect both collisionless and collisional fluids to reconnect at
the same rate.

\subsection{Model of flares}

If turbulence can drive reconnection, which in turn transforms magnetic energy
into kinetic energy, then it seems appropriate to wonder if the process can be
self-sustaining.  That is, given a very small level of ambient turbulence, how
likely is it that reconnection will speed up as it progresses, without any
further input from the surrounding medium? The corresponding problem was in
LV99. Below we follow simple arguments provided in \cite{lazarian09a}, which
clarify the related physics by presenting an idealized model of a reconnection
flare.

Let's consider a reconnection region of length $L$ and thickness $\Delta$.  The
thickness is determined by the diffusion of field lines, which is in turn
determined by the strength of the turbulence in the volume.  Reconnection will
allow the magnetic field to relax, creating a bulk flow.  However, since
stochastic reconnection is expected to proceed unevenly, with large variations
in the current sheet, we can expect that some unknown fraction of this energy
will be deposited inhomogeneously, generating waves and adding energy to the
local turbulent cascade.  We take the plasma density to be approximately uniform
so that the Alfv\'en speed and the magnetic field strength are interchangeable.
The nonlinear dissipation rate for waves is
\begin{equation}
\tau_{nonlinear}^{-1}\sim\max\left[ {k_\perp^2 v_{wave}^2\over k_\|V_A},k_\perp^2 v_{turb}\lambda_{\bot, turb}\right],
\end{equation}
where the first rate is the self-interaction rate for the waves and the second
is the dissipation rate by the ambient turbulence \citep[see][]{beresnyak08}.
The important point here is that $k_\perp$ for the waves falls somewhere in the
inertial range of the strong turbulence.  Eddies at that wavenumber will disrupt
the waves in one eddy turnover time, which is necessarily less than $L/V_A$. The
bulk of the wave energy will go into the tubulent cascade before escaping from
the reconnection zone.  (This zone will radiate waves, for the same reason that
turbulence in general radiates waves, but it will not significantly impact that
energy budget of the reconnection region.)

We can therefore simplify our model for the energy budget in the reconnection
zone by assuming that some fraction $\epsilon$ of the energy liberated by
stochastic reconnection is fed into the local turbulent cascade.  The evolution
of the  turbulent energy density per area is
\begin{equation}
{d\over dt}\left(\Delta v_{turb}^2\right)=\epsilon V_A^2 V_{rec}-v_{turb}^2\Delta {V_A\over L},
\end{equation}
where the loss term covers both the local dissipation of turbulent energy, and
its advection out of the reconnection zone.  Since $V_{rec}\sim v_{turb}$  and
$\Delta\sim L(v_{turb}/V_A)$,  we can rewrite this by defining ${\cal M}_A\equiv
v_{turb}/V_A$ and $\tau\equiv L/V_A$ so that
\begin{equation}
{d\over d\tau}{\cal M}_A^3\approx \epsilon {\cal M}_A-{\cal M}^3_A.
\end{equation}
If $\epsilon$ is a constant then
\begin{equation}
v_{turb}\approx V_A\epsilon^{1/2}\left[1-\left(1-{{\cal M}_A^2\over\epsilon}\right)e^{-2\tau/3}\right]^{1/2}.
\end{equation}
This implies that the reconnection rate rises to $\epsilon^{1/2}V_A$ is a time
comparable to the ejection time from the reconnection region ($\sim L/V_A$).
Given that reconnection events in the solar corona seem to be episodic, with
longer periods of quiescence, this implies that either $\epsilon$ is very small,
for example - dependent on the ratio of the  thickness of the current sheet to
$\Delta$, or is a steep function of ${\cal M}_A$.  If it scales as ${\cal M}_A$
to some power greater than two then initial conditions dominate the early time
evolution.

An alternative route by which stochastic reconnection might be self-sustaining
would be in the context of a series of topological knots in the magnetic field,
each of which is undergoing reconnection.  Now the problem is sensitive to
geometry.  Let's assume that as each knot undergoes reconnection it releases a
characteristic energy into a volume which has the same linear dimension as the
distance to the next knot.  The density of the energy input into this volume is
roughly $\epsilon V_A^2 v_{turb}/L$, where $\epsilon$ is the efficiency with
which the magnetic energy is transformed into turbulent energy.  We have
\begin{equation}
\epsilon V_A^2{v_{turb}\over L}\sim {v'^3\over L_k},
\end{equation}
where $L_k$ is the distance between knots and $v'$ is the turbulent velocity
created by the reconnection of the first knot.  This process will proceed
explosively if $v'>v_{turb}$ or
\begin{equation}
V_A^2 L_k\epsilon> v_{turb}^2 L.
\end{equation}
This condition is almost trivial to fulfill.  The bulk motions created by
reconnection will unavoidably generate significant turbulence as they interact
with their surrounding, so $\epsilon$ should be of order unity.  Moreover the
length of any current sheet should be at most comparable to the distance to the
nearest distinct magnetic knot.  The implication is that each magnetic
reconnection event will set off its neighbors, boosting their reconnection rates
from $v_{turb}$, set by the environment, to $\epsilon^{1/2}V_A(L_k/L)^{1/2}$ (as
long as this is less than $V_A$).  The process will take a time comparable to
$L/v_{turb}$ to begin, but once initiated will propagate through the medium with
a speed comparable to speed of reconnection in the individual knots.  In a more
realistic situation,  the net effect will be a kind of  modified sandpile model
for magnetic reconnection in the solar corona and chromosphere.  As the density
of knots increases, and the energy available through magnetic reconnection
increases, the chance of a successfully propagating reconnection front will
increase.

\begin{figure*}
\center
\includegraphics[width=0.42\textwidth]{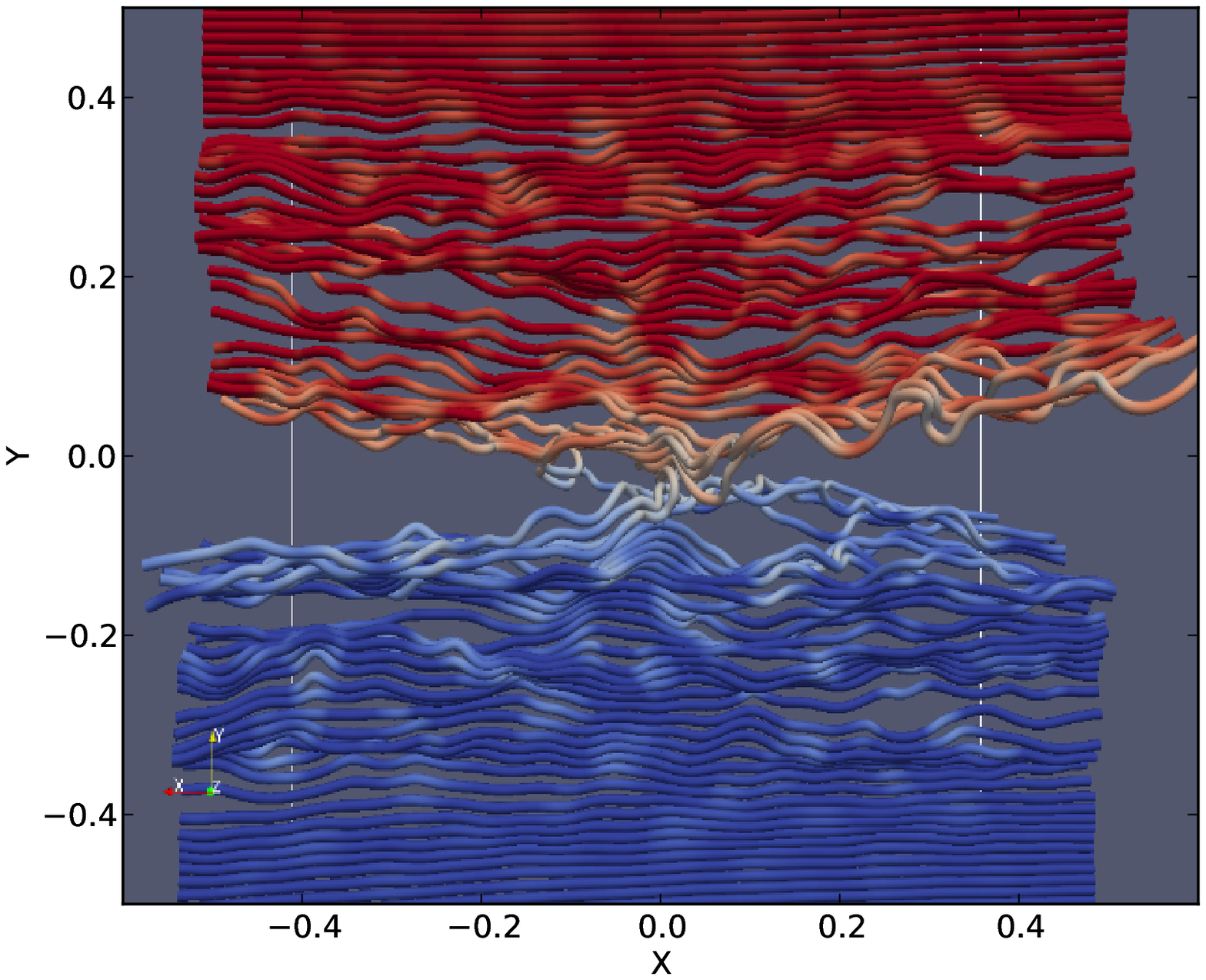}
\includegraphics[width=0.25\textwidth]{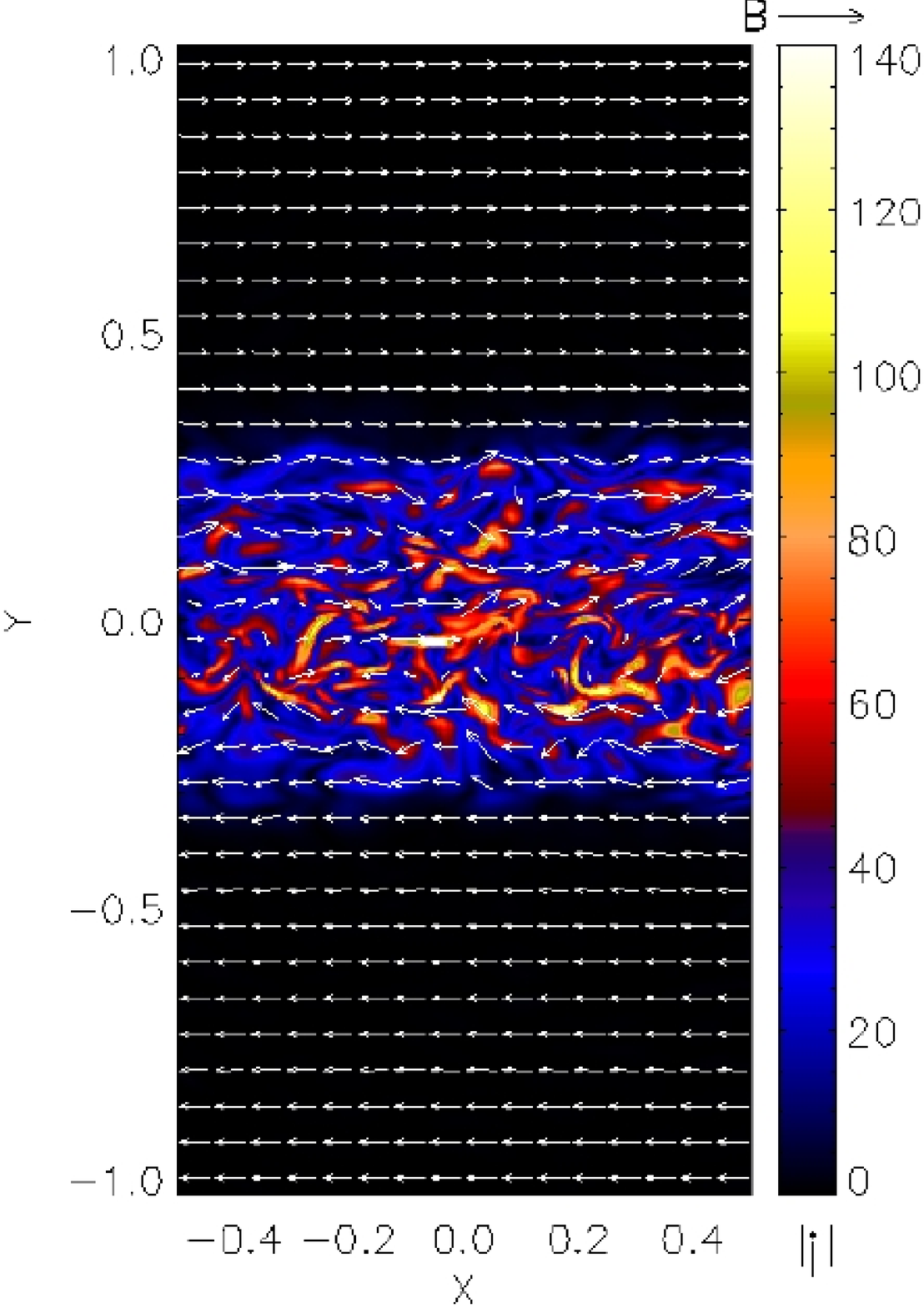}
\includegraphics[width=0.25\textwidth]{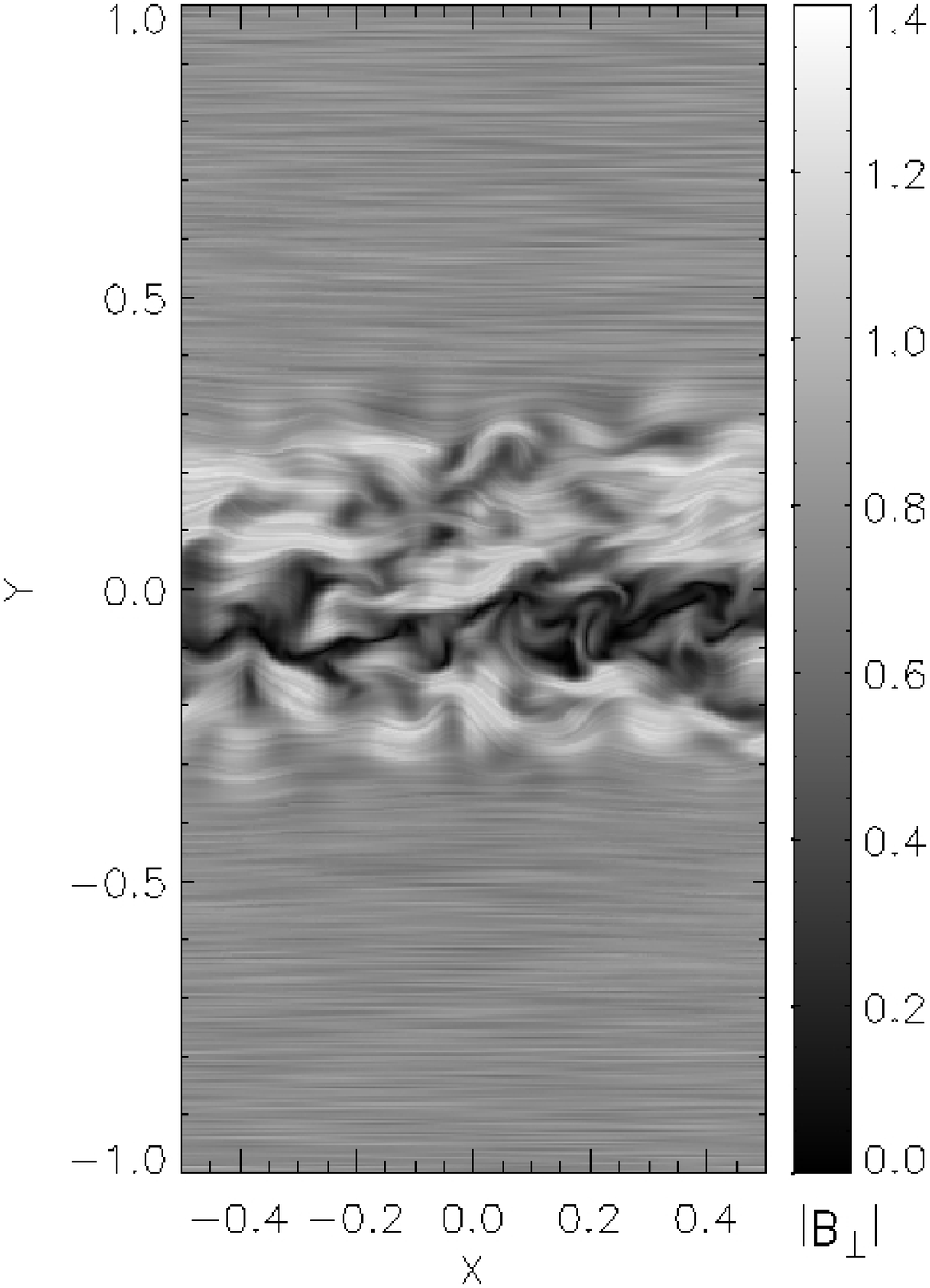}
\caption{ Visualization of reconnection simulations in \cite{kowal12b}.
{\it Left panel}: Magnetic field in the reconnection region.  Large
perturbations of magnetic field lines arise from reconnection rather than
driving; the latter is subAlfv\'enic.    The color corresponds to the direction
of magnetic lines.
{\it Central panel}: Current intensity and magnetic field configuration during
stochastic reconnection.  We show a slice through the middle of the
computational box in the xy plane after twelve dynamical times for a typical
run.  The guide field is perpendicular to the page. The intensity and direction
of the magnetic field is represented by the length and direction of the arrows.
The color bar gives the intensity of the current. The reversal in $B_x$  is
confined to the vicinity of y=0 but the current sheet is strongly disordered
with features that extend far from the zone of reversal.
{\it Right panel}: Representation of the magnetic field in the reconnection zone
with textures. \label{fig:top_turb}}
\end{figure*}

The flare of reconnection presents a reconnection instability. In the absence of
external turbulence the original outflow, e.g. originated through tearing
instability (see Loureiro et al. 2007, Bhattacharjee et al. 2009) gets turbulent
and triggers the mechanism described above. This shows that the tearing and
turbulent mechanisms may be complementary\footnote{When turbulence develops the
LV99 mechanism can provide much faster reconnection compared to tearing and
tearing may become a subdominant process. In fact, emerging turbulence may
suppress the tearing instability.}

\section{Testing of LV99 model}
\label{sec:lv99_test}

Here we describe the results of a series of three dimensional numerical
simulations aimed at adding turbulence to the simplest reconnection scenario and
testing equation (\ref{recon2}).  We divide the domain in the middle into two
equal regions along the vertical direction (along the Y axis in our case) with
equal magnetic field strength, but opposite sign of its X component.  In
addition, a constant guide field is added in the Z direction.  The domain is
periodic along the guide field (i.e. the Z axis) and open in the other
directions (the X and Y axis).  The external gas pressure is uniform and the
magnetic fields at the top and bottom of the box are taken to be the specified
external fields plus small perturbations to allow for outgoing waves.  The grid
size in the simulations varied from 256x512x256 to 512x1028x512 so that the top
and bottom of the box are far away from the current sheet and the region of
driven turbulence around it.   At the sides of the box where outflow is expected
the derivatives of the dynamical variables are set to zero.  A complete
description of the numerical methodology can be found in \cite{kowal09}. All
simulations are allowed to evolve for seven Alfv\'en crossing times without
turbulent forcing. During this time they develop the expected Sweet-Parker
current sheet configuration with slow reconnection.  Subsequently the isotropic
turbulent forcing is turned on inside a volume centered in the midplane (in the
XZ plane) of the simulation box and extending outwards by a quarter of the box
size.  The turbulence reaches its full amplitude around eight crossing times and
is stationary thereafter.

The speed of reconnection in three dimensions can be hard to define without
explicit evaluation of the magnetic field topology.  However, in this simple
case we can define it as the rate at which the $x$ component of the magnetic
field disappears.  More precisely, we consider a yz slice of the simulation,
passing through the center of the box.  The rate of change of the area integral
of  |$B_x$| is its flux across the boundaries of the box minus the rate at which
flux is annihilated through reconnection \citep[see more discussion in
][]{kowal09}
\begin{equation}
\partial_t\left(\int|B_x|dzdy\right)=\oint sign(B_x)\vec{E}d\vec{l}-2V_{rec}B_{x,ext}L_z
\label{measure}
\end{equation}
where electric field is $\vec{E}=\vec{v}\times \vec{B} -\eta \vec{j}$,
$B_{x,ext}$ is the absolute value of $B_x$  far from the current sheet and $L_z$
is the width of the box in the $z$ direction.  This follows from the induction
equation under the assumption that the turbulence is weak to lead to local field
reversals and that the stresses at the boundaries are weak to produce
significant field bending there.  In other words, fields in the $x$ direction
are advected through the top and bottom of the box, and disappear only through
reconnection.  Since periodic boundary conditions are adopted in the $z$
direction the boundary integral on the right hand side is only taken over the
top and bottom of the box.  By design this definition includes contributions to
the reconnection speed from contracting loops, where Ohmic reconnection has
occurred elsewhere in the box and $|B_x|$ decreases as the end of a reconnected
loop is pulled through the plane of integration.  It is worth noting that this
estimate is roughly consistent with simply measuring the average influx of
magnetic field lines through the top and bottom of the computational box and
equating the mean inflow velocity with the reconnection speed. Following
equation (\ref{measure}) we can evaluate the reconnection speed for varying
strengths and scales of turbulence and varying resistivity.

\begin{figure}
\center
\includegraphics[width=0.9\columnwidth]{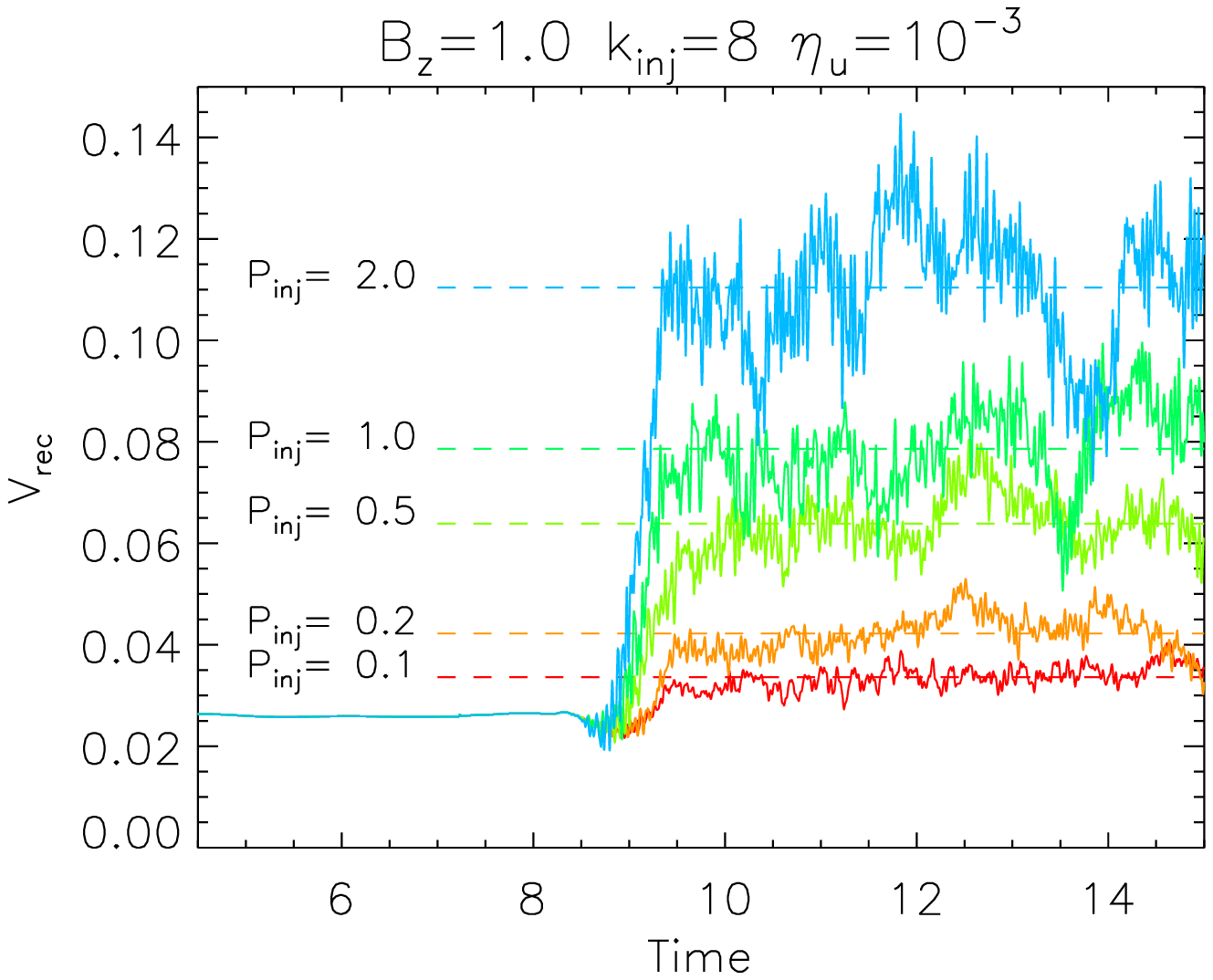}
\includegraphics[width=0.9\columnwidth]{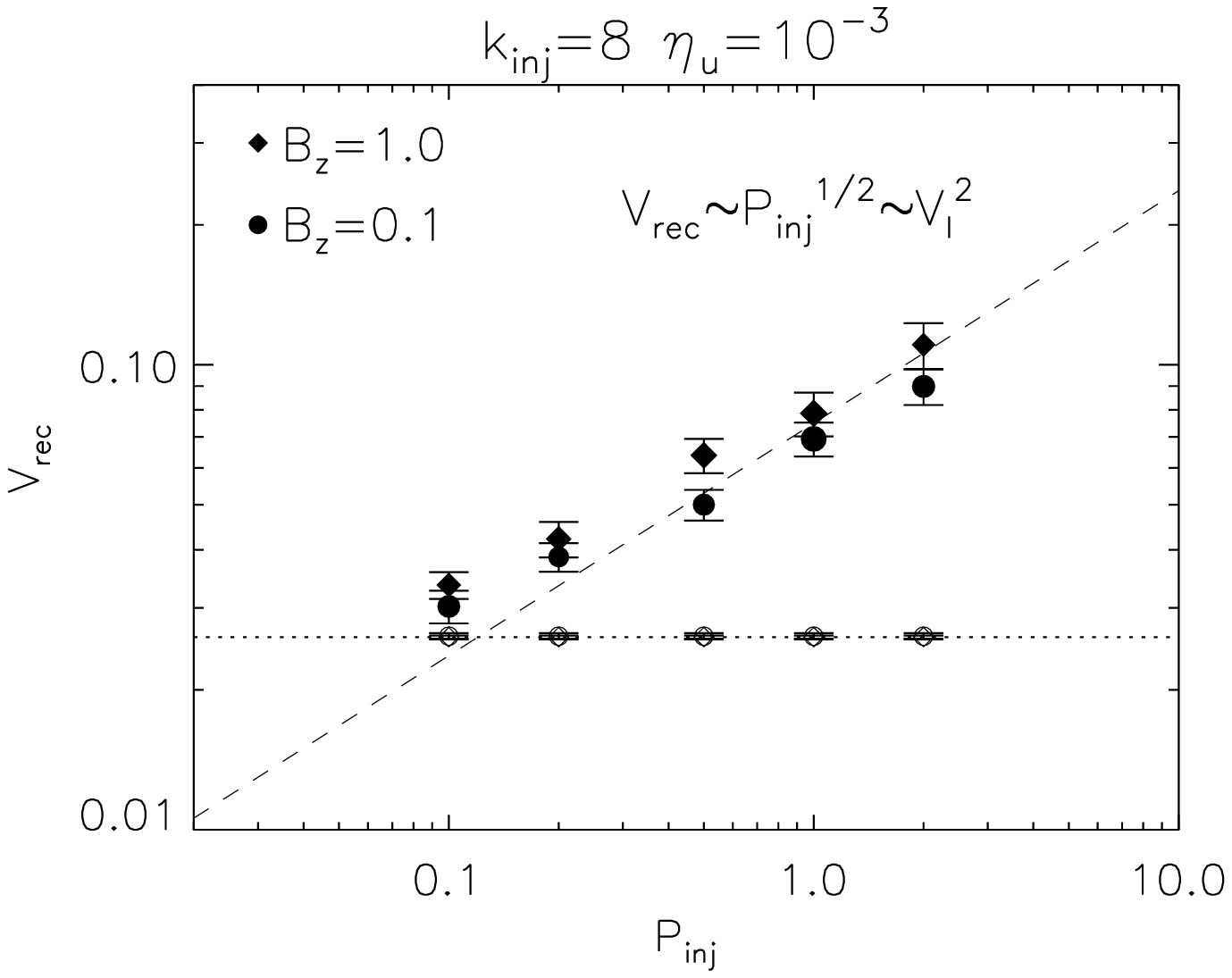}
\caption{Reconnection speed versus input power $P_{inj}$ for the driven
turbulence.  {\it Upper plot}: Variations of reconnection speed in time for
different levels of turbulent driving. {\it Lower Plot}: Reconnection speed,
plotted against the input power for an injection wavenumber equal to 8 (i.e. a
wavelength equal to one eighth of the box size) and a resistivity $\nu_u$.  The
dashed line is a fit to the predicted dependence of  $P^{1/2}$.  The horizontal
line shows the laminar reconnection rates for each of the simulations before the
turbulent forcing started.  Here the   uncertainty in the time averages are
indicated by the size of the symbols and the variances are shown by the error
bars. From \cite{kowal09}. \label{pow_dep}}
\end{figure}

In Figure~\ref{pow_dep} we see the results for varying amounts of input power,
for fixed resistivity and injection scale as well as for the case of no
turbulence at all.  The line drawn through the simulation points is for the
predicted scaling with the square root of the input power. The agreement between
equation (\ref{recon2}) and Figure~\ref{pow_dep} is encouraging but does not
address the most important aspect of stochastic reconnection, i.e. its
insensitivity to $\eta$.

\begin{figure}
\center
\includegraphics[width=0.9\columnwidth]{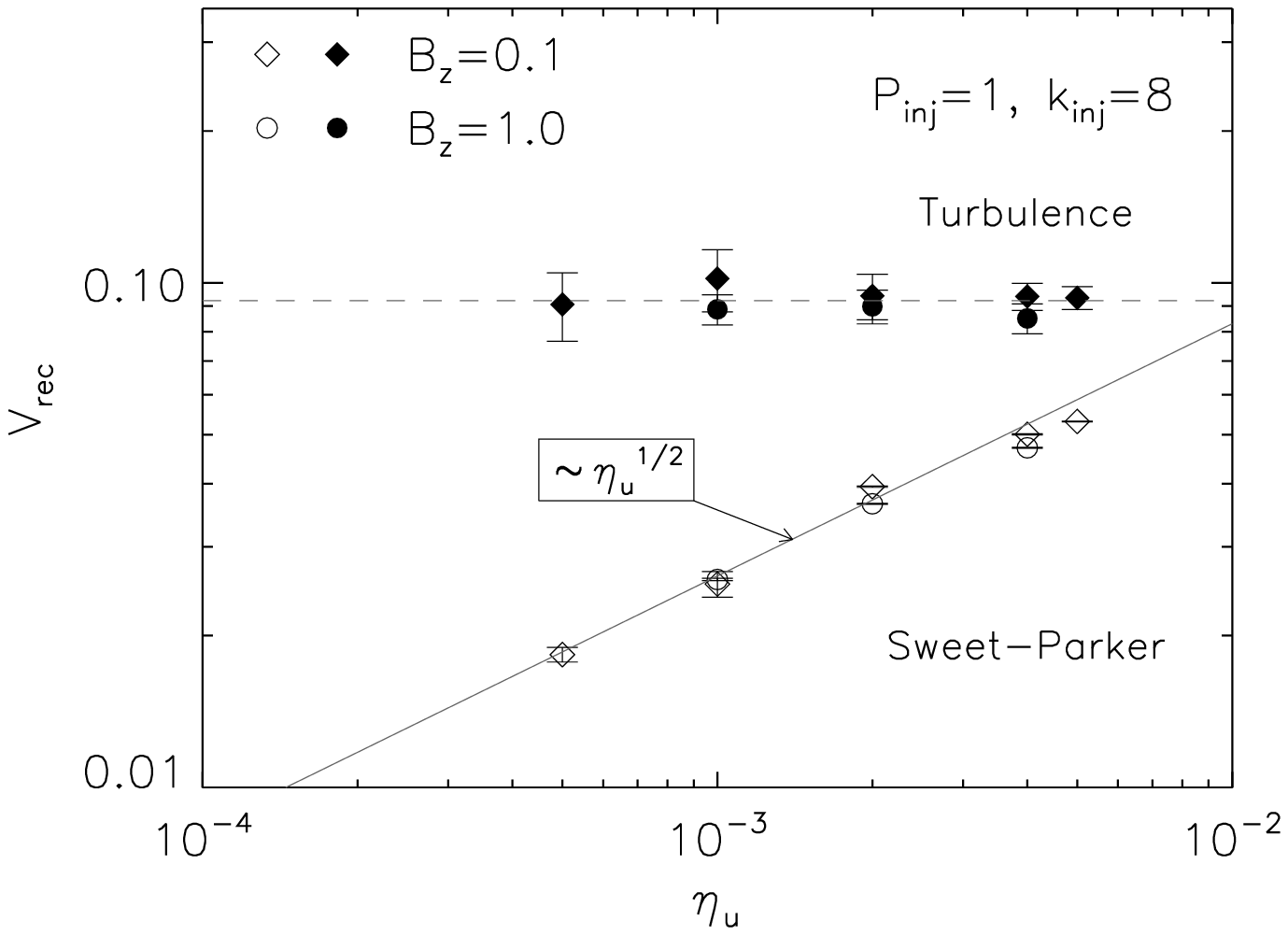}
\caption{Reconnection speed versus resistivity.  We show the reconnection speed
plotted against the uniform resistivity of the simulation for an injection
wavenumber of 8 and an injected power of one.  We include both the laminar
reconnection speeds, using the hollow symbols, fit to the expected dependence of
$\eta_u$, and the stochastic reconnection speeds, using the filled symbols.  As
before the symbol sizes indicate the uncertainty in the average reconnection
speeds and the error bars indicate the variance.  $B_x=1$ and simulations with
large, $B_z=1$, and small, $B_z=0.1$, guide fields are shown. From
\cite{kowal09}. \label{ueta_dep}}
\end{figure}

In Figure~\ref{ueta_dep} we plot the results for fixed input power and scale,
while varying the background resistivity.  In this case $\eta$  is taken to be
uniform, except near the edges of the computational grid where it falls to zero
over five grid points.  This was done to eliminate edge effects for large values
of the resistivity. We see from the Figure~\ref{ueta_dep} that the points for
laminar reconnection scale as $\sqrt{\eta}$, the expected scaling for
Sweet-Parker reconnection.  In contrast, the points for reconnection in a
turbulent medium do not depend on the resistivity at all. In summary, we have
tested the model of stochastic reconnection in a simple geometry meant to
approximate the circumstances of generic magnetic reconnection in the universe.
Our results are consistent with the mechanism described by LV99.  The
implication is that turbulent fluids in the universe including the interstellar
medium, the convection zones of stars, and accretion disks, have reconnection
speeds close to the local turbulent velocity, regardless of the local value of
resistivity.  Magnetic fields in turbulent fluids can change their topology on a
dynamical time scale.

\begin{figure}
\center
\includegraphics[width=0.9\columnwidth]{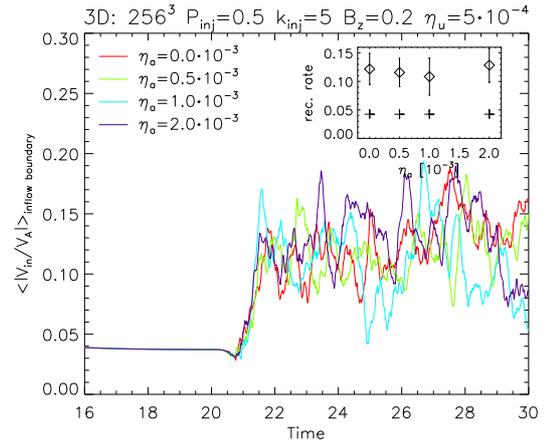}
\caption{Effect of anomalous resistivity on the reconnection speed. No
dependence is observed. From \cite{kowal09}. \label{anom_dep}}
\end{figure}

Figure~\ref{anom_dep} shows results of numerical experiments in which the
dependence of the reconnection on the anomalous resistivity was studied. The
anomalous resistivity increases effective resistivity for high current
densities. It is frequently used as a proxy for plasma effects, e.g.
collisionless effects in reconnection. While this type of resitistivity enhances
the local speed of individual reconnection events, the study in \cite{kowal09}
testifies that the total reconnection rate does not change.

\begin{figure}
\center
\includegraphics[width=0.9\columnwidth]{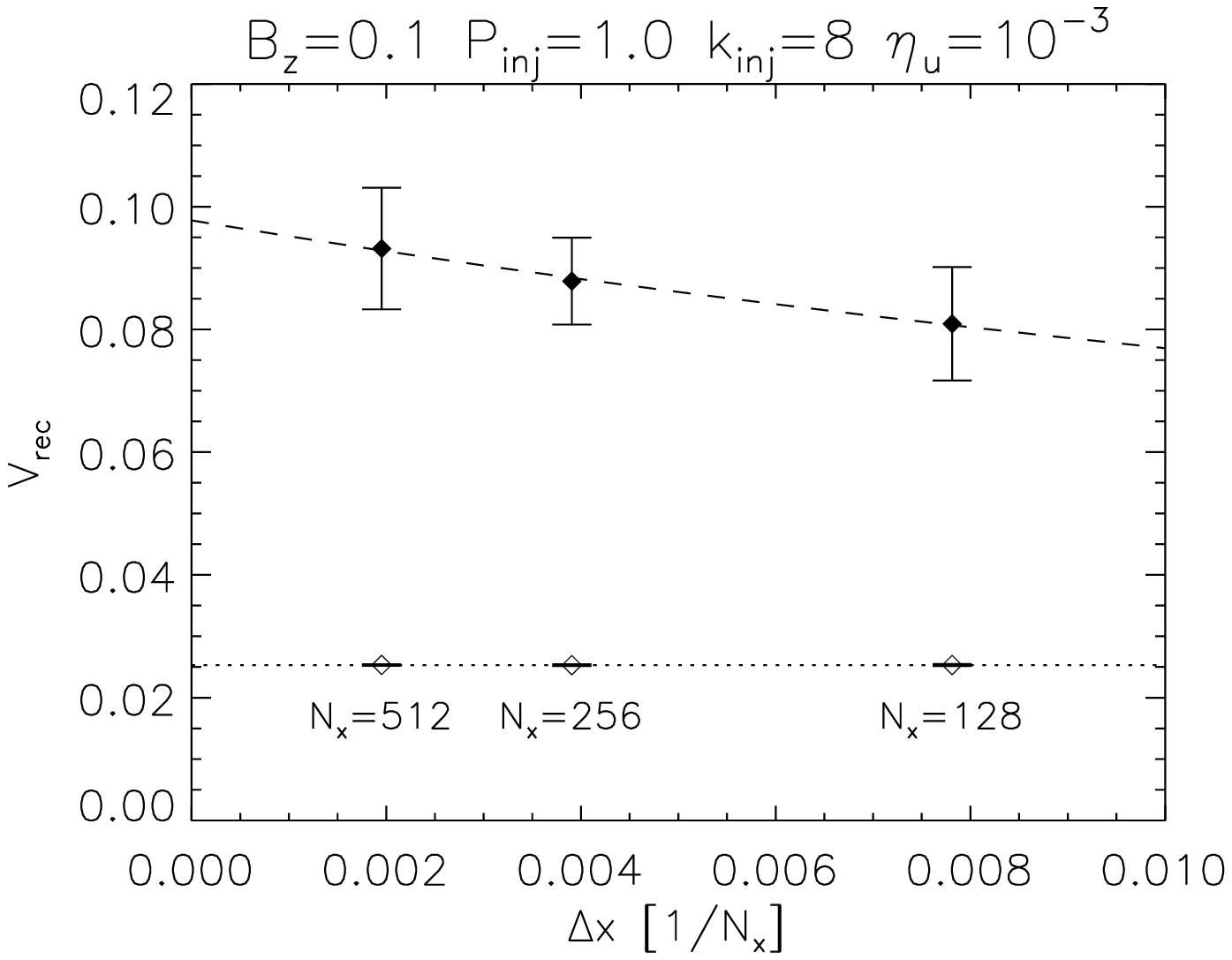}
\caption{Dependence of the reconnection rate on the numerical resolution. If the
fast reconnection were due to yet unclear numerical effects on small scales, we
would expect to see the increase of the reconnection rate with the decrease of
the numerical box. If anything, the actual dependence of the reconnection rate
on the box size shows the opposite dependence. From \cite{kowal12b}.
\label{fig:reso_dep}}
\end{figure}

Any numerical study has to address the issue of the possible numerical effects
on the results. We show the dependence of the reconnection rate on the numerical
resolution in Figure~\ref{fig:reso_dep}. The reconnection rate increases with
the increase of the resolution, which testifies that the fast reconnection is
not due to numerical effects. Indeed, higher numerical reconnection is expected
for lower resolution simulations.

Finally, it is important to give a few words in relation to our turbulence
driving. We drive our turbulence solenoidally to minimize the effects of
compression, which does not play a role in LV99 model. The turbulence driven in
the volume around the reconnection layer corresponds to the case of
astrophysical turbulence, which is also volume-driven. On the contrary, the case
of the turbulence driven at the box boundaries would produce spatially
inhomogeneous imbalanced turbulence for which we do not have analytical
predictions \citep[see discussion of such turbulence in][]{beresnyak09}.  We
stress, that it is not the shear size of our numerical simulations, but the
correspondence of the observed scalings to those predicted in LV99 that allows
us to claim that we proved that the 3D reconnection is fast in the presence of
turbulence.

\section{Turbulence and cosmic rays acceleration}
\label{sec:acceleration}

\subsection{Shock acceleration in the presence of turbulence}

Cosmic rays (CRs), relativistic charged particles with energies
$10^8-10^{22}$eV, constitute an essential part of astrophysical systems
\citep[see][]{schlickeiser03}.  The origin of CRs has been a subject of debate
from the beginning of research in the field \citep{ginzburg64}.  By now, it is
accepted that galactic CRs at least up to the ``knee'' in the spectrum, just
above $10^{15}$eV, are most likely generated primarily by strong supernova
shocks. However, it is easy to show that typical interstellar upstream magnetic
field of around $5\mu G$ is too weak to provide the return of the cosmic rays
with energies $\sim 10^{15}$ GeV back to the shock to continue the acceleration.

To overcome the problem it is important to increase the value of the magnetic
field both in the preshock and postshock regions.  The latter problem has a
straightforward solution, as postshock gas is known to be turbulent.  Turbulence
can amplify the magnetic field
\cite[see][]{kazantsev68,kulsrud92,cho09,beresnyak11b} and therefore an enhanced
magnetic field in the postshock region is expected \citep[see][]{giacalone07}. A
more challenging issue is to have the magnetic field amplified in  preshock
plasmas.  One known way is to appeal to the  streaming instability.  The latter
evolves to the nonlinear stage with $\delta B \gg B_0$ making the original
classical treatment of the instability not applicable.  What happens in the
non-linear regime has been discussed a lot in recent years
\cite[e.g.][]{blasi08,lucek00,diamond07,zirakashvili08,riquelme09} with many
researchers pointing out that the instability may become much suppressed.  In
addition, even at its linear regime the instability is likely to be suppressed
in the presence of ambient turbulence \citep{yan02,farmer04,beresnyak08}.  In
this situation, the current-driven instability proposed by \cite{bell04} became
the central idea being discussed in order to solve the problem of the
acceleration of high energy cosmic rays.

Beresnyak, Jones \& Lazarian (\citeyear[][henceforth BJL09]{beresnyak09a})
suggested an alternative model for enhancing the magnetic field in the preshock
region. This model is based on the magnetic field generation within the shock
precursor. The precursor originates from cosmic rays that can be returned back
to the shock by the original interstellar magnetic field. The fluid is stirred
within the precursor by the combination of fluid density inhomogeneities and CR
pressure. This drives vorticity and the precursor turbulence. The latter
amplifies magnetic energy through a turbulent dynamo. In other words, BJL09
claims that both the enhancement of magnetic field in the preshock and postshock
regions are due to turbulence draining energy from the shock.

The upstream diffusion of CRs leading to precursor formation is associated with
a number of possibilities, such as the Drury acoustic instability
\citep{drury84,dorfi85,kang92}, which is the enhancement of compressible
perturbations by the CR pressure gradient. Such instabilities, however, will
only operate on perturbations during the limited time that a fluid element
crosses the precursor.  This crossing time is rather short $\tau_c\approx
l_p/U_s$ where $l_p$ is the precursor thickness and $U_s$ is the shock velocity
and the growth of the density perturbation is constrained.

\begin{figure}
\includegraphics[width=0.9\columnwidth]{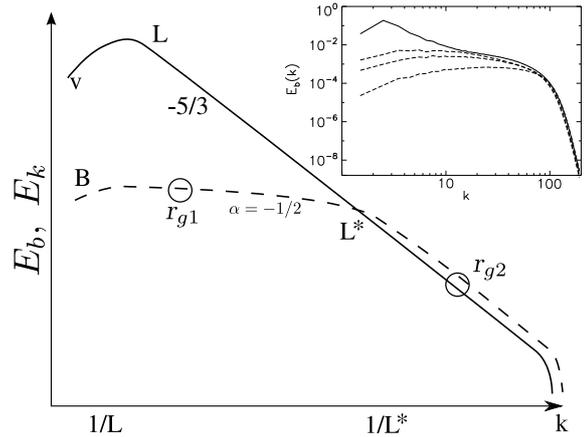}
\caption{Magnetic field spectrum (dashed), generated by a small-scale dynamo
induced by solenoidal velocity motions (solid). $L^*$ is an equipartition scale
of magnetic and kinetic motions, it plays a central role in particle scattering.
Upper panel: magnetic and velocity fields from simulations \citep{cho09} with
different dashed lines corresponding to magnetic spectra at different times.
From BJL09.} \label{spectrum2}
\end{figure}

The difference of the velocities of the fluid elements crossing the shock arises
from the difference of the inertia of dense and rarefied gas entering the shock.
When an inhomogeneous fluid enters the precursor with speed $u_0$ it gets
decelerated by the CR pressure gradient until it reaches the speed $u_1$ at the
dissipative shock front. Let us denote $A_s(u_0-u_1)$ as the difference between
the ballistic velocity of the high-density region and the full decelerated one
of the low density regions. The existence of such regions in interstellar medium
or solar wind plasma, is expected to arise from the pre-existing turbulence not
related to the shock (see \S\ref{sec:intro} , Figure~\ref{fig:big_power_law}).
The interaction of the precursor with the density inhomogeneities is expected to
result in intensive turbulence in the precursor. The turbulence in conducting
fluid is known to amplify magnetic fields as depicted in Figure~\ref{spectrum2}.

After extremely short kinematic stage, the energy is growing linearly with time
$dE_B/dt\approx \alpha_d \epsilon$, where $\epsilon\sim\rho u_s^3/L$ is the
energy transfer rate, and $\alpha_d$ is the efficiency factor $\sim 0.06$
\citep{cho09}.  The turbulent dynamo gets magnetic energy in the equipartition
with the kinetic energy at a scale $L^*$.  This scale grows with time. On scales
smaller than $L^*$ MHD turbulence exists. On scales larger than $L^*$, the
turbulence is superAlfv\'enic and velocity has a Kolmogorov spectrum.

\begin{figure}
\includegraphics[width=0.9\columnwidth]{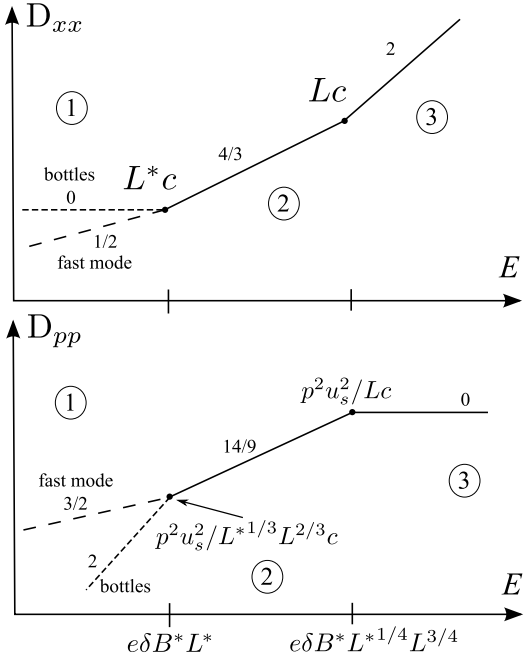}
\caption{Diffusion coefficients. (1) is low-energy scattering, which depends on
the properties of small-scale MHD turbulence (dashed: fast modes are present;
dotted: fast modes are absent); (2) is medium-energy scattering, by the magnetic
fields generated by turbulent dynamo; (3) is high-energy scattering. From
BJL09.} \label{dcoeff}
\end{figure}

As the crossing time of the precursor $\tau_c$ is short the generation of
magnetic field in the precursors does not get into the saturation regime. The
system of equations suggested in BJL09 for the evolutions of magnetic field and
$L^*$ is as follows:
\begin{equation}
\delta B^2(L^*,x_1)=8\pi A_d \epsilon \tau(x_1);
\label{sh1}
\end{equation}
\begin{equation}
\frac{\delta B^*}{\sqrt{4\pi\rho}} = u_s\left(\frac{L^*(x_1)}{L}\right)^{1/3};
\label{sh2}
\end{equation}
\begin{equation}
\tau(x_1)=\int^{x_0}_{x_1}\frac{dx}{u(x)};
\label{sh3}
\end{equation}
\begin{equation}
L^*(x_1)=(2A_du_s\tau(x_1))^{3/2}L^{-1/2}.
\label{sh4}
\end{equation}
This approach was used to calculate the diffusion coefficients shown in Figure~\ref{dcoeff}.

It was shown by \cite{bell04} that magnetic fields in the preshock region could
be amplified significantly compared to the background field. The model is based
on a current-driven instability. It was successfully tested numerically in a
number of papers \citep{vladimirov06,zirakashvili08,riquelme09}.  In
\cite{bell04} the growth of magnetic field at the linear stage is fastest on the
characteristic scale that is  determined by the initial field $B_0$, and the
current $j_d$,
\begin{equation}
l=1/k_c=\frac{cB_0}{4\pi j_d},
\end{equation}
where $j_d$ arises from high-energy CRs that correspond to $qB_0/pc<<k_c$. The
linear growth rate depends only on the current according to the relation
\begin{equation}
\gamma =\frac{j_d}{c\sqrt{\frac{\rho_0}{\pi}}}.
\end{equation}

BLJ09 demonstrated that, conservatively assuming that the linear growth rate is
applicable to the current-driven instability is in the stage with $\delta B\sim
B_0$, one still gets that the dynamo generation of magnetic fields is dominant,
i.e.
\begin{eqnarray}
\frac{dB^2_{cur}}{dB^2_{dyn}}&=&1.6\times 10^{-4}\bratio{10^{15}\mbox{eV}}{E_{esc}}
\bratio{\eta_{esc}}{0.05}\bratio{L}{1\mbox{pc}}\nonumber\\
&\times&\bratio{B_0}{5\mu\mbox{G}} \bratio{v_{A0}}{12\mbox{km/s}}\bratio{0.5 u_{sh}}{A_s(u_0-u_1)}^3.
\end{eqnarray}
where $\eta_{esc}\approx 0.05$ between the flux of CR energy emitted by the
shock and the flux of energy of the incoming fluid $\rho u_{sh}^3/2$
\citep{riquelme09}.

The higher efficiency of the BJL09 model arises from the fact that in this model
the full pressure ($\sim$ energy density) of the CRs is driving the instability,
while in the \cite{bell04} instability arises only from those CRs that are able
to freely stream. The efficiency can be reduced and \cite{bell04} instability
wins if $A_s$ parameter is much less than unity. Further research should clarify
the relative importance of the two ways of generating magnetic fields in the
precursor. BJL09 estimate that the maximum energy that can be obtained by the
acceleration of the shocks with the small-scale dynamo in the precursor is $\sim
10^{16}$~eV, which is consistent with the galactic cosmic ray measurements.

\subsection{Second order Fermi acceleration and turbulence}
\label{sec:2nd_fermi}

To deal with the problem of propagation and acceleration of CRs the so-called
diffusive approximation is frequently used. Within this approximation it is
assumed that the particle scatter or gain energy in small steps and the dynamics
is averaged to obtain the spatial diffusion coefficient, $D_{xx}$, and the
momentum diffusion  coefficient, $D_{pp}$. The resulting advection-diffusion
equation for the evolution of quasi-isotropic CR distribution function $f$ is
\def\pder#1#2{\frac{\partial #1}{\partial #2}}

\begin{eqnarray}
\pder ft+u\pder fx & = &\pder{}{x}\left(D_{xx}\pder fx\right)\nonumber\\
                   & + & \frac p3 \pder ux \pder fp
+ \frac 1{p^2} \pder{}{p}\left(p^2 D_{pp} \pder  fp\right),
\label{adv_diff}
\end{eqnarray}
\citep[e.g.][]{skilling75}, where for simplicity it is assumed that $f(x,p)$
depends only on one spatial coordinate $x$ and the magnitude of CR momentum,
$p$. In Eq. (\ref{adv_diff}) the momentum diffusion corresponds to the second
order Fermi acceleration \citep[see][]{longair10}.

The diffusion coefficients in Eq. (\ref{adv_diff}) depend on the statistical
properties of magnetic turbulence that interacts with the particles.  Adopting
the decomposition of compressible MHD turbulence in modes
(\S\ref{sec:mhd_waves}) \cite{yan02,yan04} identified the fast modes as the
principal modes responsible for scattering and turbulent acceleration of CRs in
galactic environment.  Later analogous conclusions were reached for the CR
acceleration in clusters of galaxies \citep{brunetti07}.

The above conclusion follows from Alfv\'en and slow modes being inefficient for
resonance interaction with cosmic rays \citep{chandran00,yan02}.  Indeed, as we
discussed in \S\ref{sec:mhd_waves} the two modes are anisotropic with the
anisotropy increasing with the decrease of the scale.  The resonant interaction
of the CRs and Alfv\'enic perturbations happens when the CR Larmor radius is of
the order of the parallel scale of the eddy.  As the eddies at scales much less
than injection scale are very elongated, the CR samples many uncorrelated
eddies, which reduces the interaction efficiency.

\begin{figure*} [h!t]
{\centering \leavevmode
\includegraphics[width=2.5in]{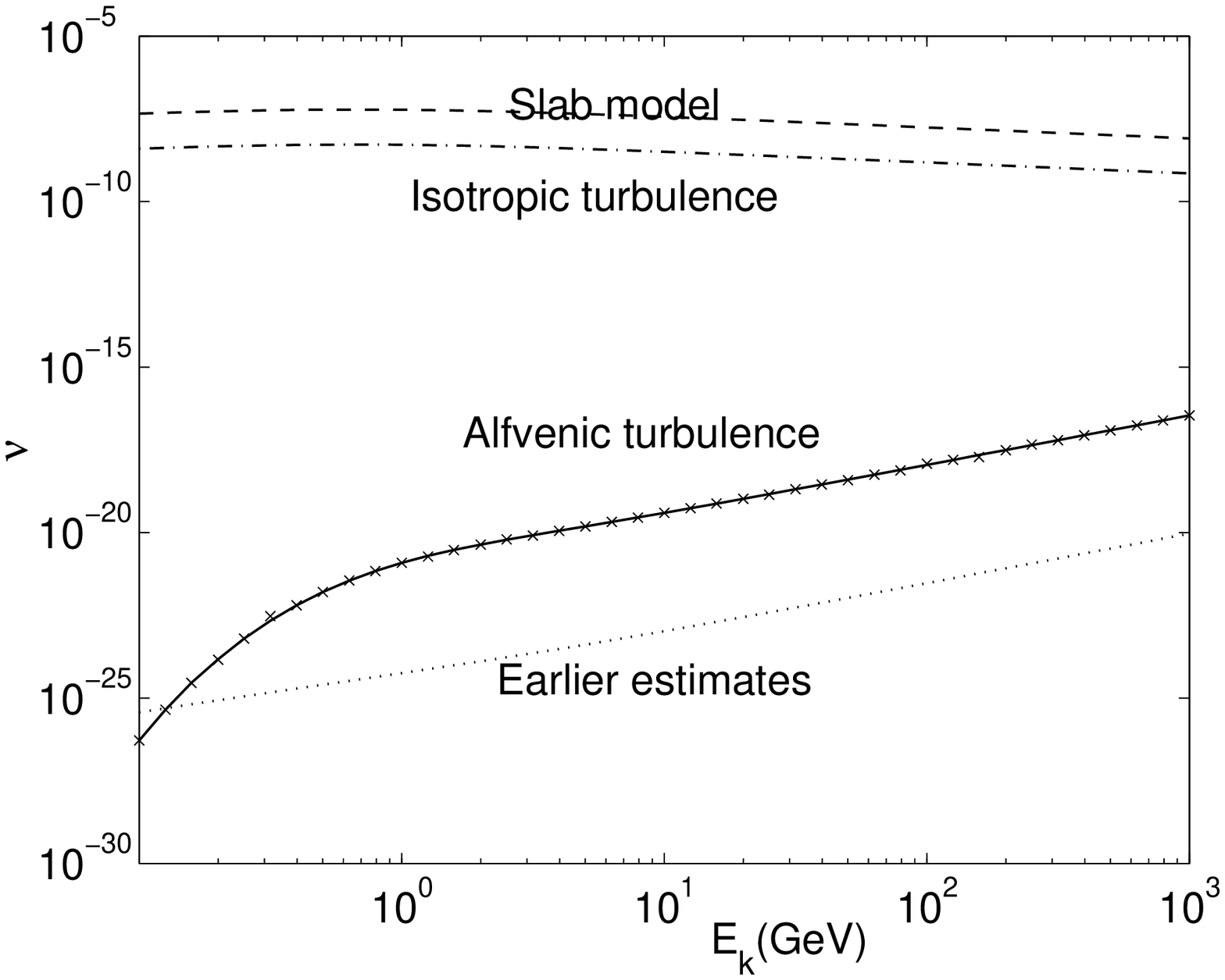}
\hfil
\includegraphics[width=2.5in]{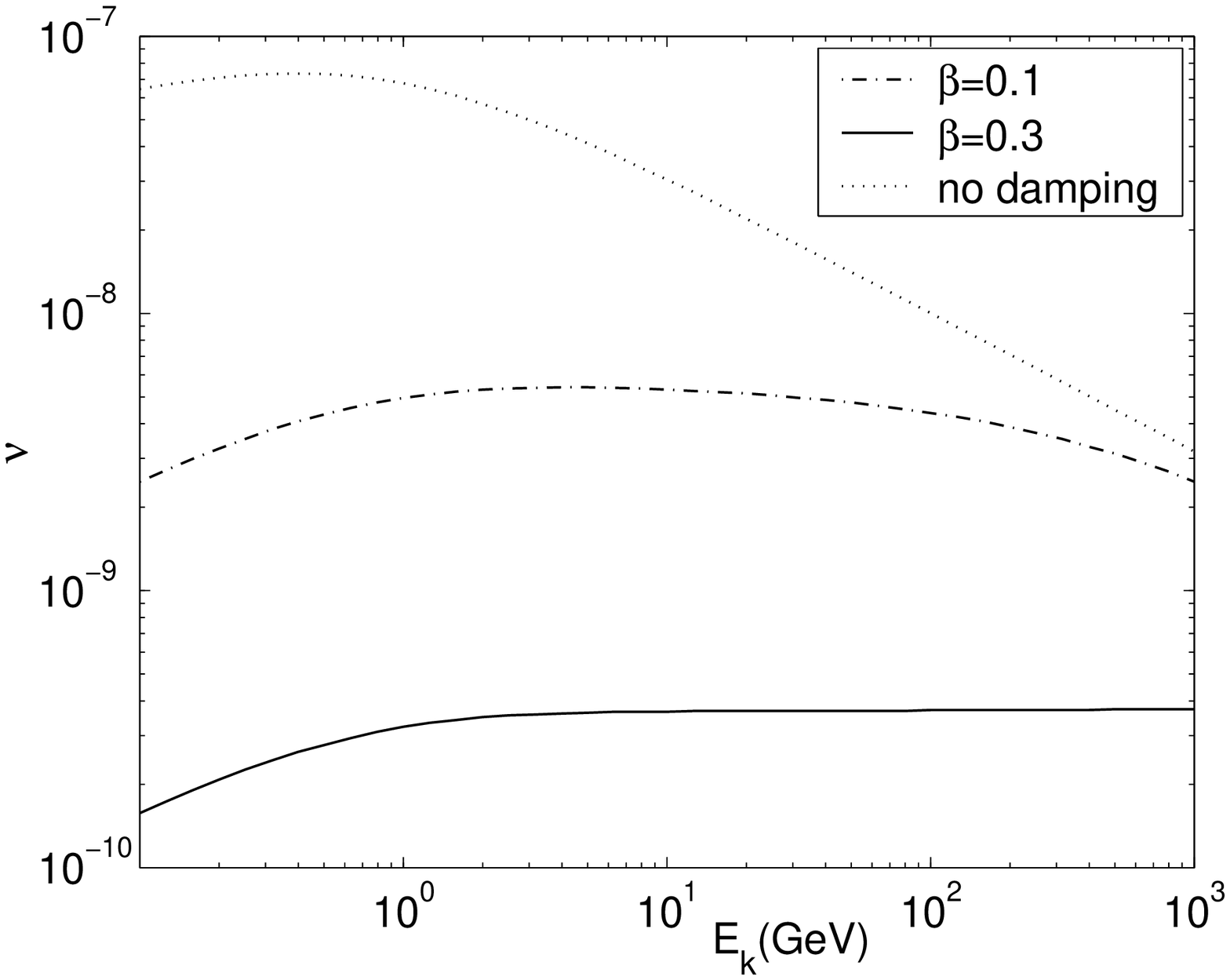}
}
{\centering \leavevmode
\includegraphics[width=2.5in]{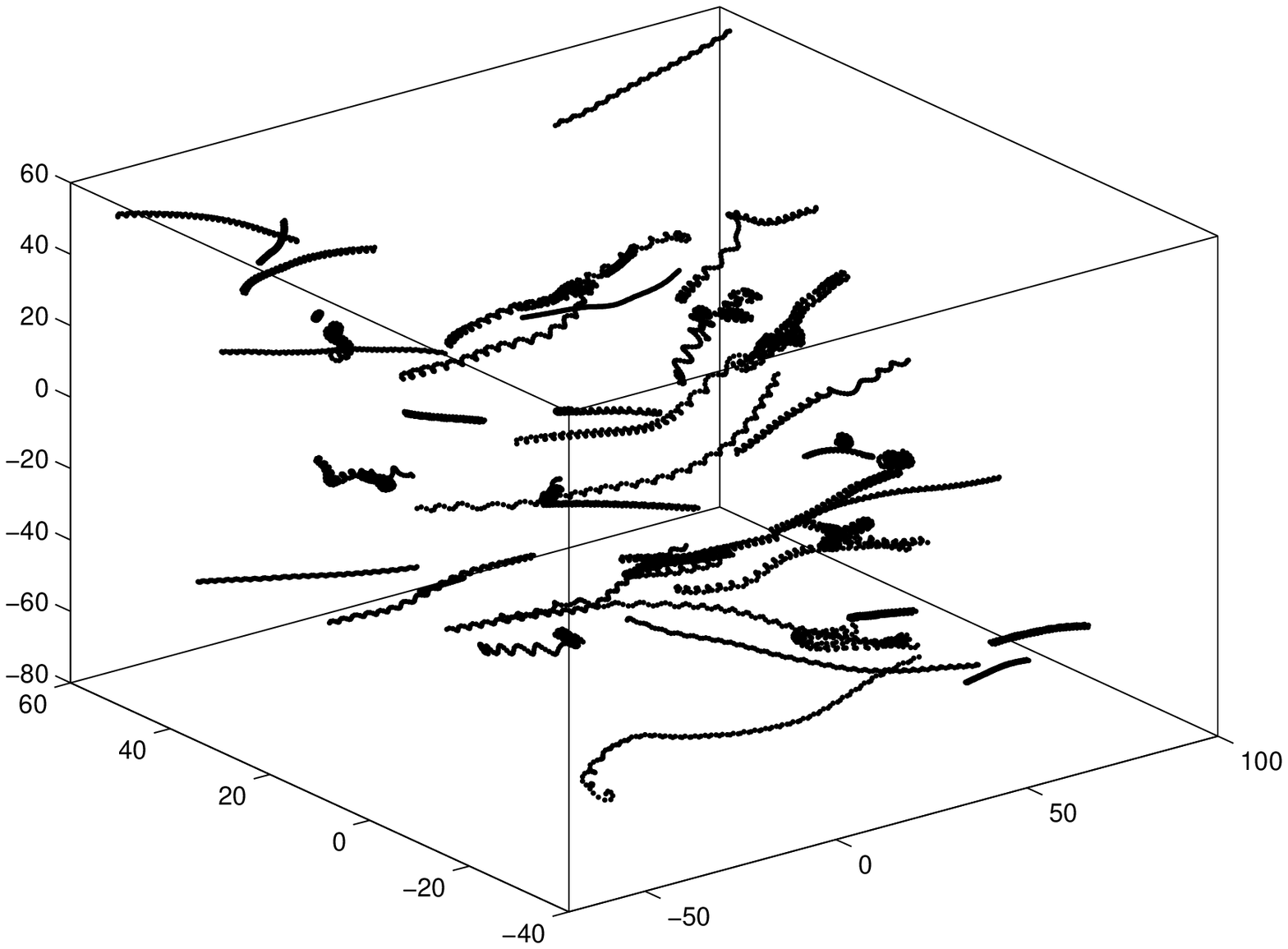}
\hfil
\includegraphics[width=2.5in]{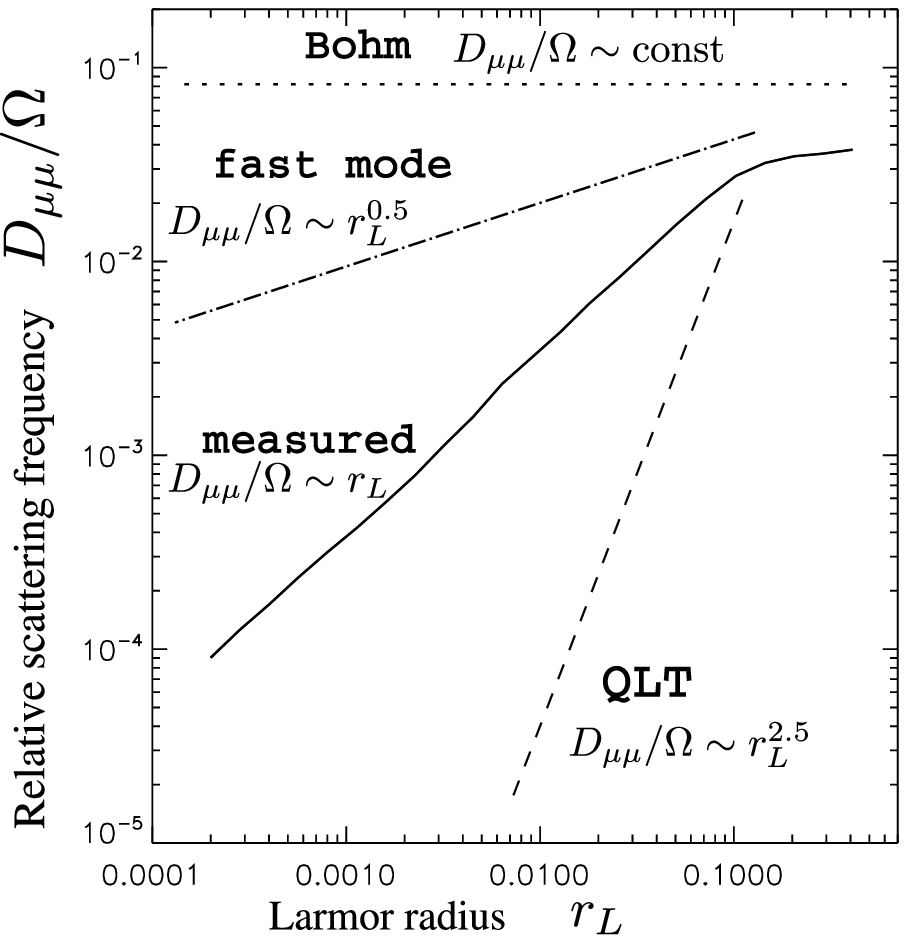}
}
\caption{ {\it Implications of turbulence for cosmic rays.} {\it Upper left:}
Rate of CR scattering by Alfv\'en waves versus CR energy.  The lines at the top
of the figure are the accepted estimates obtained for Kolmogorov turbulence. The
dotted curve is from \cite{chandran00}. The analytical calculations are given by
the solid line with our numerical calculations given by crosses. {\it Upper
right:} The rate of CR scattering ($\nu$) by fast modes in magnetically
dominated plasma. The rate of scattering depends on damping of the fast waves
(see , which in turn depends on the ratio of gaseous to magnetic pressure
($\beta= P_{gas}/P_{mag}$). {\it Lower left:} Individual trajectories of CRs
tracked by the  Monte Carlo scattering code. {\bf B} is obtained through 3-D
simulations of MHD turbulence. These calculations provide estimates of CR
diffusion. From \cite{yan02,yan04}. {\it Lower right}: Testing of the
\cite{yan08a} non-linear theory with numerical simulations in
\cite{beresnyak09}.} \label{impl}
\end{figure*}

\cite{yan02} showed that the scattering by fast modes, which are isotropic
(CL02), dominates (see Fig.~\ref{impl}). However, fast modes are subject to both
collisional and collisionless damping\footnote{On the basis of weak turbulence
theory, \citep{chandran05} has argued that high-frequency fast waves, which move
mostly parallel to magnetic field, generate Alfv\'en waves also moving mostly
parallel to magnetic field. We expect that the scattering by these generated
Alfv\'en modes to be similar to the scattering by the fast modes created by
them. Therefore we expect that the simplified approach adopted in \cite{yan04}
and the papers that followed this one to hold.}, which was taken into account in
\cite{yan04}.

More recent studies of cosmic ray propagation and acceleration that explicitly
appeal to the effect of the fast modes include
\cite{cassano05,brunetti06,brunetti07,yan08a,yan08b}.

The question of the effect of Alfv\'en modes is not trivial, however. The
calculations that were discussed above use a quasi-linear theory (QLT), whose
domain of applicability is limited to very strong fields with small
perturbations. The key assumption of QLT is that the particle's orbit is
unperturbed, significantly limits its applicability. Additionally, QLT has
problems in treating scattering of particles with momentum nearly perpendicular
to the magnetic field
\citep[see][]{jones73,jones78,volk73,volk75,owens74,goldstein76,felice01} and
perpendicular transport \cite[see][]{kota00,matthaeus03}.

Various non-linear theories have been proposed to improve the QLT
\citep[see][]{dupree66,volk73,volk75,jones73,goldstein76}.  In the recent paper
of \citet[][henceforth YL08]{yan08a}, a nonlinear theory (NLT) based on the
\cite{volk75} suggestion was developed. The new formalism was applied to two
major processes of CR acceleration in turbulence, namely, to gyroresonance and
to the transient time damping (TTD) \cite{fisk74, goldstein75}.

The efficient particle-wave interactions happen when the condition
\begin{equation}
\omega-k_{\parallel}v\mu=n\Omega ~~~~~~~ n=0, \pm 1,2...
\label{gyrores}
\end{equation}
is satisfied. In Eq. (\ref{gyrores})  where $\omega$ is the wave frequency,
$\Omega=\Omega_{0}/\gamma$ is the relativistic gyration frequency,
$\mu=\cos\theta$, $\theta$ is the pitch angle of particles.  TTD corresponds to
$n=0$ and it requires compressible perturbations. The most important for the
gyroresonance are the interactions with $n=1$.

Contrary to QLT which assumes that the magnitude of the magnetic field stay
constant, NLT relaxes this assumption and allows the magnetic field to change in
a smooth way. Due to conservation of adiabatic invariant $p_\bot^2/B$
\cite[see][]{landau75} the pitch angle will
gradually vary, resulting in resonance broadening \citep{volk75}.  Nonlinear
transport (NLT) formalism is based on the replacement of the sharp resonance
between waves and particles $\delta(k_{\parallel}v_{\parallel}-\omega\pm
n\Omega)$ from QLT to the ``resonance function'' $R_n$ (YL08):
\begin{eqnarray}
R_n&=&\Re\int_0^\infty dt e^{i(k_\|v_\|+n\Omega-\omega) t-\frac{1}{2}k_\|^2v_\bot^2t^2 \left(\frac{<\delta B_\parallel^2>}{B_0^2}\right)^\frac{1}{2}}\nonumber\\
&=&\frac{\sqrt{\pi}}{|k_\|\Delta v_\||}\exp\left[-\frac{(k_\|v
    \mu-\omega+n\Omega)^2}{k_\|^2\Delta v_\|^2}\right],
\label{resfunc}
\end{eqnarray}

The width of the resonance function depends on the strength of the turbulence
$\Delta \mu=\Delta v_\|/v_\bot\simeq \sqrt{\delta B/B}=\sqrt{M_A}$). For
gyroresonance ($n=\pm 1,2,...$) the result depends on whether $\mu$ is strongly
or weakly perturbed by regular field. If $\mu\gg \Delta \mu$, the result is
similar to QLT, because the exponents in Eq.(\ref{resfunc}) is close to the
$\delta$-functions.  For $\mu<\Delta \mu$ the result is different. For instance,
for the case of $90^\circ$ scattering $\mu\rightarrow 0$ and the resonance
happens mostly at $k_{\|,res}\sim \Omega/\Delta v$, while in QLT
$k_{\|,res}\sim\Omega/v_\|\rightarrow \infty$.

The change of the efficiency of the gyroresonance scattering and acceleration by
Alfv\'enic modes is not sufficient to affect the conclusions about the
inefficiency of the process for low energy cosmic ray scattering by turbulence
driven at large injection scales. However, \cite{yan08a} showed that TTD gets
appreciably more efficient, in contrast with the QLT-based result.  TTD due to
nonlinear scattering can be understood as a scattering by large-scale magnetic
compressions arising from the slow mode perturbations.

To test the NLT results particle tracing simulations were performed in
\cite{beresnyak11b}.  Test particle simulation is the tool extensively used to
study CR scattering and transport \cite[e.g.][]{giacalone99,mace00,qin02}. The
aforementioned studies, however, used synthetic data for turbulent fields which
is problematic due to a number of reason. First of all, creating synthetic
turbulence data which has scale-dependent anisotropy with respect to the local
magnetic field as it is required both by theory and simulations has not been
done so far. In addition, synthetic data has used Gaussian statistics and
delta-correlated fields, which is hardly appropriate for description of strong
turbulence. Therefore \cite{beresnyak11b} used the magnetic field obtained as a
result of numerical MHD simulations. The results of this testing is shown in
Figure~\ref{impl}.

The difference between QLT and NLT has important astrophysical consequences.
Indeed, in some phases, such as hot ISM, the fast mode is strongly damped, which
makes vital the issue of the interaction of CRs with Alfv\'en and slow modes.
QLT predicts the marginal interaction of these modes with CR. One can view the
NLT result for the TTD as the lower limit for the turbulence interaction with
CRs.

\subsection{Complexity of turbulent acceleration}

\subsubsection{A network of acceleration sites}

We can view the fragmented acceleration sites as a synthesis of local
accelerators (turbulent quasi-perpendicular shocks, Strong turbulence and multi
island reconnection) all of them share the same transport characteristics i.e.
particles visit sporadically the localized and strong electric fields were the
particles ``trapped'' for a short time (called ``stickiness'' in non-equilibrium
statistical mechanics) and undergo sudden energy jump (called in statistical
mechanics ``Levy flight").  \cite{vlahos04}, studied the evolution of particles
inside  a \emph{Network of Acceleration Sites} (non-linear structures with
random size)  associated with (random strength) electric fields. They use three
density probabilities mentioned already above to analyze the transport
properties of the energetic particles.  $P1(\ell_i)$ represented the density
probability for the spacial transport between the ``scattering centers or
Nodes''  of the accelerators.

\begin{figure}[ht]
\centering
\includegraphics[width=\columnwidth]{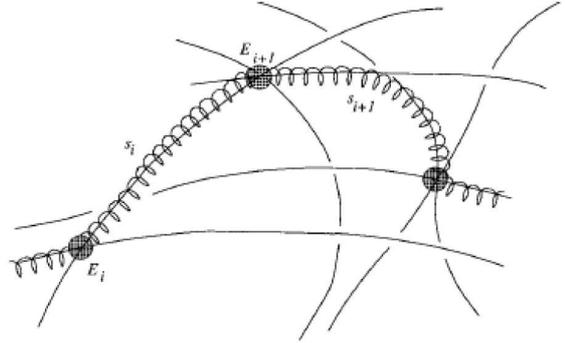}\\
\caption{Sketch of the basic elements of the Network of particle accelerators. A
particle (spiraling line) basically follows the magnetic field lines (solid
lines), although also undergoing drifts and travel freely a distance $\ell_i$
until it enters the "scattering center or acceleration Node" (filled circle),
where it is accelerated by the local "electric field $E_i$". After spending a
time  $\tau_i$ inside the acceleration node it move freely again till it meets
the new "Acceleration Node". \label{fig:network}}
\end{figure}

The $P(E)$  is the density probability for the energy gain of the particle
interacting with the the acceleration sites (``scattering center or the
Nodes''), and finally the density probability $P(\tau_i)$ of the time the
particle spend inside the ``scattering center or the Node''.

Particle acceleration in astrophysics is a multi-scale process. A number of
multi-scale systems starting from  large scale "turbulent shocks" or "strong
turbulence" and go down to the small scale structures (multiple islands and/or
current sheets). The fundamental question which we address next is: Is it
possible, by following the dynamics of accelerated particle inside the
environments discussed above, to reconstruct their transport properties?

\begin{figure}[ht]
\centering
\includegraphics[width=\columnwidth]{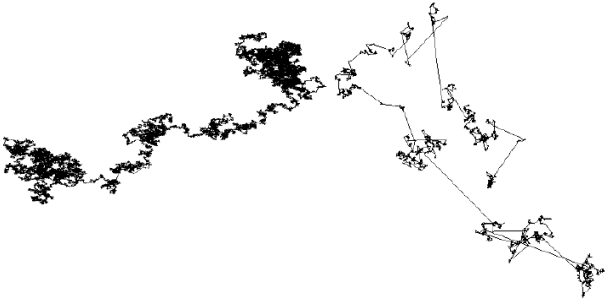}\\
\caption{(a) Random walk in dynamical systems close to equilibrium (normal
diffusion:trajectory in the left) (b) Random walk in dynamical systems far from
equilibrium (anomalous diffusion: trajectory on the right).
\label{fig:random_walk}}
\end{figure}

{\it Transport properties of the unified accelerator.}\\
The key problem on the transport properties of particles acceleration is the
characteristics of the orbits in a dynamical system close to equilibrium (see
Fig. \ref{fig:random_walk}a) and non-equilibrium systems (see Fig.
\ref{fig:random_walk}b).

In the one hand we have the well known Brownian motion which is very well
studied and analyzed (called in the literature as Normal Diffusion) and in the
other hand we see orbits following a mixture of short scale random walks
interrupted with large "flights" called in statistical mechanics ``Levy
flights'' and represent in the literature as Anomalous Diffusion (see the
tutorial \cite{vlahos08} for more details). We present bellow the man
characteristics and the dynamics of the orbits and the transport equations for
these types of orbits.

{\it Normal diffusion and the Fokker Planck equation.}\\
Deriving the the Fokker Planck equation (or Kolmogorov forward equation) from
the characteristics of the density probabilities discussed already is very
instructive, that defines its limitations and shows its connection with systems
close to equilibrium. We start from a description of diffusion in terms of a
random walk, assuming that: (i) the mean value of the random walk steps can be
different from zero, which corresponds to a systematic motion of the particles,
and (ii) we assume that both the mean and the variance can be spatially
dependent, which means that the distribution of increments depends on the
spatial location, i.e.\ it is of the form $q_{\Delta z,z}(\Delta z,z)$. The
probability distribution function for a particle at time t at the position z is
\beq
P(z,t) = \int_\infty^\infty P(z-\D z,t-\D t)
q_{\D z,z}(\D z,z-\D z) \,d\D z  ,
\label{eq:chapkol}
\eeq
which is the {\it Chapman-Kolmogorov equation}, and where now $q_{\D z, z}(\D
z,z)$ is the probability density for being at position $z$ and making a step $\D
z$ in time $\D t$. The FP equation can be derived as follows: we expand the
integrand of Eq.\ (\ref{eq:chapkol}) in a Taylor-series in terms of $z$, so that
$P(z,t) = \int_\infty^\infty A B \,d\D z$, with
\beq
 A=P(z,t)-\p_t P(z,t)\D t-\p_z P(z,t)\D z+ \frac{1}{2}\p_z^2 P(z,t)\D z^2 + ... ,
\label{eq:Aeq}
\eeq
where we have also expanded to first order in $t$,
\beq
B=q_{\D z,z}(\D z,z)-\p_z q_{\D z,z}(\D z,z)\D z+ \frac{1}{2}\p_z^2
q_{\D z,z}(\D z,z)\D z^2 + ...
\label{eq:Beq}
\eeq
(note that the Taylor expansion is with respect to the second argument of $q_{\D
z,z}$, we expand only with respect to $z$, not though with respect to $\D z$).
Keeping all terms up to second order in $\D z$, we find the {\it FP
equation} \citep{vlahos08}
\beq
\p_t P(z,t) = -\p_z[V(z)P(z,t)]+\p_z^2 [D(z)P(z,t)] ,
\label{eq:FokkerP}
\eeq
with $V(z)$ systematic or drift term, and $D(z)\equiv \langle\D z^2\rangle
(z)/2\Delta t$ the diffusion coefficient \citep{gardiner94}. The FP equation is
also applied to velocity space, e.g.\ in plasma physics in order to treat
collisional effects, or to position and velocity space together. It has the
advantage of being a deterministic differential equations that allows to
describe the evolution of stochastic systems, as long as the diffusivities and
drift velocities are known, and as long as the conditions for its applicability
are met.

From its derivation it is clear that the FP equation is suited only for systems
close to equilibrium, with just small deviations of some particles from
equilibrium, or, in the random walk sense, with just small steps of the
particles performing the random walk.  A further natural generalization for a
diffusion equation in the approach followed here would be not to stop the Taylor
expansion in Eqs.\ (\ref{eq:Aeq}) and (\ref{eq:Beq}) at second order in $z$, but
to keep all terms, which would lead to the so-called {\it Kramers-Moyal
expansion}. More details about the Fokker-Planck equation can be found in the
literature \cite{gardiner94}.

Assuming that $A(z)=0$ and $D(z)=D$ (constant) the FP equation is simple
\beq
\partial_t P{z,t} = \frac{\sigma^2_{\Delta z}}{2\Delta t} \partial_z^2 P(z,t)
\label{eq:einsdiff}
\eeq
and the solution to it in infinite space is again the Gaussian
\begin{equation}
    P(z,t)=\frac{1}{\sqrt{4 \pi D t}}e^{-z^2/4Dt}.
\label{eq:solDiff}
\end{equation}
The square mean displacement is
\begin{equation}
    <z^2(t)>=\int z^2 P(z,t) \, dz=2Dt ,
\label{eq:normalDiff}
\end{equation}
which is characteristic for the normal diffusion.

{\it Anomalous diffusion and the Fractional Diffusion Equation}\\
Normal diffusion has as basic characteristic the linear scaling of the mean
square displacement of the particles with time, $\langle r^2 \rangle\sim Dt$.
Many different experiments though, including the one shown in the previous
section, reveal deviations from normal diffusion, in that diffusion is either
faster or slower, and which is termed anomalous diffusion. A useful
characterization of the diffusion process is again through the scaling of the
mean square displacement with time, where though now we are looking for a more
general scaling of the form
\beq
\langle r^2(t) \rangle \sim t^\gamma  .
\label{eq:asca}
\eeq
Diffusion is then classified through the scaling index $\gamma$. The case
$\gamma=1$ is normal diffusion, all other cases are termed anomalous. The cases
$\gamma>1$ form the family of super-diffusive processes, including the
particular case $\gamma=2$, which is called ballistic diffusion, and the cases
$\gamma<1$ are the sub-diffusive processes. If the trajectories of a sufficient
number of particles inside a system are known, then plotting $\log <r^2>$ vs
$\log t$ is an experimental way to determine the type of diffusion occurring in
a given system.

As an illustration, let us consider a particle that is moving with constant
velocity $v$ and undergoes no collisions and experiences no friction forces. It
then obviously holds that $r=vt$, so that $\langle r^2(t)\rangle\sim t^2$. Free
particles are thus super-diffusive in the terminology used here, which is also
the origin of the name ballistic for the case $\gamma=2$. Accelerated particles
would even diffuse faster. The difference between normal and a anomalous
diffusion is also illustrated in Fig. \ref{fig:random_walk}, where in the case
of anomalous diffusion long "flights" are followed by efficient "trapping" of
particles in localized spatial regions, in contrast to the more homogeneous
picture of normal diffusion.

It is to note that anomalous diffusion manifests itself not only in the scaling
of Eq.\ (\ref{eq:asca}) with $\gamma\ne 1$ (which experimentally may also be
difficult to be measured), but also in 'strange' and 'anomalous' phenomena such
as 'uphill' diffusion, where particles or heat diffuse in the direction of
higher concentration, or the appearance of non-Maxwellian distributed particle
velocities very often of power-law shape, and it is the main interest in this
proposal.

\cite{bian08} developed a simple model for particle acceleration based on the
ideas mentioned above and using the Continuous Random Walk developed a more
general the fractional diffusion equation \citep{metzler00}. It is demonstrated
clearly \cite[see also the review][]{vlahos08} that depending on the form of the
probabilities $P_(\ell_i),P_2(E), P_3(\tau_i)$ we can build a transport equation
which tailored on the characteristics of the above probabilities e.g. assuming
that $P_1, P_2, P_3$ are Gaussian we recover the FP equation listed above. In
the more general case were the probabilities are more general, as we have
demonstrated for the case of the quasi-perpendicular shock, the strong
turbulence and the multi-island reconnection the transport equation for the
accelerated particles becomes fractional \citep{bian08}.

Finally, we would like to mention that while in GS95 model of turbulence
subdiffusion is difficult to realize \cite{yan08a}, the superdiffusion related
to the accelerated divergence of magnetic field lines in space is the expected
and confirmed property \cite[LV99, see also][]{lazarian04a}.

\section{Acceleration of cosmic rays within LV99 model of reconnection}

\subsection{Model of first order Fermi acceleration}
\label{sec:1st_fermi}

\begin{figure}[!t]
\includegraphics[width=\columnwidth]{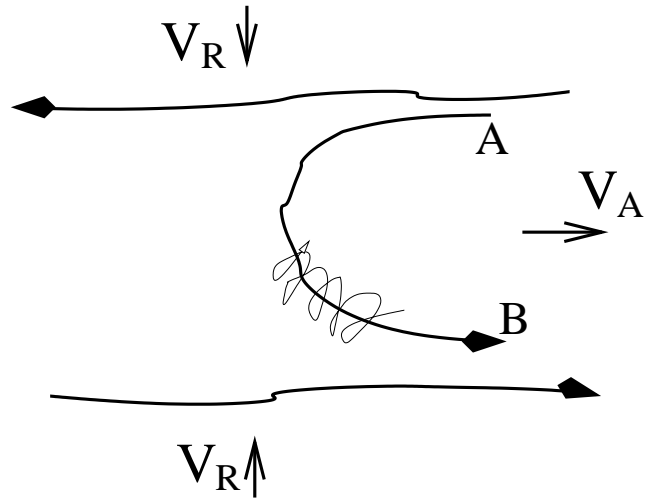}
\caption{ Cosmic rays spiral about a reconnected magnetic field line and bounce
back at points A and B. The reconnected regions move towards each other with the
reconnection velocity $V_R$. The advection of cosmic rays entrained on magnetic
field lines happens at the outflow velocity, which is in most cases of the order
of $V_A$. Bouncing at points A and B happens because either of streaming
instability induced by energetic particles or magnetic turbulence in the
reconnection region. In reality, the outflow region gets filled in by the
oppositely moving tubes of reconnected flux which collide only to repeat on a
smaller scale the pattern of the larger scale reconnection.  From
\cite{lazarian05}.} \label{fig_recon}
\end{figure}

In what follows we discuss the first order Fermi acceleration which arises from
volume-filling reconnection\footnote{We would like to stress that
Figure~\ref{fig_rec} exemplifies only the first moment of reconnection when the
fluxes are just brought together. As the reconnection develops the volume of
thickness $\Delta$ gets filled with the reconnected 3D flux ropes moving in the
opposite directions.}. The LV99 presented such a model of reconnection and
observations of the Solar magnetic field reconnection support the volume-filled
idea \citep{ciaravella08}.

Figure~\ref{fig_recon} exemplifies the simplest realization of the acceleration
within the reconnection region expected within LV99 model. As a particle bounces
back and forth between converging magnetic fluxes, it gains energy through the
first order Fermi acceleration described in \citet[][henceforth
GL05]{degouveia03,degouveia05} \cite[see also][]{lazarian05}.

To derive the energy spectrum of particles one can use the routine way of
dealing with the first order Fermi acceleration in shocks
\cite[see][]{longair92}.  Consider the process of acceleration of $M_0$
particles with the initial energy $E_0$.  If a particle gets energy $\beta E_0$
after a collision, its energy after $m$ collisions is $\beta^m E_0$.  At the
same time if the probability of a particle to remain within the accelerating
region is $P$, after $m$ collisions the number of particles gets $P^m M_0$. Thus
$\ln (M/M_0)/\ln(E/E_0)=\ln P/\ln\beta$ and
\begin{equation}
\frac{M}{M_0}=\left(\frac{E}{E_0}\right)^{\ln P/\ln\beta}
\end{equation}
For the stationary state of accelerated particles the number $M$ is the number
of particles having energy equal or larger than $E$, as some of these particles
are not lost and are accelerated further. Therefore:
\begin{equation}
N(E)dE=const\times E^{-1+(\ln P/\ln\beta)} dE
\label{NE}
\end{equation}

To determine $P$ and $\beta$ consider the following process.  The particles from
the upper reconnection region see the lower reconnection region moving toward
them with the velocity $2V_{R}$ (see Figure~\ref{fig_recon}). If a particle from
the upper region enters at an angle $\theta$ into the lower region the expected
energy gain of the particle is $\delta E/E=2V_{R}\cos\theta/c$. For isotropic
distribution of particles their probability function is
$p(\theta)=2\sin\theta\cos\theta d\theta$ and therefore the average energy gain
per crossing of the reconnection region is
\begin{equation}
\langle \delta E/E \rangle =\frac{V_{R}}{c}\int^{\pi/2}_{0} 2\cos^2\theta \sin\theta d\theta=4/3\frac{V_{R}}{c}
\end{equation}
An acceleration cycle is when the particles return back to the upper
reconnection region. Being in the lower reconnection region the particles see
the upper reconnection region moving with the speed $V_{R}$. As a result, the
reconnection cycle provides the energy increase $\langle \delta E/E
\rangle_{cycle}=8/3(V_{R}/c)$ and
\begin{equation}
\beta=E/E_0=1+8/3(V_{R}/c)
\label{beta}
\end{equation}

Consider the case of $V_{diff}\ll V_R$. The total number of particles crossing
the boundaries of the upper and lower fluxes is $2\times 1/4 (n c)$, where $n$
is the number density of particles. With our assumption that the particles are
advected out of the reconnection region with the magnetized plasma outflow the
loss of the energetic particles is $2\times V_{R}n$. Therefore the fraction of
energetic particles lost in a cycle is $V_{R} n/[1/4(nc)]=4V_{R}/c$ and
\begin{equation}
P=1-4V_{R}/c.
\label{P}
\end{equation}

Combining Eq.~(\ref{NE}), (\ref{beta}), (\ref{P}) one gets
\begin{equation}
N(E)dE=const_1 E^{-5/2}dE,
\label{-5/2}
\end{equation}
which is the spectrum of accelerated energetic particles for the case when the
back-reaction is negligible \cite[see GL05, cf.][]{drury12}. \footnote{The
obtained spectral index is similar to the one of Galactic cosmic rays.}.

\begin{figure}[!t]
\includegraphics[width=\columnwidth]{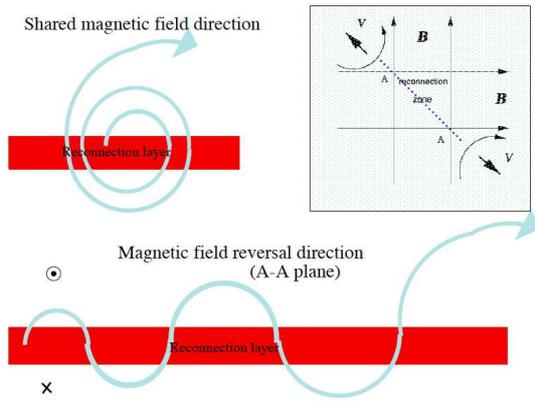}
\caption{Particles with a large Larmor radius gyrate about the magnetic field
shared by two reconnecting fluxes (the latter is frequently referred to as
``guide field''. As the particle interacts with converging magnetized flow
corresponding to the reconnecting components of magnetic field, the particle
gets energy gain during every gyration. \label{fig_accel2}}
\end{figure}

The first order acceleration of particles entrained on the contracting magnetic
loop can be understood from the Liouville theorem. In the process of the
magnetic tubes contraction a regular increase of the particle's energies is
expected. The requirement for the process to proceed efficiently is to keep the
accelerated particles within the contracting magnetic loop. This introduces
limitations on the particle diffusivity perpendicular to the magnetic field
direction. The subtlety of the point above is related to the fact that while in
the first-order Fermi acceleration in shocks magnetic compression is important,
the acceleration via the LV99 reconnection process is applicable even to
incompressible fluids. Thus, unlike shocks, it is not the entire volume that
shrinks for the acceleration, but only the volume of the magnetic flux tube.
Thus high perpendicular diffusion of particles may decouple them from the
magnetic field. Indeed, it is easy to see that while the particles within a
magnetic flux rope depicted in Figure~\ref{fig_recon} bounce back and forth
between the converging mirrors and get accelerated, if these particles leave the
flux rope fast, they may start bouncing between the magnetic fields of different
flux ropes which may sometimes decrease their energy. Thus it is important that
the particle diffusion both in the parallel and perpendicular directions to the
magnetic field stay different. The particle anisotropy which arises from
particles preferentially getting acceleration in terms of the parallel momentum
may also be important.  Similarly, the first order Fermi acceleration can happen
in terms of the perpendicular momentum.  This is illustrated in
Figure~\ref{fig_accel2}.  There the particle with a large Larmour radius is
bouncing back and forth between converging mirrors of reconnecting magnetic
field systematically getting an increase of the perpendicular component of its
momentum.  Both processes take place in reconnection layers.

\subsection{Simulations of the acceleration of cosmic rays by reconnection}

The numerical studies of the cosmic ray acceleration in reconnection regions
were performed in  \cite{kowal11,kowal12a}, where to test the mechanism the data
cubes obtained from the models of weakly stochastic magnetic reconnection were
used.  For a given snapshot we obtain a full configuration of the plasma flow
variables (density and velocity) and magnetic field.  We inject test particles
in such an environment and integrate their trajectories solving the motion
equation for relativistic charged particles
\begin{equation}
 \frac{d}{d t} \left( \gamma m \vec{u} \right) = q \left( \vec{E} + \vec{u} \times \vec{B} \right) ,
\end{equation}
where $\vec{u}$ is the particle velocity, $\gamma \equiv \left( 1 - u^2 / c^2
\right)^{-1}$ is the Lorentz factor, $m$ and $q$ are particle mass and electric
charge, respectively, and $c$ is the speed of light.

The study of the magnetic reconnection is done using the magnetohydrodynamic
fluid approximation, thus we do not specify the electric field $\vec{E}$
explicitly.  Nevertheless, the electric field is generated either by the flow of
magnetized plasma or by the resistivity effect and can be obtained from the
Ohm's equation
\begin{equation}
 \vec{E} = - \vec{v} \times \vec{B} + \eta \vec{j} ,
\end{equation}
where $\vec{v}$ is the plasma velocity and $\vec{j} \equiv \nabla \times
\vec{B}$ is the current density.

In our studies we are not interested in the acceleration by the electric field
resulting from the resistivity effects, thus we neglect the last term.  After
incorporating the Ohm's law, the motion equation can be rewritten as
\begin{equation}
 \frac{d}{d t} \left( \gamma m \vec{u} \right) = q \left[ \left( \vec{u} - \vec{v} \right) \times \vec{B} \right] . \label{eq:trajectory}
\end{equation}

\begin{figure}[ht]
 \center
 \includegraphics[width=0.48\textwidth]{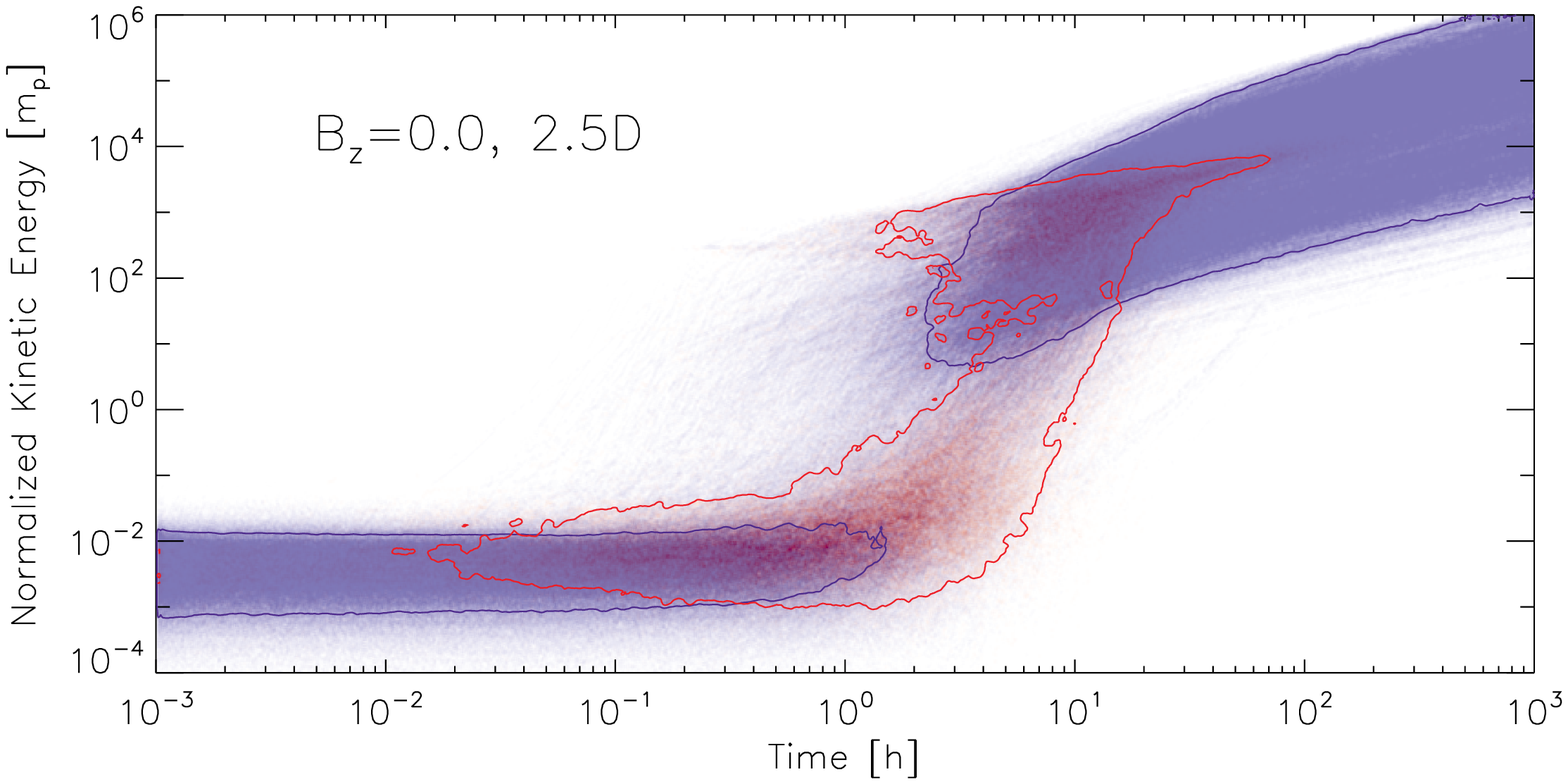}
 \includegraphics[width=0.48\textwidth]{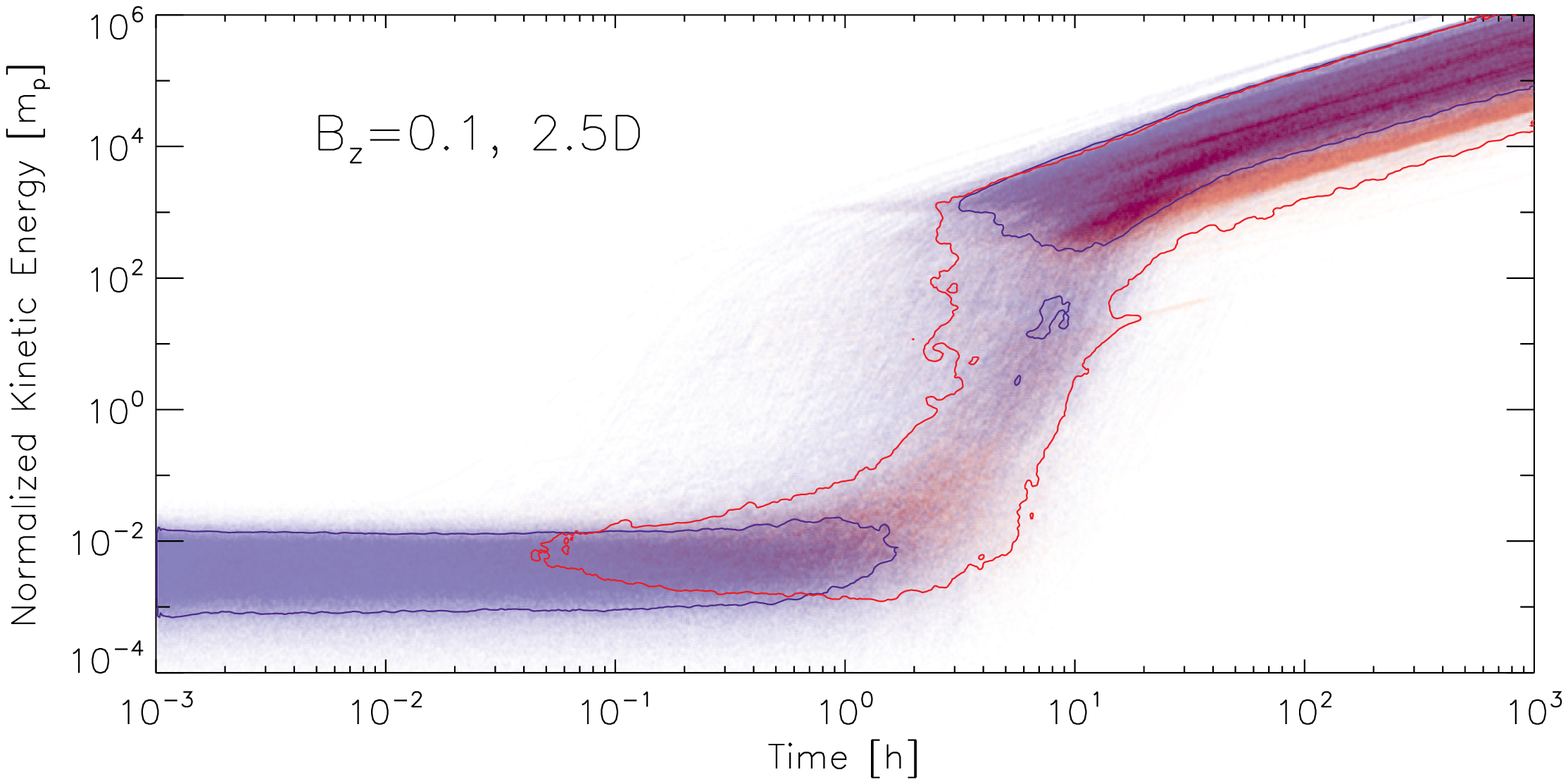}
 \includegraphics[width=0.48\textwidth]{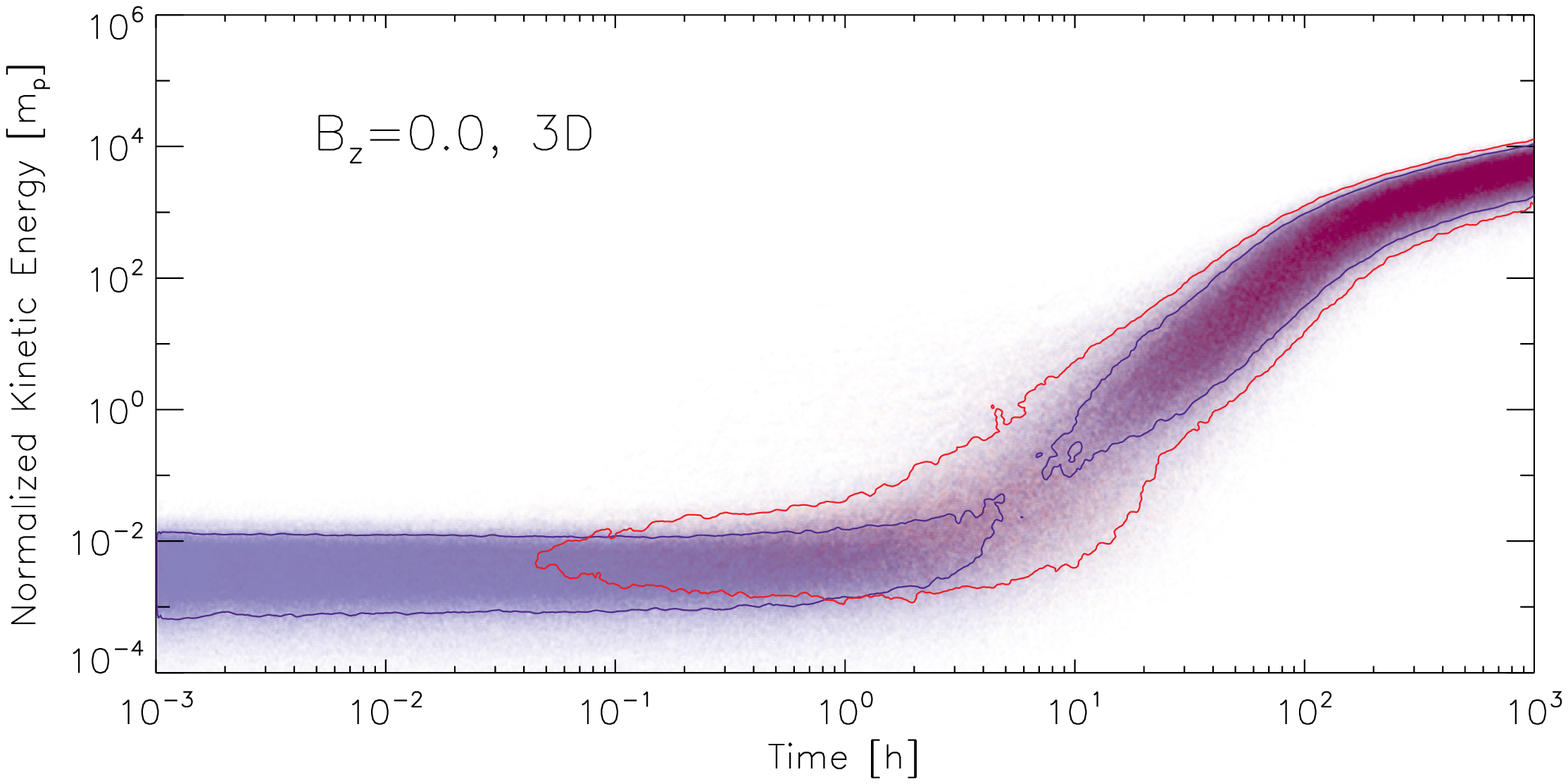}
 \caption{Kinetic energy evolution of a group of 10$^4$ protons in 2D models of
reconnection with a guide field $B_z$=0.0 and 0.1 (top panels, respectively). In
the bottom panel a fully 3D model with initial $B_z$=0.0 is presented.  The
colors show how the parallel (red) and perpendicular (blue) components of the
particle velocities increase with time. The contours correspond to values 0.1
and 0.6 of the maximum number of particles for the parallel and perpendicular
accelerations, respectively.  The energy is normalized by the rest proton mass
energy.  The background magnetized flow with multiple current sheet layers is at
time 4.0 in Alfv\'en time units in all models. From \cite{kowal11}.
\label{fig:energy_2d_3d}}
\end{figure}

In Figure~\ref{fig:energy_2d_3d}, we present the time evolution of the kinetic
energy of the particles which have their parallel and perpendicular (red and
blue points, respectively) velocity components accelerated for three models of
reconnection.  The upper left panel shows the energy evolution for a 2D model
without the guide field (as in the models studied in the previous sections).
Initially, the particles pre-accelerate by increasing their perpendicular
velocity component only.  Later we observe an exponential growth of energy
mostly due to the acceleration of the parallel component which stops after the
energy reaches values of 10$^3$--10$^4$~$m_p$ (where $m_p$ is the proton rest
mass energy).  Further on, particles accelerate their perpendicular component
only with smaller linear rate in a log-log diagram.  In 2.5D case, there is also
an initial slow acceleration of the perpendicular component followed by the
exponential acceleration of the parallel velocity component.  Due to the
presence of a weak guide field, the parallel component accelerates further to
higher energies at a similar rate as the perpendicular one.  This implies that
the presence of a guide field removes the restriction seen in the 2D model
without a guide field and allows the particles to increase their parallel
velocity components as they travel along the guide field, in open loops rather
than in confined 2D islands.  This result is reconfirmed by the 3D model in the
bottom panel of Figure~\ref{fig:energy_2d_3d}, where no guide field is necessary
as the MHD domain in fully three-dimensional.  In this case, we clearly see a
continuous increase of both components, which suggests that the particle
acceleration behavior changes significantly when 3D effects are considered,
where open loops replace the closed 2D reconnecting islands.

\subsection{Implications of the acceleration via reconnection}

\subsubsection{Origin of the anomalous cosmic rays}

The processes of the energetic particle acceleration in the process of turbulent
reconnection can pre-accelerate particles to the intermediate energies helping
to solve the problem of particle injection into shocks. It can also act as the
principal process of acceleration. Below we present the case where we believe
that the latter takes place.

Since the crossing of the termination shock (TS) by Voyager 1 (V1) in late 2004
and by Voyager 2 (V2) in mid 2007 it became clear that several paradigms needed
to be revised. Among them was the acceleration of particles. Prior to the
encounter of the termination shock by V1 the prevailing view was that anomalous
cosmic rays (ACRs) were accelerated at the TS by diffusive shock acceleration
(DSA) to energies 1-300 MeV/nuc \cite[e.g.][]{jokipii98,cummings98}.  However,
with the crossing of the TS by V1 the energy spectrum of ACR did not unroll to
the expected source shape: a power-law at lower energies with a roll off at
higher energies.  After 2004, both the V1 spectrum in the heliosheath and the V2
spectrum upstream the TS, continued to evolve toward the expected source shape.

To explain this paradox several models were proposed. Among them,
\cite{mccomas06} suggested that at a blunt shock the acceleration site for
higher energy ACRs would be at the flanks of the TS, where the injection
efficiency would be higher for DSA and connection times of the magnetic field
lines to the shock would be longer, allowing acceleration to higher energies.
\cite{fisk06} on the other hand suggested that stochastic acceleration in the
turbulent heliosheath would continue to accelerate ACRs and that the high-energy
source region would thus be beyond the TS. Other works, such as \cite{jokipii06}
and \cite{florinski06} try to explain the deficit of ACRs based on a dynamic
termination shock.  \cite{jokipii06} pointed out that a shock in motion on time
scales of the acceleration time of the ACRs, days to months, would cause the
spectrum to differ from the expected DSA shape.  \cite{florinski06} calculated
the effect of Magnetic Interacting Regions (MIRs) with the Termination Shock on
the ACR spectral shape.  They show that there is a prolonged period of depressed
intensity in mid-energies from a single MIR.  Other recent works have included
stochastic acceleration, as well as other effects
\citep{moraal06,zhang06,langner06,ferreira07}.  It became clear after the
crossing of the TS by V2 that these models would require adjustments.  The
observations by V2 indicate for example that a transient did not cause the
modulation shape of the V2 spectrum at the time of its TS crossing.  When both
spacecraft were in the heliosheath in late 2007, the radial gradient in the
13-19 MeV/nuc ions did not appear to be caused by a transient.  The 60-74
MeV/nuc ions have no gradient, so no north-south or longitudinal asymmetry is
observed in the ACR intensities at the higher energies.

In \citet[][henceforth LO09]{lazarian09b} we propose an alternative model, which
explains the source of ACRs as being in the heliosheath and we appeal to
magnetic reconnection as a process that can accelerate particles. LO09 explained
the origin of the magnetic field reversals that induce magnetic reconnection in
heliosheath and heliopause.

Indeed, it is well known that magnetic field in the heliosphere change polarity
and induce reconnection.  For instance, as the Sun rotates magnetic field twists
into a Parker spiral \citep{parker58} with magnetic fields separated by a
current sheet \cite[see][]{schatten71}. The changes of magnetic field are also
expected due to the Solar cycle activity.

\begin{figure}[ht]
 \center
 \includegraphics[width=0.7\columnwidth]{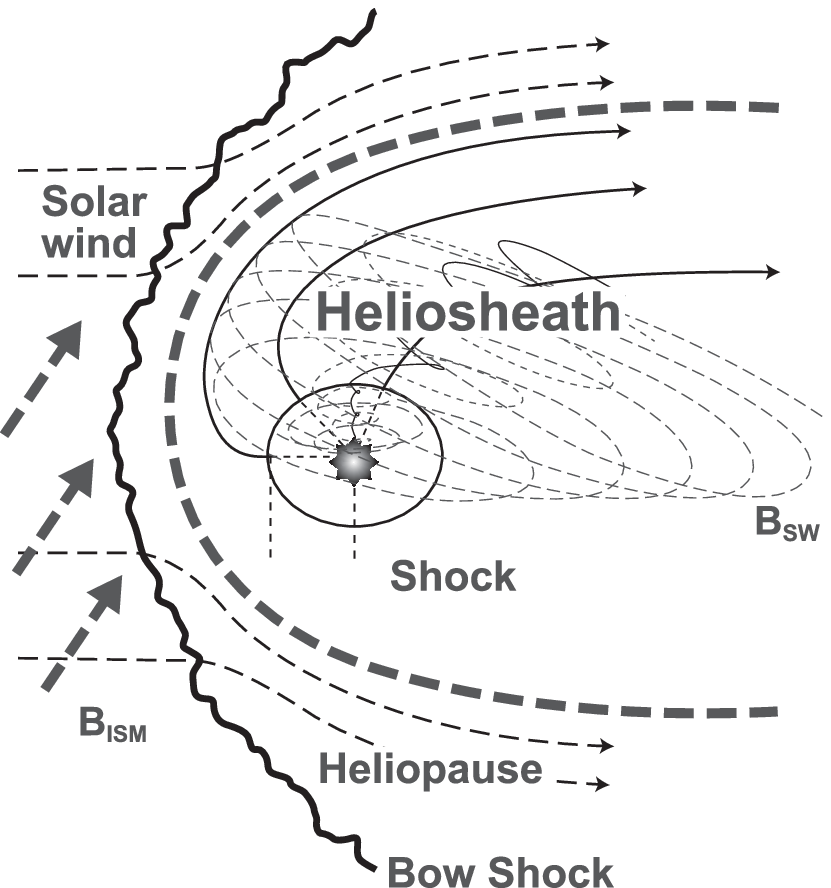}
 \includegraphics[width=0.7\columnwidth]{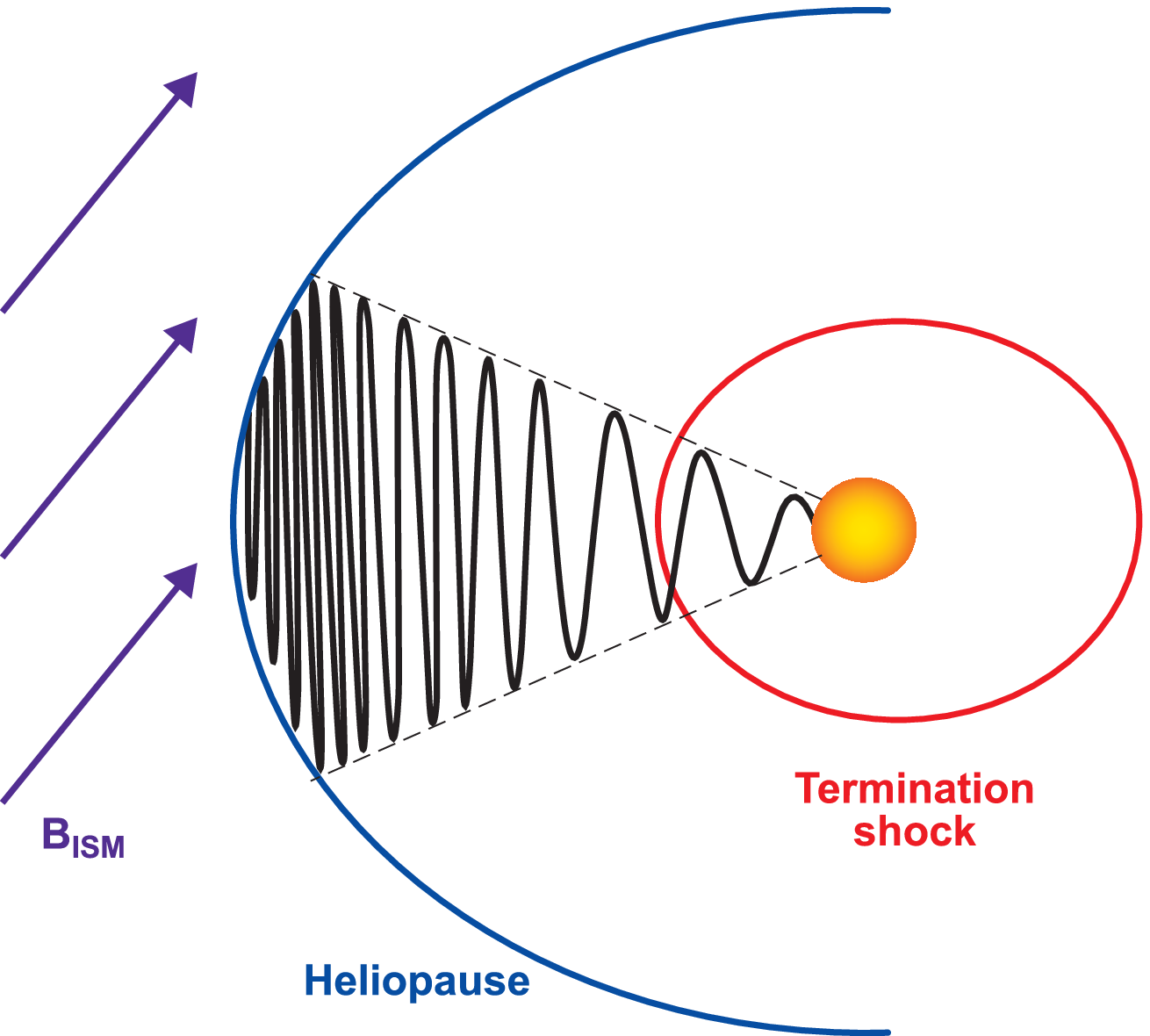}
 \caption{{\it Upper plot}. Global view of the interaction of the solar wind
with the interstellar wind. The spiral solar magnetic field (shown in dark
dashed lines) is  shown being deflected at the heliopause. The heliopause itself
is being deflected by the interstellar magnetic field. \citep[figure adapted
from][]{suess06}. {\it Lower plot}. A meridional view of the boundary sectors of
the heliospheric current sheet  and how the opposite sectors get tighter closer
to the heliopause. The thickness of the outflow regions in the reconnection
region depends on the level of turbulence. From LO09. \label{fig_anomal}}
\end{figure}

The question now is at what part of the heliosheath we expect to see reversals.
The structure of the magnetic field in the solar wind is complex. The solar
magnetic field lines near the termination shock are azimuthal and form a spiral
(see Figure~\ref{fig_anomal}). We expect the reconnection and the corresponding
energetic particle acceleration to happen at the heliosheath closer to the
heliopause. This explains why Voyagers do not see the signatures of anomalous
cosmic ray acceleration as they pass the termination shock. Appealing to their
model of collisionless reconnection, \cite{drake10} provided a similar
explanation of the origin of the anomalous cosmic rays.

\subsubsection{Acceleration of cosmic rays in heliototail}

It is known that cosmic rays arrival direction has an energy dependent large
angular scale anisotropy with an amplitude of order $10^{-4}-10^{-3}$. The first
comprehensive observation of this anisotropy was provided by a network of muon
telescopes sensitive to sub-TeV energies and located at different latitudes
\citep{nagashima98}. More recently, an anisotropy was also observed in the
multi-TeV energy range by the Tibet AS$\gamma$ array \citep{amenomori06},
Super-Kamiokande \citep{guillian07} and by MILAGRO \citep{abdo09}, and the first
high statistics observation in the southern hemisphere in the 10 TeV region, is
being reported by IceCube \citep{abbasi10}. The origin of the large angular
scale anisotropy in the cosmic rays arrival direction is still unknown. The
structure of the local interstellar magnetic field is likely to have an
important role. However the combined study of the anisotropy energy and angular
dependency, its time modulation and angular scale structure seem to suggest that
the observation might be a combination of multiple superimposed effects, caused
by phenomenologies at different distances from Earth.

In this context, particular interest is derived from the observation of a broad
excess of sub-TeV cosmic rays in a portion of the sky compatible with the
direction of the heliospheric tail (or heliotail) \citep{nagashima98,hall99}.
The heliotail is the region of the heliosphere downstream the interstellar
matter wind delimited within the heliopause, i.e. the boundary that separates
the solar wind and interstellar plasmas \citep{izmodenov06}. The observed excess
was attributed to some unknown anisotropic process connected with the heliotail
(thus called tail-in excess). The gyro-radius of sub-TeV cosmic protons is less
than about 200 AU (in a $\sim$1 $\mu$G interstellar magnetic field), which is
approximately the size of the heliosphere and, most likely, smaller than the
width and length of the heliotail. The persistence of the cosmic ray anisotropy
structure in the multi-TeV energy range makes it challenging to link this
observation to the heliosphere. Although the unknown size and extension of the
heliotail contributes to the uncertainty on the energy scale at which
heliospheric influence on cosmic rays starts to be negligible. However, we know
that the observations of multi-TeV cosmic rays anisotropy show small angular
scale patterns superimposed to the smooth broad structure of the tail-in excess,
which is suggestive of a local origin, i.e. within the heliotail. With the same
technique used in gamma ray detection to estimate the background and search for
sources of gamma rays, the MILAGRO collaboration discovered two localized excess
regions in the cosmic rays arrival direction distribution \citep{abdo08}. The
same excess regions were reported by the ARGO-YBJ air shower array
\citep{vernetto09}. The strongest and more localized of them (with an angular
size of about 10$^{\circ}$) coincides with the direction of the heliotail.

\begin{figure}[!t]
\includegraphics[width=\columnwidth]{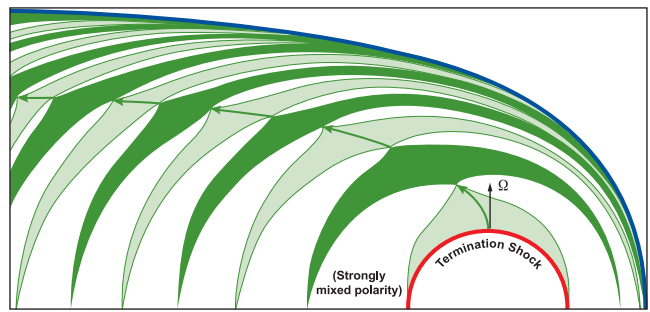}
\caption{ A meridional view of the boundary sectors of the heliospheric current
sheet  and how the opposite sectors get tighter closer to the heliopause and
into the heliotail. The thickness of the outflow regions in the reconnection
region depends on the level of turbulence. The length of the outflow regions $L$
depends on the mean geometry of magnetic field and turbulence. From
\cite{lazarian10}.} \label{structure1}
\end{figure}

The explanation of this excess related to the acceleration of energetic
particles in reconnection region was suggested by \cite{lazarian10}.
Figure~\ref{structure1} represents the possible structure of the heliotail which
arises from the solar magnetic field cycles \citep{parker79}. The magnetic
fields of the opposite polarities emerge as the result of 11 year solar dynamo
cycle. As the magnetic field is carried away by solar wind, the reversed
magnetic field regions get accumulated in the heliotail region. This is where
reconnection is expected to occur.

Naturally, the actual heliotail is going to be turbulent, which is not
represented by the idealized drawing in Figure~\ref{structure1}. As the Alfv\'en
speed is smaller than the Solar wind speed, magnetic reconnection does not
change the overall magnetic field structure. Nevertheless, the effects of
turbulence are expected to be very important from the point of view of magnetic
reconnection and the particle acceleration that it entails.

The simulations of the heliotail are extremely challenging
\cite[see][]{pogorelov09a,pogorelov09b} and have not been done with the
sufficient resolution and extent. While we believe that future research will
provide details necessary for quantitative modeling, the schematic
representation of the heliotail structure depicted in Figure~\ref{structure1} is
true in terms of major features.

Galactic cosmic rays entering the heliotail will get re-accelerated, which would
affect their spectrum. The estimate of the maximal energy of protons which can
be accelerated through this process can be obtained through the usual arguments
that the Larmor radius should not be larger than the size of the magnetized
region $L_{zone}$ \citep{longair97}
\begin{equation}
E_{max}\approx 0.5~{\rm TeV} \left(\frac{B}{1~{\rm \mu G}}\right) \left(\frac{L_{zone}}{100~AU}\right),
\end{equation}
which sets the limit of energies, i.e. $E_{max}$, through appealing to the fact
that protons of larger energies cannot be confined by magnetic fields to
experience the acceleration through multiple bouncing back and forth between the
reconnecting magnetic fluxes. The magnetic field estimates vary and can be
expected of the order of $\sim 4 \mu$G. $L_{zone}$ is in the range from 100 to
400 AU \citep{pogorelov09a}. Therefore cosmic rays with energies up to 10TeV can
be re-accelerated in the heliotail, explaining the observations\footnote{An
alternative explanation based on scattering cosmic rays in the heliotail was
suggested in \cite{desiati11}}.

\subsubsection{Acceleration of particles by reconnection in solar flares and
galaxy clusters}

Acceleration is known to accompany solar flares. LV99 explains solar flares and
predicts that only insubstantial portion of energy goes into heating during the
reconnection. The rest goes into turbulence and the acceleration of energetic
particles. The resulting turbulence can accelerate the energetic particles
\citep{petrosian08} via the second order Fermi process (see
\S\ref{sec:2nd_fermi}), but the first order Fermi acceleration arising from
reconnection (see \S\ref{sec:1st_fermi}.) is also present and may dominate. The
relative role of the two processes requires more studies.

LV99 reconnection was discussed as the source of the observed energetic
particles in \cite{lazarian06a} and \cite{lazarian11a}. Magnetic fields of
different polarity get into contact as magnetized gas is being collected into
the potential well of the galaxy cluster. In addition, superAlfv\'enic
turbulence in galaxy clusters would produce magnetic field reversals of magnetic
fields which will also dissipate through reconnection. Therefore we believe that
the acceleration may be efficient. A quantitative study of the process is
presented in Brunetti \& Lazarian (2012, in prep.).

\subsubsection{Acceleration in AGNs}

Cosmic ray acceleration in the high energy range is still not fully understood.
The spectrum of the  highest energy cosmic rays (UHECRs) is consistent with an
origin in extragalactic astrophysical sources and candidates range from the
birth of compact objects to explosions related to gamma-ray bursts (GRBs),  or
to events in active galaxies  (AGNs) \citep[e.g.][]{melrose09,kotera11}. Very
high energy observations of AGNs and GRBs with the Fermi and Swift satellites
and  ground based gamma ray observatories (HESS, VERITAS and MAGIC) on the other
hand, are challenging current theories of particle acceleration - mostly based
on stochastic acceleration in shocks -  which have to explain how particles are
accelerated to  $>$ TeV energies in regions relatively small compared to the
fiducial scale of their sources (Sol et al. 2012, in prep.).

Traditionally discussed predominantly in the context of solar flares
\citep[e.g.][]{drake06,drake09,gordovskyy10,nishizuka10,gordovskyy11,zharkova11}
, the Earth magnetotail \citep{lazarian09b,drake10}, and the solar wind
\citep[e.g.][]{lazarian10}, particle acceleration in magnetic reconnection sites
is currently also being explored in relativistic  astrophysical environments. It
has been  invoked in the production of ultra high energy cosmic rays
\citep[e.g.][]{degouveia00,degouveia01,kotera11}, in jet-accretion disk systems
\citep{degouveia10,degouveia10a,giannios10,delvalle11,ding10}, and in the
general framework of AGNs and GRBs
\citep{ostrowski02,lazarian03,degouveia10,giannios10,zhang11,uzdensky11a,
uzdensky11b,degouveia11,mckinney12}.

Magnetic reconnection events like those associated to solar flares can be a very
powerful mechanism operating on accretion disks (G05). In fact, the magnetic
power released in fast reconnection flares has been found to be more than
sufficient to accelerate relativistic plasmons and produce the observed radio
luminosity of the nuclear jets associated both to microquasars and low luminous
AGNs. The observed correlation between the radio luminosity and the black hole
mass in these sources, which spans $10^9$ orders of magnitude in mass
\citep{falcke04}, is naturally explained in this model as simply due to the
magnetic reconnection activity at the jet launching region of the accretion disk
coronae of these sources \citep{degouveia10,degouveia10a}. A similar process may
also explain the observed x-ray flares in YSOs.

\subsection{Acceleration and Gamma Ray Bursts}

In the field of Gamma Ray Bursts (GRBs),  the new observations that followed the
launching of Swift and Fermi satellites, while solving some old problems, have
raised new questions. The most important unknown parameter is the ratio
($\sigma$) between the Poynting flux and the matter (baryonic $+$ leptonic)
flux. In the standard fireball internal shock (IS) scenario
\citep{paczynski86,shemi90}, magnetic fields are assumed not to play a
dynamically important role, i.e. $\sigma << 1$. An alternative view is that the
GRB outflow carries a dynamically important magnetic field component, i.e.
$\sigma >> 1$. The GRB radiation in this case would be powered by dissipation of
the magnetic field energy in the ejecta
\cite[e.g.][]{usov92,thompson94,meszaros97,piran99,piran05,lyutikov03}. Several
recent developments suggest that we should consider seriously magnetic dominated
jets as a viable option for the source of GRBs. First, the analogy with AGNs
jets  in which it is quite certain that the inner engine cannot accelerate
baryonic dominated jets suggest that similarly in GRBs at least in its inner
region the jet is Poynting flux dominated.

Recent Fermi observation of GRB 080916C shows that the bright photosphere
emission associated with a putative fireball is missing, which  raises a
challenge to the traditional fireball IS model and suggests that the central
engine likely launches a Poynting-flux-dominated outflow at least for this
burst.

In this case, magnetic energy may be sufficient to feed GRBs. Magnetic
reconnection was suggested as a component for GRBs more than a decade ago
\citep{thompson94}. The problem lay, however, in the intrinsic difficulty of
reconnection as it is a very slow process in ordered fields. As with the case
for solar flares, both a slow phase of accumulation of the oppositely directed
flux and a fast bursty phase are required for reconnection. Essential progress
was made by \citep{lazarian03}, who proposed a new GRB model of self-adjusted
reconnection based on the findings of fast reconnection in 3D turbulent magnetic
fields \citep{lazarian99}. As a result of the increased turbulence the
reconnection rate increases, inducing a positive feedback which results in the
explosive reconnection. Such a process provides an alternative explanation to
gamma ray bursts (Lazarian et al. 2003). More recently the model was elaborated
and connected with observational data in \cite{zhang11}.

Similar to the internal shock model, the mini-shells interact internally at the
radius $R_{\rm IS}\sim \Gamma^2 c \Delta t$.  Most of these early collisions,
however, have little energy dissipation, but serve to distort the ordered
magnetic field lines entrained in the ejecta. As in other astrophysical objects,
the system is prone to turbulence because of high Reynolds and magnetic Reynolds
numbers\footnote{This is a nontrivial statement given the fact that the system
is highly collisionless. Taking into account, nevertheless, the high
magnetization factor, most kinetic motions are concentrated perpendicular to
magnetic field so that it is the much suppressed perpendicular viscosity and
resistivity that should be adopted.} Since in GRBs, the outflow is highly
relativistic, it is impractical for the magnetic field to be distorted. In any
case, fast reconnection can be triggered in weakly stochastic magnetic field
(LV99). The details require further studies especially in the regime of
relativistic turbulence.

\begin{figure*}
\centering
\includegraphics[width=\columnwidth]{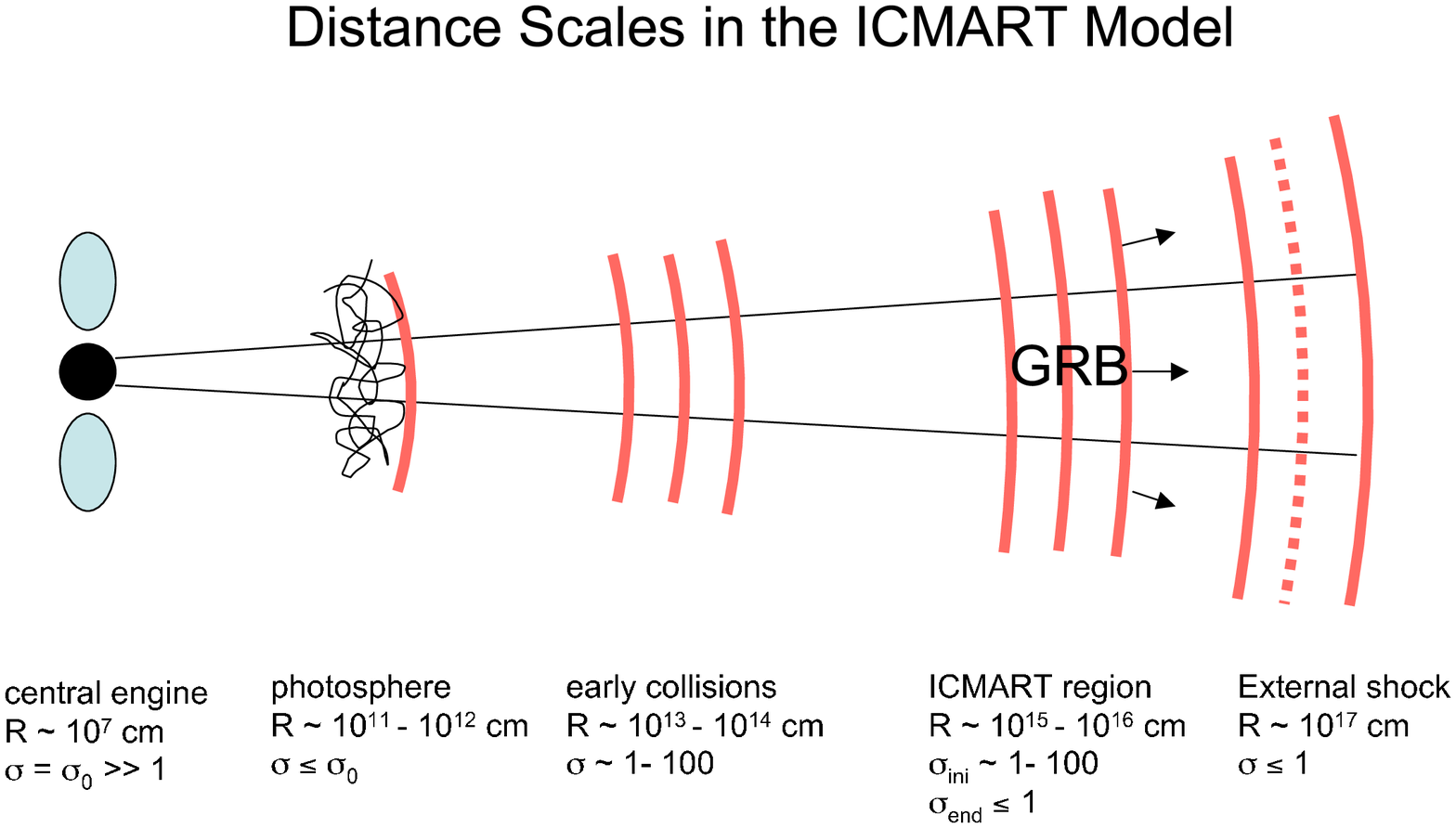}
\includegraphics[width=\columnwidth]{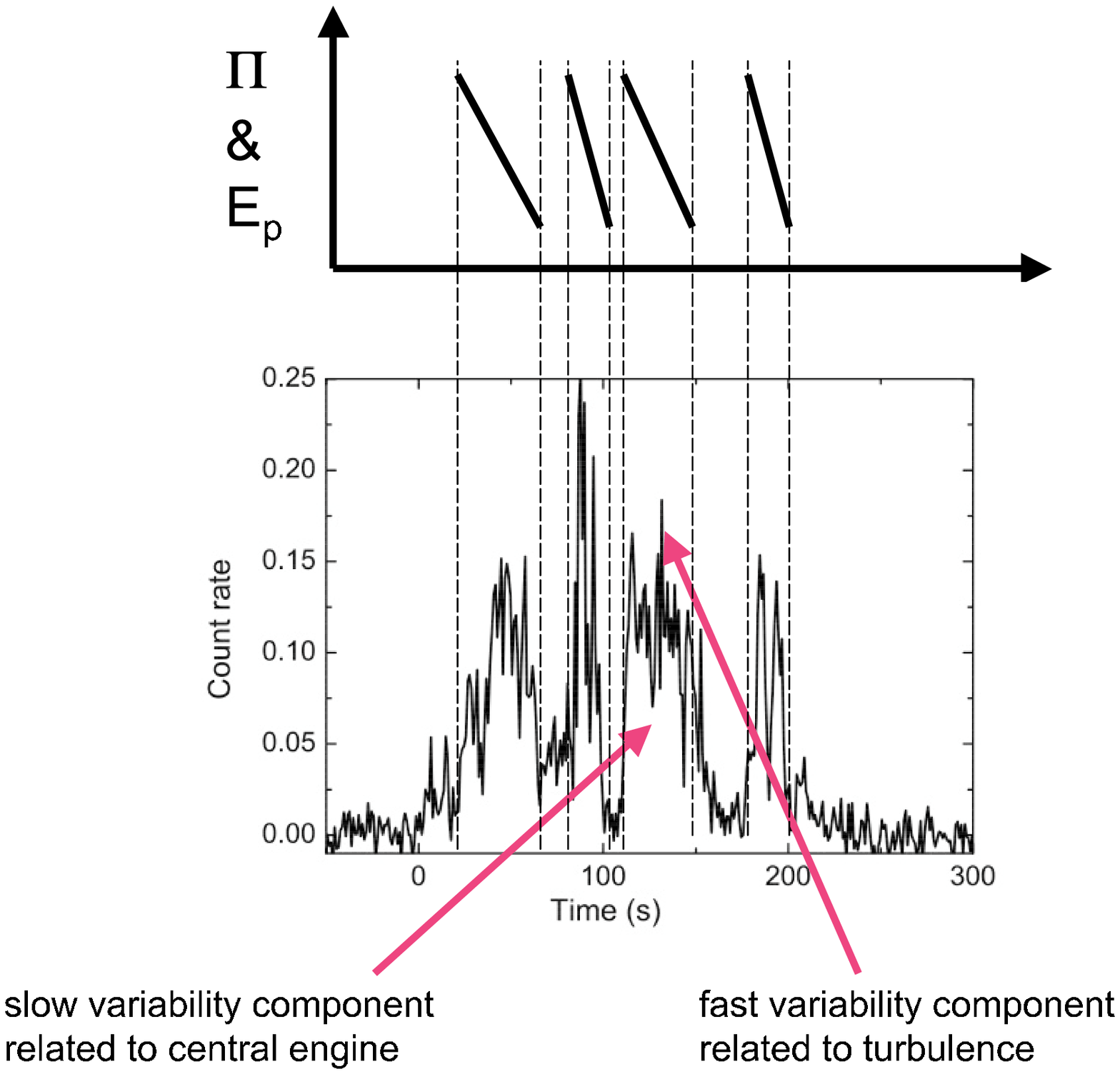}
\caption{\small {\it Left panel}: A cartoon picture of the ICMART model. The
typical distances and $\sigma$ values of various events are marked. {\it Right
panel}: An example of GRB light curve that shows two variability time scales.
The light curve of GRB 050607 is taken from the NASA Swift GRB archival data web
site http://swift.gsfc.nasa.gov/docs/swift/archive/grb\_table/
grb\_lookup.php?grb\_name=050607. The predictions of decreasing gamma-ray
polarization degree $\Pi$ and the spectral peak energy $E_p$ within individual
pulses are indicatively presented. Detailed decaying functions would be
different depending on the details of evolution of magnetic field configuration,
$\sigma$ value, as well as balance between heating and cooling of electrons. The
general decreasing trend is robust. \cite[from][]{zhang11}} \label{icmart}
\end{figure*}

The perturbations build up as the mini-shells propagate outward (see
Fig.\ref{icmart} {\em right}). At a certain point, the turbulence reaches the
critical for a fast reconnection to take place.  Reconnection events rapidly
eject outflow, which further increase the turbulence intensity. This results in
a run-away discharge of the magnetic field energy in a reconnection/turbulence
avalanche. This is one ICMART event, which corresponds to one GRB pulse. During
the magnetic field energy discharge, the $\sigma$ value drops from the original
value to around unity.

A GRB is composed of several ICMART events (i.e. broad pulses). The peak energy
$E_p$ is expected to drop from high to low across each pulse. The $\gamma$-ray
polarization degree is also expected to drop from $\sim 50-60\%$ to $\sim$ a few
$\%$ during each pulse. The magnetic field configuration at the end of prompt
emission is largely randomized, but still has an ordered component. The GRB
light curves should have two variability components, a broad (slow) component
related to the central engine activity, and a narrow (fast) component associated
with the relativistic magnetic turbulence  (see Fig.\ref{icmart}). More detailed
discussions can be found in \cite{zhang11}.

\section{Discussion}

\subsection{Turbulence and cosmic ray acceleration in astrophysical fluid}

Turbulence is the essential part of astrophysical fluid dynamics. As
astrophysical turbulence is magnetized, MHD turbulence presents the most
important process to be accounted for. The progress in understanding MHD
turbulence calls for the adequate representation of the turbulence in the
process of cosmic ray propagation and acceleration. It is regretful that a
number of outdated ad hoc ideas about MHD turbulence are still being used while
describing cosmic ray interactions with turbulence.

The decomposition of MHD turbulence into modes (Alfv\'en, fast and slow)
provides a way to describe the cosmic ray -- turbulence interactions in a
theory-motivated way. For instance, for the cascade driven at large scales the
Alfv\'en modes get marginally important for scattering of low energy cosmic
rays. This is contrary to the claims in most of the textbooks where Alfv\'en
modes are described as the principal scattering agent. On the contrary, the
scattering by fast modes is efficient, but the damping of fast modes must be
taken into account. This changes the treatment of acceleration compared to the
accepted one. At the same time, the ranges of sonic and Alfv\'en Mach number for
which the decomposition works require further studies.

Turbulence is also important for the acceleration of cosmic rays in shocks. The
generation of magnetic fields both in the pre-shock and post-shock plasmas are
important ingredients of forming strong enough converging mirrors between which
cosmic rays bounce and get accelerated. In addition, turbulence makes
reconnection of magnetic fluxes fast and this induces the acceleration of cosmic
rays via through yet another process of the First order Fermi acceleration type.

A sober note is also due. One has to accept that magnetic turbulence is far from
being completely understood phenomenon. There are different regimes of
turbulence, e.g. turbulence in partially ionized gas presents a pronouncedly
different behavior compared to the standard MHD turbulence picture. Similarly,
turbulence in the presence of localized sources and sinks of turbulent energy,
i.e. imbalanced turbulence, is rather different from the balanced counterpart.
In addition, the instabilities of cosmic ray fluid and the back-reaction of
cosmic rays on turbulence may also be important. The consequences of these
changes on the first order Fermi acceleration and the second order Fermi
acceleration require further studies.

\subsection{Models of reconnection}

Since the introduction of the LV99 model, more traditional approaches to
reconnection have been changed considerably. At the time of its introduction,
the models competing with LV99 were some modifications of a single X-point
collisionless reconnection. Those models had point-wise localized reconnection
region and inevitably prescribed opening of the reconnection region upon the
scales comparable to $L$ (see Figure~\ref{fig_rec}). Such reconnection was
difficult to realize in astrophysical conditions in the presence of random
forcing which at high probability would collapse the opening of the reconnection
layer. Single X-point reconnection were rejected in observations of solar flares
by \cite{ciaravella08}.

Modern models of collisionless reconnection resemble the original LV99 model in
a number of respects. For instance, they discuss, similarly to the LV99, the
volume filled reconnection, although one may still wonder how this volume
filling is being achieved in the presence of a single reconnection layer
\citep[see][]{drake06}.  While the authors still talk about islands produced in
the reconnection, in three dimensions these islands are expected to evolve into
contracting 3D loops or ropes \citep{daughton08}, which is similar to what is
depicted in Figure~\ref{fig_recon}. Thus we do not expect to see a cardinal
difference between the first order Fermi processes of the acceleration described
in GL03 and later in \cite{drake06}. This suggests that the back-reaction of the
particles calculated in \cite{drake06} considering the firehose instability may
be employed as a part of the acceleration process described in GL03.

The departure from the idea of regular reconnection and introduction of magnetic
stochasticity is also obvious in a number of the recent papers appealing to the
tearing mode instability\footnote{The idea of appealing to the tearing mode as a
means of enhancing the reconnection speed can be traced back to
\cite{strauss88}, \cite{waelbroeck89} and \cite{shibata01}. LV99 showed that the
linear growth of tearing modes is insufficient to obtain fast reconnection. The
new attack on the problem assumes that the non-linear growth of the islands due
to merging provides their growth rates at the large scales that are larger than
the direct growth of the tearing modes at those scales. This situation when the
non-linear growth is faster than the linear one is rather unusual and requires
further investigation.} as the process of enhancing reconnection
\citep{loureiro09,bhattacharjee09}. The 3D loops that should arise as a result
of this process should be able to accelerate energetic particles via the process
described in GL03. As tearing modes can happen in a collisional fluid, this may
potentially open another channel of reconnection in such fluid. The limitation
of this process is that the tearing mode reconnection should not be too fast as
this would present problems with explaining the accumulation of the flux prior
to the flare. At the same time the idea of tearing reconnection does not have
the natural valve of enhancing the reconnection speed, which is contrary to the
LV model where the degree of reconnection is determined by the level of
turbulence. Thus the periods of slow reconnection in LV99 model are ensured by
the low level of turbulence prior to the flare. We believe that tearing
reconnection can act to destabilize the laminar Sweet-Parker reconnection layer,
inducing turbulence. In this sense the new ideas on tearing instability may be
complementary to the LV99 model.

We note, however, that in most astrophysical situations one has to deal with the
{\it pre-existing} turbulence, which is the consequence of high Reynolds number
of the fluid. Such turbulence may modify or suppress instabilities, including
the tearing mode instability. We claim that it, by itself, induces fast
reconnection. We may claim that even if the astrophysical fluid is kept
initially laminar the fluid a thick outflow from the reconnection region caused
by tearing is expected to become turbulent. It was argued in \cite{lapenta12}
that this may be the cause of the reconnection explosions reported recently
within MHD simulations \citep{lapenta08,bettarini10}.

\section{Summary}

The results of this review can be briefly summarized as follows:\\
$\bullet$  Astrophysical turbulence is ubiquitous and it plays the essential
role for both acceleration of cosmic rays and magnetic reconnection.\\
$\bullet$  Advances in understanding of MHD turbulence help understanding both
cosmic ray acceleration and reconnection. Our understanding of
MHD turbulence in its different regimes is still incomplete, however.\\
$\bullet$ Turbulence makes magnetic reconnection fast. The LV99 model that
describes the process has been successfully tested numerically.\\
$\bullet$ Preexisting in the media density fluctuations generate magnetic field
while interacting with the shock precursor. This increases the efficiency of
cosmic ray acceleration and in some instances can even bootstrap the process of
acceleration.\\
$\bullet$ The turbulent second order Fermi acceleration is important for many
environments. The standard Quasi Linear Theory (QLT) is not accurate
to describe the process, but it can be improved; the improved theory corresponds
to numerical simulations.\\
$\bullet$ The LV99 model of reconnection entails the first order Fermi
acceleration of energetic particles. Numerical simulations support this
conclusion.\\
$\bullet$ A number of processes, e.g. the acceleration of anomalous cosmic rays
within the magnetosphere, acceleration of cosmic rays in magnetotail
as well as in galaxy clusters can be attributed to the acceleration caused by
magnetic reconnection.

\begin{acknowledgements}
A. Lazarian's research is supported by the NASA Grant NNX09AH78G, as well as the
support of the NSF Center for Magnetic Self-Organization.  The Humboldt Award at
the Universities of Cologne and Bochum, as well as Vilas Associate Award and the
hospitality of the International Institute of Physics (Brazil) are acknowledged.
G. Kowal's research is supported by FAPESP (Brazil) grant no. 2009/50053-8.  E.
de Gouveia Dal Pino also acknowledges partial supported from FAPESP (Brazil:
grant no. 2006/50654-3) and from CNPq (Brazil; grant no. 300083/94-7).  We thank
the anonymous referee for comments and suggestions that improved the paper.
\end{acknowledgements}

%
\bibliographystyle{aps-nameyear}
\bibliography{bibliography}
\nocite{*}

\end{document}